\documentstyle[12pt,epsf]{article}

\newcommand{\be}{\begin{equation}}
\newcommand{\ee}{\end{equation}}
\newcommand{\ba}{\begin{eqnarray}}
\newcommand{\ea}{\end{eqnarray}}
\newcommand{\non}{\nonumber\\}
\newcommand{\nom}{\nonumber}
\newcommand{\del}{\partial}
\newcommand{\dis}{\displaystyle}

\newcommand{\hp}{hep-th/}
\newcommand{\Kappa}{\mbox{\Large $\kappa$}}
\newcommand{\Eta}{\mbox{\large $\eta$}}
\newcommand{\kankaku}{\,\,\,\,\,\,}

\newcommand{\co}{{\cal O}}

\newcommand{\ket}[1]{{|#1 \rangle}}

\newcommand{\braket}[2]{{\langle #1|#2 \rangle}}
\newcommand{\ketbra}[1]{{\langle #1\rangle}}

\newcommand{\cleqn}{\setcounter{equation}{0}}

\begin{document}

%%%%%%%%%%%%%%%%%%% titlepage %%%%%%%%%%%%%%%%%%%%%
\begin{titlepage}
\nopagebreak
%%%%%%%%%%%%%%%%%%%%%%%%%%%%%%%%%%%%%%%%%%%%%%%%%%%%
\begin{flushright}
May 1998\hfill 
KUCP-0117\\
%hep-th/9805058
hep-th/9805058
\end{flushright}

\vfill
\begin{center}
{\LARGE Disk Amplitudes in Topological A-Model}

~

{\LARGE on Calabi-Yau Spaces and Fano Manifolds}

\vskip 12mm

{\large Katsuyuki~Sugiyama${}^{\sharp}$
}
\vskip 10mm
${}^{\sharp}$
{\sl Department of Fundamental Sciences}\\
{\sl Faculty of Integrated Human Studies, Kyoto University}\\
{\sl Yoshida-Nihon-Matsu cho, Sakyo-ku, Kyoto 606-8501, Japan}\\
%{\sl E-mail\,:\,sugiyama@phys.h.kyoto-u.ac.jp, FAX:+81-075-753-6804}\\
\end{center}
\vfill

%%%%%%%%%%%%%%%%%%%%%%%%%%%%%%%%%%%%%%%%%%%%%%%%
\begin{abstract}

We study topological A-model disk amplitudes with Calabi-Yau
%$d$-fold 
target spaces by mirror symmetries. 
This allows us to calculate holomorphic instantons of 
Riemann surfaces with boundaries that are mapped into susy cycles
in Calabi-Yau $d$-folds. Also we analyse disk amplitudes in
Fano manifold cases by considering fusion relations between A-model operators.
\end{abstract}

%PACS codes; 11.25.-w, 11.25.Mj, 11.25.Sq, 02.\\ 
%Keywords; Calabi-Yau, Fano Manifold, Mirror Symmetry,
%topological field theory, \\
%\mbox{}\hspace{21mm}non-linear sigma model,  holomorphic instanton
\vfill 
\noindent
\rule{7cm}{0.5pt}\par
\vskip 1mm
{\small \noindent ${}^{\sharp}$  
E-mail : sugiyama@phys.h.kyoto-u.ac.jp}
\end{titlepage}

%%%%%%%%%%%%%%%%%%%%%%%%%%%%%%%%%%%%%%%%%%%%%%%%%%%%
%   Introduction
%%%%%%%%%%%%%%%%%%%%%%%%%%%%%%%%%%%%%%%%%%%%%%%%%%%%
\section{Introduction}
D-branes \cite{Dbrane} play important roles to describe the solitonic modes
in string theory and could make clear 
dynamics in strong coupling regions. 
They are applied to many non-perturbative 
situations in string theory, such as Black Brane physics \cite{BH}
and susy Yang-Mills theories \cite{MQCD}. 
The calculations of degeneracy of BPS states shed light on profound
problems of Black Hole's entropy and give us the microscopic description of 
its properties. As an attempt to unify various string
dualities, the M-theory \cite{M} 
is expected to be formulated by taking D-branes as
fundamental objects. In some limit of large solitonic charges
in supergravity theories (SUGRA), the dynamics of the system are effectively 
controlled by large $N$ super Yang-Mills theories associated to
D-brane world volumes \cite{largeN}. 
It might be possible to study
strong coupling regions of SUGRA by dimensionally reduced susy Yang-Mills
actions because open string sectors govern the dynamics in short distances.
Also MQCD contributes to our understanding of dynamics and moduli spaces
of SQCD \cite{SW} schematically from geometrical configurations of branes.

The physical observables of D-brane's effective theories have
dependences on moduli of compactified strings or wrapping D-branes.
We expect that properties of compactified internal spaces essentially
control non-perturbative effects in susy Yang-Mills theories.
One of various remarkable successes is calculation of 
prepotential in susy Yang-Mills in a geometric
engineering technique \cite{engineer}
based on analyses of moduli spaces in compactified string.
It determines non-perturbative parts of the effective theory completely
and leads us to open some geometrical methods in understanding
the susy Yang-Mills and string effective theories.

String inspired effective theories in various dimensions 
give a hope that some non-renormalizable field theories might be
formulated in non-perturbative ways.
These Yang-Mills theories are related to fluctuations of
open strings and are closely connected with D-branes.

In this paper, 
we focus on the type II superstring compactified on Calabi-Yau
manifold and study disk amplitudes of its topological sector
to study properties of moduli spaces.
Strings can couple to the D-branes through boundary of the disk.
The amplitudes allow us to obtain
world sheet instanton properties in the open string sector. 
The complex structure moduli sector is described by topological
B-model and receive no world sheet quantum corrections. 
One can analyse the B-model moduli by period integrals .
On the other hand, the K{\"a}hler structure have (world sheet )
instanton corrections and we study them by topological A-model.
For Calabi-Yau cases, there are 
mirror symmetries \cite{mirror0,mirror1,mirror2}
between these two models.
Transforming former information to that in A-model side,
we can obtain instanton corrections in K{\"a}hler moduli spaces.

Motivated with this consideration,
we intend to examine the amplitudes 
by means of mirror techniques. 
Some considerations
are given in \cite{Oo} for disk amplitudes 
for Calabi-Yau cases.
But there still remains several uncertain points.

Our aim is to develop a concrete method to calculate 
disk amplitudes for the A-model.
We present disk amplitudes for $d$ dimensional 
Calabi-Yau manifolds explicitly in order to clarify
relations with three point couplings on sphere 
and generalize the calculation to 
Fano manifold cases by studying fusion products of operators.

The paper is organized as follows. In section2, we review an $N=2$
supersymmetric non-linear sigma model and its possible two
(A-, B-type) boundary conditions. We also explain the results in ref.
\cite{Oo} about boundary states and associated disk amplitudes
in topological A-, B-models in order to fix notations for
later sections. In section 3, we consider fusion relations of
A-model operators $\co^{(l)}$'s and their disk amplitudes $c^{(l)}$.
Taking derivatives of the amplitude with respect to a moduli parameter
is equivalent to insertion of an extra associated operator in the
correlator. Operator products are represented linearly and lead us
to a closed set of differential equations about the $c^{(l)}$'s.
In general it is difficult to evaluate quantum corrections in $c^{(l)}$'s
directly in the A-model side. But the corresponding amplitudes
in the B-model case are period integrals themselves.
By using mirror symmetries to evaluate fusion couplings 
for Calabi-Yau cases, we can obtain
the $c^{(l)}$'s exactly in the A-model as a set of solutions of the above
differential equations. As concrete examples, we analyse
two $d$-fold cases and argue on geometrical meanings of
expansion coefficients with respect to K{\"a}hler parameters 
in the amplitude.

Although this result is very satisfactory, we should still keep in mind
the fact that this mirror technique requires the existence of B-model
and the story is necessarily restricted to Calabi-Yau cases 
for consistency of the model.
If one can discusses the analyses within the A-model side only, 
there still exist many K{\"a}hler manifolds which can be used as
target spaces in the model.
For examples, in the cases of K{\"a}hler manifolds with positive first
Chern classes (Fano manifolds), the B-model cannot be defined consistently
and the trick of mirror symmetries cannot be applied.

To defeat this difficulty, we discuss, in section 4, relations between the 
amplitudes $c^{(l)}$'s in Fano manifold cases and the three point 
couplings $\{ \Kappa_m \}$ in the tree level of closed string.
There we obtain general formulae for $c^{(l)}$'s as some linear
combinations of multiple integrals.
Also it is known that the closed string amplitudes $\{ \Kappa_m \}$
can be calculated recursively \cite{rational1, rational2, rational3}
by associativity relations of operators for Fano manifolds.
In our case these associativities appear naturally as integrable conditions
for a set of differential equations satisfied by the 
$c^{(l)}$'s in open string theory.
Applying these recipes developed in the section, we evaluate the open 
string amplitudes with disklike topology for 
projective spaces $CP^N$ $(N=1,2,3)$, a Grassmann manifold
$Gr(2,4)$ and a degree $3$ hypersurface $M_{4,3}$ in $CP^4$ 
concretely.
In these Fano cases, 
to avoid complexity and to make the analyses clear,
we restrict ourselves to switch on only marginal 
K{\"a}hler parameters and write down their detailed results.

Section 5 is devoted to conclusions and comments. 
In appendix A, we collect several results of the expansion coefficients
for Calabi-Yau cases in lower dimensions.
We summarize calculations for a Fermat type Calabi-Yau $d$-fold with
one K{\"a}hler modulus in appendix B.

~

%%%%%%%%%%%%%%%%%%%%%%%%%%%%%%%%%%%%%%%%%%%%%%%%%%%%%%%%%%%%%
%        Section 2
%%%%%%%%%%%%%%%%%%%%%%%%%%%%%%%%%%%%%%%%%%%%%%%%%%%%%%%%%%%%%

\section{Topological Sigma Models}\label{sigma model}
\cleqn

In this section, we shall review non-linear sigma models 
\cite{mirror2} and their possible boundary conditions. 
Also we explain relations between period matrices with disk amplitudes
in topological sectors in these models.

\subsection{Boundary Conditions}
We shall analyse the moduli spaces in the type II
superstring on Calabi-Yau manifold M. 
Let us consider two dimensional $N=2$ susy sigma model with
Calabi-Yau target space. 
 Bosonic fields $X^i$, $X^{\bar{\imath}}$ in this model are maps from
Riemann surface $\Sigma$ to the target manifold M. 
The system contains four world sheet fermions 
$\psi_L^i$, $\psi_L^{\bar{\imath}}$, 
$\psi_R^i$ and $\psi_R^{\bar{\imath}}$.
The subscripts $L$, $R$ represent respectively 
left-, right-movers on the world sheet.
The superscripts $i$, $\bar{\imath}$ are
K{\"a}hler coordinate indices of target manifold with
K{\"a}hler metric $g_{i\bar{\jmath}}$.
%With these fields, the action is written
%\ba
%S&=& \int_{\Sigma} \Biggl[ \,\,\frac{1}{2} g_{i\bar{\jmath}} 
%(\del_zX^i\del_{\bar{z}}X^{\bar{\jmath}}
%+\del_{\bar{z}}X^i\del_{z}X^{\bar{\jmath}}) \nom\\
%&& + \sqrt{-1} g_{i\bar{\jmath}} \psi^{\bar{\jmath}}_L D_{\bar{z}} \psi^i_L 
%+\sqrt{-1} g_{i\bar{\jmath}} \psi^{\bar{\jmath}}_R D_z \psi^i_R
%+ R_{i\bar{\jmath}k\bar{l}} \psi^i_L \psi^{\bar{\jmath}}_L
%\psi^k_R \psi^{\bar{l}}_R \,\,\Biggr] \,\,\,.\nom
%\ea
%Here covariant derivatives $D_z$, $D_{\bar{z}}$ are defined by using a
%pull-back of a Levi-Civita connection ${\Gamma^i}_{jk}$ of M into $\Sigma$
%\ba
%&& D_z \psi^i_R := 
%\frac{\del}{\del z} \psi^i_R +\frac{\del X^j }{\del z}
%{\Gamma^i}_{jk}  \psi^k_R \,\,\,,\nom \\
%&& D_{\bar{z}} \psi^i_L := 
%\frac{\del}{\del \bar{z}} \psi^i_L +\frac{\del X^j }{\del \bar{z}}
%{\Gamma^i}_{jk}  \psi^k_L \,\,\,.\nom
%\ea
In an infrared fixed point, there are $N=2$ superconformal
symmetries in this system and we
express a set of super conformal currents as
\ba
( T_L ,G^{\pm}_L , J_L )\,\,\,,\,\,\,
( T_R ,G^{\pm}_R , J_R )\,\,\,.\nom
\ea
The $T_{L,R}$ are energy momentum tensors
and $G^{\pm}_{L,R}$ are their super partners.
The $U(1)$ currents $J_{L,R}$ are essentially pull-backs
of a K{\"a}hler form $K := k_{\mu \nu} d{X^{\mu}}\wedge d{X^{\nu}} $
\footnote{The indices $\mu$, $\nu$ are real coordinates of the
target K{\"a}hler manifold.}
of the target manifold to Riemann surface,
\ba
 J_L = k_{\mu\nu} \psi^{\mu}_L \psi^{\nu}_L \,\,\,,\,\,\,
 J_R = k_{\mu\nu} \psi^{\mu}_R \psi^{\nu}_R \,\,\,.\label{J}
\ea

When we consider open strings, these left- and right-part energy 
momentum tensors $T_{L,R}$
are not independent and related at boundary, $ T_L = T_R $.
For super stress tensors $G^{\pm}_{L,R}$ and $U(1)$ currents
$J_{L,R}$, there are two types of boundary conditions \cite{Oo}
\footnote{We use the terms ``A-type'', ``B-type'' for boundary conditions
oppositely to expressions in ref.\cite{Oo}. 
It is convenient to use
this convention in discussing disk amplitudes from the point of view of
topological A-, B-models.}
in terms of a closed string channel
\ba
&&\mbox{A-type}\,;\, G^{-}_L =\pm i G^{-}_R \,\,,\,\,
J_L =- J_R \,\,\,.\label{A}\\
&&\mbox{B-type}\,;\, G^{+}_L =\pm i G^{-}_R \,\,,\,\,
J_L =+ J_R \,\,\,,\label{B}
\ea
In the A-type conditions, the left and right $U(1)$ charges have
same absolute values but with opposite signs on the boundary.
On the other hand, the two charges are identified with same sign
for the B-type boundary cases.
These equations (\ref{A}), (\ref{B}) 
can be solved and give us relations for fields $X$
and $\psi$
\ba
&& \del_z X^{\mu} = {R^{\mu}}_{\nu} \del_{\bar{z}} X^{\nu}
\,\,\,,\,\,\, \psi^{\mu}_L =\pm i {R^{\mu}}_{\nu}  \psi^{\nu}_R \,\,\,,\non
&& g_{\mu\nu} {R^{\mu}}_{\rho} {R^{\nu}}_{\sigma} = g_{\rho\sigma}
\,\,\,.\label{R}
\ea
The matrix $R$ connects left- and right-movers on the boundary
and its eigenvalues take $\pm 1$.
It essentially contains information about open string boundary.
We impose  Neumann boundary conditions on
coordinates $X^{\mu}$
with $+1$ eigenvalue indices.
Vectors with $-1$ eigenvalue satisfy Dirichlet conditions

Let us apply the Eqs.(\ref{R}) to a homology $p$-cycle $\gamma$
of the Calabi-Yau $d$-fold. Tangential directions
of $\gamma$ have free boundary conditions and are
described by $p$ dimensional coordinates $\{ y^A \}$ $(A=1,2,\cdots ,p)$. 
Normal directions to $\gamma$ are parametrized by 
$\{ y^a \}$ $(a=p+1,p+2, \cdots ,2d)$ and have fixed boundary conditions.
As a first case we take A-type conditions.
Then the K{\"a}hler form must satisfy
a relation 
\ba
 k_{\mu\nu} {R^{\mu}}_{\rho} {R^{\nu}}_{\sigma}
= -  k_{\rho\sigma}\,\,\,.\label{kA}
\ea
 When one decomposes the $ k_{\mu\nu} $ in terms of
the $y^A$ and  $y^a$, the block off-diagonal parts 
$ k_{aB} $ and  $ k_{Ab}$ are
the only non-vanishing components that satisfy Eq.(\ref{kA}).
The non-degeneracy of the $ k_{\mu\nu} $ leads us to a restriction on $p$
as $p=d$. Thus the homology cycle with A-type conditions are
real $d$ dimensional submanifold.

Next the B-type conditions impose a relation on K{\"a}hler form
\ba
 k_{\mu\nu} {R^{\mu}}_{\rho} {R^{\nu}}_{\sigma}
= +  k_{\rho\sigma}\,\,\,.\label{kB}
\ea
Non-zero parts of $ k_{\mu\nu}$ with Eq.(\ref{kB})
are block diagonal elements and the $p$ turns out to be even 
by the non-degeneracy of $k_{\mu\nu}$.
The corresponding cycle is able to have a complex structure derived from its
K{\"a}hler form.

Mirror transformation changes the sign of the right $U(1)$ current
and exchanges the A-type and B-type conditions.
Then the transformation 
relates $d$-dimensional cycles of Calabi-Yau $d$-fold M to
 even dimensional ones of a partner manifold $W$.

\subsection{Disk Amplitudes}

In this subsection, we study boundary states \cite{Bstate1, Bstate2}
for topological sigma models with Calabi-Yau target spaces M.
The boundary state $\ket{B}$ is respectively defined by equations
for the A-type and B-type cases in the closed string channel \cite{Oo}
\ba
&&\mbox{A-type}\,;\, ( G^{-}_L \mp i G^{-}_R )\ket{B}=0\,\,\,,
\,\,\,( J_L + J_R )\ket{B}=0\,\,\,,\label{Abound}\\
&&\mbox{B-type}\,;\, ( G^{+}_L \mp i  G^{-}_R )\ket{B}=0\,\,\,,
\,\,\,( J_L - J_R )\ket{B}=0\,\,\,,\label{Bbound}
\ea
The state is a sort of source of closed strings and emits various fields.
A disk amplitude is characterized by the state $\ket{B}$
and inserted operators on Riemann surface $\Sigma$.
For topological models, the boundaries of $\Sigma$'s
are mapped into homology cycles $\gamma \subset M$ and inserted operators
are topological observables associated with some cohomology elements.
We will show details of topological models.

In order to obtain topological versions of the $N=2$ non-linear
sigma model, let us consider alternations (twistings) 
of bundles on which fermions 
take values. We change spins of fermions by an amount depending on
their $U(1)$ charges. As a result fermions take values not on
spin bundles but on (anti-)canonical bundles.
We are interested in two cases (A-, B-models).
In the A-model case, the fermions $\psi^i_L$ and $\psi^{\bar{\imath}}_R$ 
become spin zero fields on the Riemann surface $\Sigma$. The remaining
fields  $\psi^{\bar{\imath}}_L$ ($\psi^i_R$) come to be respectively
holomorphic (anti-holomorphic) one forms on $\Sigma$.
When we consider a B-type twisting, each right mover $\psi^i_R$,
$\psi^{\bar{\imath}}_R$ has the same spin as that in the A-model. But
the roles of $\psi^i_L$, $\psi^{\bar{\imath}}_L$ are exchanged in compared to
the A-type case. We summarize spin zero fields on $\Sigma$ for 
these topological models and their associated super charges $Q$'s
\ba
\begin{array}{lccc}
               &   & \mbox{spin zero fields} & \mbox{supercharges} \\
\mbox{A-model} & ; & \psi^i_L \,\,,\,\, \psi^{\bar{\imath}}_R &
Q^{+}_L \,\,,\,\, Q^{-}_R  \\
\mbox{B-model} & ; & \psi^{\bar{\imath}}_L \,\,,\,\, \psi^{\bar{\imath}}_R &
Q^{-}_L \,\,,\,\, Q^{-}_R 
\end{array}\,\,\,\,.\nom
\ea
In the A-model, local observables are defined as BRST cohomologies 
of a BRST charge $ Q_A := Q^{+}_L + Q^{-}_R $.
We can associate an arbitrary de~Rham cohomology element with a
physical observable in the A-model. 
Similarly B-model observables are associated to Dolbeault cohomology
elements induced by a BRST charge
$ Q_B := Q^{-}_L + Q^{-}_R $.
Topological sectors in the disk amplitudes are described respectively
by A-, B-models depending on A-, B-type boundary conditions.

Let us first consider the B-type boundary conditions.
The requirement for $U(1)$ currents $ J_L = J_R $ at the boundary
means the identification of left and right $U(1)$ charges 
$q_L = q_R$.
Physical observables of B-model with charges $(q,q)$ 
correspond to 
middle cohomology elements $ v_q \in \mbox{H}^{q,d-q}$(M)
and we write the associated operators as $\tilde{\phi}_q$'s.
Disk amplitudes with one inserted operator $\tilde{\phi}_q$ is
calculated as an inner product $\ket{\tilde{\phi}_q}$ and $\ket{B}$
\ba
\tilde{c}_q :=\braket{\tilde{\phi}_q}{B}\,\,\,.\nom
\ea
These B-model amplitudes are known to be independent of K{\"a}hler moduli and
we can take large volume limit. Then classical calculation is
exact and there are no quantum corrections.
The non-zero contributions in the topological sector
from the $\ket{B}$ are described by 
$d$ dimensional homology cycle $\tilde{\gamma}$ 
because of charge conservation.
Poicar{\'e} dual of this $\tilde{\gamma}$ 
is some middle cohomology element and
one can expand it by a set of basis $\{ v_q \}$.
Considering a state $\ket{v_q}_B$ for each $ v_q $
with $\ket{B}_{top}=\sum_q \ket{v_q}_B {}_B\braket{v_q}{B} $,
we define a part of boundary state $\ket{\tilde{\gamma}}$
associated with the cycle $\tilde{\gamma}$ as some linear combination of
$ \ket{v_q }_B $'s.
Then disk amplitudes are calculated
as period integrals 
for Calabi-Yau $d$-fold M
\ba
&& \tilde{c}_q (\tilde{\gamma})=\braket{\tilde{\phi}_q}{\tilde{\gamma}}
= \int_{\tilde{\gamma}} v_q\,\,\,\,,\nom \\
&& v_q \in \mbox{H}^{q,d-q} (\mbox{M})
\,\,,\,\,\tilde{\gamma}\in \mbox{H}_d (\mbox{M})\,\,\,.\label{Bamp}
\ea
When we choose a canonical homology basis $\{\tilde{\gamma}_k \}$ for 
$\mbox{H}_d$(M),
the amplitudes are collected into a matrix $\tilde{\Pi}$ 
\ba
{\Pi}  =
%\left( 
%\begin{array}{ccc}
\bordermatrix{
 & & {\scriptstyle \tilde{\gamma}_m } & \cr
 &  &  & \cr
{\scriptstyle \tilde{\phi}_l} &  &  \tilde{c}_l (\tilde{\gamma}_m) & \cr
 &  &  & }
%\end{array}
%\right)
\,\,\,.\label{Bperiod}
\ea
This is the usual period matrix of complex moduli spaces of Calabi-Yau
manifolds.

Next we shall turn to the A-type boundary conditions.
Because of the identification of $U(1)$ charges $ q_L =- q_R \equiv q$ 
on the boundary,
physical operators $\co^{(l)}$'s 
for this A-model are associated with even dimensional cohomology elements
$e_l \in \mbox{H}^{l,l}$. More precisely the operators 
are elements of cohomology of moduli space of holomorphic maps 
and contain information about world sheet instantons.
In this model, disk amplitude ${c}_l$ 
is defined for each inserted operator $\co^{(l)}$
\ba
{c}_l =\braket{ \co^{(l)} }{B} \,\,\,.
\ea
That is independent of complex structure moduli 
but depends on K{\"a}hler moduli of M.
When we introduce boundary state 
$\ket{{\gamma}}$
associated with even dimensional 
homology cycle ${\gamma} \in \oplus \mbox{H}_{2l} $,
only non-vanishing contributions to the $c_l$ come from
$\ket{{\gamma}}$ in $\ket{B}$.
We put together 
$ {c}_l ({\gamma_m})= \braket{ \co^{(l)} }{ {\gamma}_m }$
into one matrix ${\Pi}$ by using a set of canonical basis 
$ {\gamma}_m $ in $ \oplus \mbox{H}_{2k} $, 
\ba
{\Pi}=
%\left(
%\begin{array}{ccc}
\bordermatrix{
 & &{\scriptstyle {\gamma}_m } & \cr
 & & & \cr
{\scriptstyle \co^{(l)} } 
& & \braket{ \co^{(l)} }{ {\gamma}_m } & \cr
 & & & }
%\end{array}
%\right) 
\,\,\,.\nom
\ea
Components of this ${\Pi}$ contain world sheet instanton corrections.
We will analyse these amplitudes in the next section by
using mirror symmetry.

\section{Disk Amplitudes of A-Model}

In this section we investigate disk amplitudes in the A-model
for Calabi-Yau target cases.
We analyse fusion structures of A-model operators and
study relations between disk amplitudes and fusion couplings.

\subsection{Calabi-Yau $3$-fold}\label{3fold}

Let M be an arbitrary Calabi-Yau $3$-fold with a Hodge number
$h_{11}=\varrho$. We focus on a set of cohomology elements 
$\{  e^{(0)} , e^{(1)}_i , e^{(2)}_j , e^{(3)} \}$ $(i, j= 1,2,\cdots , 
\varrho)$
in the vertical part
$ e^{(l)} \in \mbox{H}^{l,l}(\mbox{M},{\bf Z})$ $(l=0,1,2,3)$ 
and write an A-model operator associated with $e^{(l)}$ as
${\cal O}^{(l)}$.
We choose a canonical basis of the homology cycles 
\[
\gamma= \{ \alpha^0 , \alpha^{i} , \beta_{i} , \beta_0 \}
\in \oplus_{\ell} \mbox{H}_{2\ell},
\]
and consider disk amplitudes $c^{(l)} (\gamma)$ associated with
operators $\co^{(l)}$ $(l=0,1,2,3)$.

Correlators for A-model is defined as
a path integral over holomorphic maps from $\Sigma$ to M
\ba
\ketbra{\cdots}:=\int_{\Sigma} {\cal D}X\,
e^{\sum_{l\geq 0} \sum_{i_l }
t^{(l)}_{i_l}  \co^{(l)}_{i_l}}
\cdots %\Big|_{t^{(l)}_{i_l}=0\,\,(l\neq 1)} 
\,\,\,.
\ea
(We use an abbreviated notation about zero modes of fermions in the measure.
In section (\ref{geom}), we write down precise forms
about correlators including fermion zero modes.)
The set of parameters $\{t^{(\ell)}\}$ 
is a background source and
each $t^{(\ell)}_{i_{\ell}}$ 
\mbox{}\footnote{The superscript $(\ell)$ for 
the $t^{(\ell)}_{i_{\ell}}$ represents a degree $2 \ell$ of a cohomology
element $e^{(\ell)}_{i_{\ell}} \in \mbox{H}^{2\ell}$(M) associated with
an A-model operator $\co^{(\ell)}_{i_{\ell}}$.
The subscript $i_{\ell}$ of the letter $t^{(\ell)}_{i_{\ell}}$
labels each individual operator with a fixed degree $2 \ell$.}
couples with an associated operator $\co^{(\ell)}_{i_{\ell}}$
with a definite degree $2\ell$ of the BRST cohomology.

In considering Calabi-Yau cases we switch off all background sources
except for marginal ones and we write these non-vanishing
(K{\"a}hler) parameters $t^i :=t^{(1)}_i$ in an abbreviated form.
Operator products are represented linearly in the correlator 
$\ketbra{ \co^{(l)} \co^{(m)} \cdots}$
\ba
&&\co^{(1)}_i \co^{(0)} = \co^{(1)}_i \,\,\,,\nom \\
&&\co^{(1)}_i \co^{(1)}_j = {\Kappa_{ij}}^{k} \co^{(2)}_k \,\,\,,\nom \\
&&\co^{(1)}_i \co^{(2)}_j = {\Eta_{ij}} \co^{(3)} \,\,\,,\nom \\
&&\co^{(1)}_i \co^{(3)} = 0\,\,\,.\label{3fusion}
\ea
The $\Eta_{ij}$ is topological metric $\Eta_{ij} :=
\langle {\co^{(1)}_i \co^{(2)}_j } \rangle$ and is independent of moduli
parameters.
The ${\Kappa_{ij}}^k$ is three point function on sphere
and we regard it as an $(i,j)$ component of a matrix $\Kappa_i$.
When we define an inverse matrix $\Kappa^{-1}_i$ of the
$\Kappa_i$, the $\co^{(2)}$ can be expressed as an operator product of two
$\co^{(1)}$s
\ba
\co^{(2)}_l = \sum_j {(\kappa^{-1}_i)_l}^j \co^{(1)}_i \co^{(1)}_j \,\,\,.
\label{op2}
\ea
Let us first pick the amplitude $c^{(0)}(\gamma) =\braket{\co^{(0)}}{\gamma}$
and take its derivative with respect to a K{\"a}hler parameter $t^i$.
The action of this derivative on the amplitude is 
insertion of an operator $\co^{(1)}_i$ in the correlator
\ba
\del_{t^i} c^{(0)} (\gamma) = \braket{ \co^{(1)}_i \co^{(0)} }{\gamma}\,\,\,.
\label{c1}
\ea
Further the right hand side in Eq.(\ref{c1}) turns out to be  
$c^{(1)}_i (\gamma):=\braket{\co^{(1)}_i}{\gamma}$
because of fusion relations Eq.(\ref{3fusion}).
In applying the same method as $c^{(2)}_l (\gamma)$ case, 
we can relate the $c^{(2)}_l$ to $c^{(1)}_i$
\ba
\del_{t^j} c^{(1)}_i (\gamma) &=& \del_{t^i} \del_{t^j} c^{(0)} (\gamma)\nom \\
&=& \braket{\co^{(1)}_i \co^{(1)}_j}{\gamma} \nom \\
&=& \sum_l {(\Kappa_i)_j}^l \braket{\co^{(2)}_l }{\gamma} \nom  \\
&=& \sum_l {(\Kappa_i)_j}^l c^{(2)}_l (\gamma) \,\,\,.\nom 
\ea
That is, the $c^{(2)}$ is written as a
linear combination of second order derivatives
of $c^{(0)}$
\ba
c^{(2)}_l (\gamma) =\sum_j {(\Kappa^{-1}_i)_l}^j \del_{t^i}\del_{t^j}
c^{(0)} (\gamma) \,\,\,.\label{c2}
\ea
Also a derivative of l.h.s. of Eq.(\ref{c2}) is associated to 
$c^{(3)}(\gamma)$
\ba
%\del_{t^i} c^{(2)}_j (\gamma)&=& \braket{\co^{(1)}_i \co^{(2)}_j}
%{\gamma}\nom \\
%&=& \Eta_{ij} \braket{\co^{(3)}}{\gamma}\nom \\
%&=& \Eta_{ij} c^{(3)}\,\,\,,\nom \\
%{i.e.} \,\,\,\, 
c^{(3)} (\gamma) &=&\sum_i \Eta^{mi} \braket{\co^{(1)}_i \co^{(2)}_m }{\gamma}
\nom \\
&=& \sum_i \Eta^{mi} \del_{t^i} c^{(2)}_m 
(\gamma) \nom \\
&=& \sum_{i,j} \Eta^{mi} \del_{t^i} {(\Kappa^{-1}_l)_m}^j \del_{t^l} \del_{t^j}
 c^{(0)} (\gamma)\,\,\,.
\ea
Thus all disk amplitudes can be calculated if we know the $c^{(0)}(\gamma)$
for an arbitrary homology cycle $\gamma$. 
But the $\gamma$ is expressed as some linear combination of canonical 
basis $( \alpha^0 ,\alpha^i , \beta_j ,\beta_0 )$ and
all we have to do is to evaluate the functions
$c^{(0)}(\alpha)$'s and $c^{(0)}(\beta)$'s.
It is difficult to calculate these disk amplitudes in A-model directly.
But the corresponding ones in the B-model side are 
period integrals themselves and we can estimate them classically.
Furthermore there are mirror maps between K{\"a}hler moduli parameters
$\{ t^i \}$ and complex moduli ones $\{ \psi_j \}$ in some mirror
partner W.
Using these mirror maps $t^i = t^i (\psi_j)$, we obtain the
$c^{(0)}$'s in A-model associated with M
\ba
&& c^{(0)} (\alpha^0)=1 \,\,\,,\,\,\,
c^{(0)} (\alpha^i)= t^i \,\,\,,\,\,\,\nom \\
&& c^{(0)} (\beta_i)= \del_i F \,\,\,,\,\,\,
c^{(0)} (\beta_0)= t^i \del_i F-2F \,\,\,,\,\,\,\nom \\
&& \del_i :=\frac{\del}{\del t^i}\,\,\,.
\ea
Here the $F$ is a prepotential of the K{\"a}hler moduli space
and is expanded in a power series
with respect to 
instanton multi-degrees $\{m\}:=\{ m_1 , m_2 , \cdots m_{\varrho} \}$
\ba
&&F=\frac{1}{6} \Kappa^{(0)}_{ijk} t^i t^j t^k +a
+f\,\,\,,\nom \\
&&f:=\frac{1}{(2\pi i)^3}\sum_{\{ m \}}N_{\{ m \}} \mbox{Li}_3 (q^{\{m \}})
\,\,\,,\,\,\, q^{\{ m \}} := 
\exp\left( 2\pi i \sum_{i=1}^{\varrho} m_i t^i \right)
\,\,\,,\nom \\
&& \Kappa^{(0)}_{ijk} := \int_M  e^{(1)}_i \wedge  e^{(1)}_j \wedge  e^{(1)}_k
\,\,\,\,\,\,\,\,\,(\,\, e^{(1)}_i \in \mbox{H}^2 
(\mbox{M};{\bf Z})\,\,\,)\,\,\,.
\ea
The ``$a$'' is a contribution from sigma model loops 
and is related to a Euler number $\chi$ of M, 
$a=-\frac{i}{2}\frac{\zeta (3)}{(2\pi )^3}\chi$.
We collect all these results into one matrix ${\Pi}$
\begin{eqnarray}
{\Pi}&=&
\bordermatrix{
 & \, \alpha^0 & \alpha^{i} & \beta_{j} & \beta_0  \cr
{\scriptstyle {\cal O}^{(0)}}
  & \, 1 & t^i     & \partial_j F & t^i \partial_i F-2F  \cr
{\scriptstyle {\cal O}^{(1)}_{l} }
&  \, 0 & {\delta_{l}}^i  & \partial_l \partial_j F & 
t^i \partial_l \partial_i F- \partial_l F  \cr
{\scriptstyle {\cal O}^{(2)}_{m}} &\, 0 & 0  & \eta_{m j} & \eta_{mn} t^n  \cr
{\scriptstyle {\cal O}^{(3)}} & \, 0 & 0  & 0 & 1
}\nom \\
&=&
\bordermatrix{
  &   &  {\scriptstyle k}  &  {\scriptstyle n}  &   \cr
  & 1 &  0  & \del_n f & -2f-2a \cr
{\scriptstyle l} & 0 & {\delta_l}^k & \del_n \del_l f & -\del_l f \cr
{\scriptstyle m} & 0 & 0 & \eta_{mn} & 0 \cr
  & 0 & 0 & 0 & 1 
}\nom \\
&& \times \,\,
\bordermatrix{
  &   &  {\scriptstyle i}  &  {\scriptstyle j}  &   \cr
  & 1 & t^i & \frac{1}{2}\kappa^{(0)}_{j r s} t^r t^s &  
\frac{1}{6}\kappa^{(0)}_{u r s} t^u t^r t^s \cr
{\scriptstyle k} & 0 & {\delta_k}^i & \kappa^{(0)}_{j k r} t^r & 
\frac{1}{2}\kappa^{(0)}_{k r s} t^r t^s \cr
{\scriptstyle n} & 0 & 0 & {\delta^n}_j & t^n \cr
  & 0 & 0 & 0 & 1
}\label{disk3}
\end{eqnarray}
Also we can rewrite 2nd matrix as
\ba
\exp({\cal N}({\bf t}))&=&\left(
\matrix{
    1 & t^i & \frac{1}{2}\kappa^{(0)}_{j r s} t^r t^s &  
\frac{1}{6}\kappa^{(0)}_{u r s} t^u t^r t^s \cr
 0 & {\delta_k}^i & \kappa^{(0)}_{j k r} t^r & 
\frac{1}{2}\kappa^{(0)}_{k r s} t^r t^s \cr
 0 & 0 & {\delta^n}_j & t^n \cr
 0 & 0 & 0 & 1
}\right)\,\,\,,\nom \\
{\cal N}({\bf t})&:=&
\left(
\matrix{
0 & {\bf t}^t & 0 & 0 \cr
0 & 0 & {\bf t}\cdot {\Kappa}^{(0)}& 0 \cr
0 & 0 & 0 & {\bf t} \cr
0 & 0 & 0 & 0
}\right)\,\,\,,\\
({{\bf t}\cdot {\Kappa}^{(0)}})_{jk}
&:=&\sum_i t^i \Kappa^{(0)}_{ijk} \,\,\,.\nom 
\ea
The 1st matrix in the second line in Eq.(\ref{disk3}) 
is invariant under a monodromy transformation
${\bf t}\rightarrow {\bf t}+{\bf m}$ with ${}^{\exists}{\bf m}\in 
{\bf Z}^{\otimes \varrho}$.
But the 2nd matrix $\exp({\cal N}({\bf t}))$ changes into 
$\exp({\cal N}({\bf t})) \cdot\exp({\cal N}({\bf m})) $ and
has a non-trivial monodromy property. It reflects the fact that
the homology cycles belong to mixed combinations 
$ \oplus_{m=0}^l \mbox{H}_{2m}$
of homology groups with
different dimensions.
Concretely the set of homology basis $\{\alpha^0 ,\alpha^i ,
\beta_i , \beta_0\}$ receives an effect under this transformation
and linearly changes into a form
\ba
(\alpha^0 \,\,\,\alpha^i \,\,\,
\beta_i \,\,\, \beta_0) \rightarrow
(\alpha^0 \,\,\,\alpha^i \,\,\,
\beta_i \,\,\, \beta_0)\cdot \exp \left({\cal N}({\bf m})\right)\,\,\,.\nom
\ea
In the D-brane language,
these homology cycles are wrapped by D-branes with appropriate
dimensions. The $0$-cycle $\alpha_0$ is invariant under the above
monodromy transformations and
corresponds to a D0-brane in
the IIA string theory. When we turn to the mirror W,
the corresponding cycle is a unique fundamental 3-cycle $\Gamma$
in $\mbox{H}_3$(W), which is dual to an element of $\mbox{H}^{3,0}$(W).
When we consider a patch $X_5 =1$ in the Fermat type case
and take a set of variables
$(X_1 ,X_2 ,X_3)$ as independent coordinates of the W,
the cycle $\Gamma$ is obtained
\ba
\Gamma:=\{(X_1 ,X_2 ,X_3)\in {\bf C}^3 \,;\, |X_i|=1 \,\,(i=1,2,3)\}\,\,\,
.\nom
\ea
It is a real 3-dimensional torus itself and 
its structure is universal for all
the toric cases. This fact illustrates a conjecture
that an arbitrary Calabi-Yau $3$-fold is realized
as a torus fibered space over some 3-dimensional special
Lagrangian manifold \cite{SYZ}.
Next we consider other cycles.
The $\alpha^i$, $\beta_j$, $\beta_0$ are 
wrapped respectively by D2-, D4-, D6-branes.
The monodromy transformation mixes all these cycles
with different dimensions.
Also when one shifts a complexified K{\"a}hler parameter $t^i$ 
by a constant real number,
an NS-NS 2-form B-field changes into
$B\rightarrow B+\zeta$ for
some $\zeta \in \mbox{H}^2$(M).
It implies an existence of
Chern-Simons term 
$C\wedge \exp(B)$
for even form 
Ramond-Ramond (RR) fields $C:=\sum_{m\geq 0} C^{(2m)}$'s in IIA string.

\begin{figure}%\label{fig1}
     \epsfxsize=10cm
     \centerline{\epsfbox{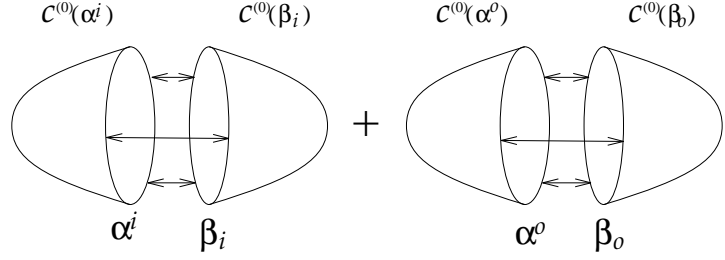}}
     \epsfxsize=10cm
     \caption{The prepotential $F$ is written as a sum of products of
      disk amplitudes associated with canonical basis $\{\alpha\}$ 
      and $\{\beta\}$. A pair of canonical cycles is glued on their disk
      boundaries and constructs a sphere.}\label{disk}
 \end{figure}

On the other hand, 
quantum corrections in the disk amplitudes are essentially encoded
in the function $f$. But $f$ is originally a generating function of instanton
numbers, more precisely Euler numbers of
instanton moduli spaces with a fixed degree,
in the closed string tree amplitudes.
Why do they appear here?
Recall the prepotential formula obtained in \cite{Hoso}.
\ba
F=\frac{1}{2}(\omega^{(1)\,i} \omega^{(2)}_i 
- \omega^{(0)} \omega^{(3)} )\,\,\,.\nom
\ea
Here the $\omega^{(0)}$, $\omega^{(1)\,i}$, $\omega^{(2)}_j$, $\omega^{(3)}$
are normalized periods with $\omega^{(0)} =1$ in mirror side.
In our results, these $\omega$'s are interpreted as
disk amplitudes $c^{(0)}$'s when they are rewritten in terms of variables
$\{ t^i \}$,
\ba
&&\omega^{(0)} =c^{(0)}(\alpha^0)\,\,\,,\,\,\,
\omega^{(1)\,i} =c^{(0)}(\alpha^i)\,\,\,,\,\,\,\nom \\
&&\omega^{(2)}_j  =c^{(0)}(\beta_j)\,\,\,,\,\,\,
\omega^{(3)} =c^{(0)}(\beta_0)\,\,\,.\,\,\,\nom 
\ea
Thus the $F$ is a sum of combinations of these disk amplitudes
\ba
F=\frac{1}{2}(c^{(0)}(\alpha^i) c^{(0)}( \beta_i )
- c^{(0)}(\alpha^0) c^{(0)}(\beta_0) )\,\,\,.\label{pre}
\ea
That is, we glue two disks at their boundaries and obtain a sphere.
It is shown in Fig \ref{disk} schematically.

Eq.(\ref{pre}) relates instantons in disk amplitudes with world sheet 
quantum corrections in sphere.
Geometrical meaning of the disk amplitudes will be explained in
section \ref{geom}.
In the next subsection we extend the recipe in the $3$-fold case
to $d$-dimensional Calabi-Yau cases.

\subsection{Calabi-Yau $d$-fold}\label{dfold}

We restrict ourselves to consider a Calabi-Yau $d$-fold M
represented as a Fermat-type hypersurface embedded in $CP^{d+1}$
\begin{eqnarray*}
&&p:=X_{1}^N +X_{2}^N +\cdots +X_{N}^N =0\,\,\,,\\
&& N:=d+2\,\,\,.
\end{eqnarray*}
Non-vanishing Hodge numbers $h^{p,q}$ of this $d$-fold appear
only in the primary vertical ($ \oplus_p \mbox{H}^{p,p}$) 
and primary horizontal ($ \oplus_p \mbox{H}^{d-p,p}$) subspaces \cite{dais}
\ba
h^{p,q} &=& \delta_{p,q}
\,\,\,\,\,(0 \leq p \leq d \,,\,0 \leq q \leq d \,,\,p+q \not= d)\,\,\,,\nom\\
h^{d-p,p} &=& \delta_{2p,d}+\sum_{l=0}^p (-1)^l
\left(\matrix{d+2 \cr l}\right)\cdot 
\left(\matrix{(p+1-l)(d+1)+p \cr d+1}\right)
\,\,\,\,\,(0 \leq p \leq d) \,\,\,.\nom
\ea
When $d$ is odd, the vertical and horizontal parts
are completely decoupled each other.
But in even dimensional case, there are middle cohomology elements
in $\mbox{H}^{d/2,d/2}$. 
Products of K{\"a}hler element $e$ analytically produce 
only one element in this middle group and
we focus on this from now on.
Let $\mbox{H}^{l,l}_J$'s be analytic subgroups in $\mbox{H}^{l,l} $
generated from the K{\"a}hler form $e$.
Each $\mbox{H}^{l,l}_J$ is spanned by one element $e^l $ and
we write a corresponding A-model operator as $\co^{(l)}$ $(l=0,1,\cdots ,d)$.
Fusion structures of these operators are investigated in 
\cite{GMP, KS3, KS4, KS5}
\ba
\co^{(1)}\co^{(j-1)} &=& \Kappa_{j-1} \co^{(j)} \,\,\,\,\,\,\,\,
(1\leq j \leq d)\,\,\,,\nom\\
\co^{(1)}\co^{(d)} &=& 0\,\,\,.\label{dfusion} 
\ea
We switch on only one background source ``$t$'' in the A-model
correlator,
which is the K{\"a}hler moduli parameter.
In the point of view of mirror symmetry, it is a mirror map itself
and is a function of a complex moduli parameter $\psi$ of a mirror W
\ba
t(\psi)=\frac{N}{2\pi i}\Biggl[\,\,\log (N\psi)^{-1} 
+\frac{\dis\sum_{n=1}^{\infty}
\frac{(Nn)!}{(n!)^N}\left(\sum_{l=n+1}^{Nn}\frac{1}{l}\right) 
\cdot (N\psi)^{-n} }
{\dis \sum_{m=0}^{\infty} \frac{(Nm)!}{(m!)^N}\cdot (N\psi)^{-m}}
\,\,\Biggr]\,\,\,.\label{map}
\ea
Also $q$ is defined as ``$q:=\exp (\,2\pi i \,t) $''.
A canonical basis of homology cycles is chosen
as $\gamma_l \in \oplus_{m=0}^l \mbox{H}_{2l}$.
Disk amplitudes $c^{(0)}(\gamma)$'s associated with
an operator $ \co^{(0)}$ are periods themselves if they are
expressed by the
variable ``$\psi$'' in the mirror side,
but one can translate the amplitudes into A-model side
by the mirror map $t=t(\psi)$.
They are collected into one matrix $u_0$
\begin{eqnarray*}
u_0 &:=&
\bordermatrix{ 
 & {\scriptstyle \gamma_0} & {\scriptstyle \gamma_1} &\cdots  & 
{\scriptstyle \gamma_d} \cr
{\scriptstyle \co^{(0)}} 
& {c^{(0)}(\gamma_0)} & {c^{(0)}(\gamma_1)} & \cdots & 
{c^{(0)}(\gamma_d)} }
\\
&=& \left(
\matrix{ a_0 & a_1 & a_2 & \cdots & a_{d} }
\right)\cdot \exp (t {\cal N}) \,\,\,,\\
%\end{eqnarray*}
%\begin{eqnarray*}
{\cal N} &:=& \left(
\matrix{
0 & 1 &   &   &   & 0 \cr
  & 0 & 1 &   &   &  \cr
  &   & 0 & 1 &   &  \cr
  &   &   &  \ddots & \ddots  & \cr
  &   &   &   & 0 & 1 \cr
0 &   &   &   &   & 0 }
\right)\,\,\,.
\end{eqnarray*}
The matrix $\exp (t {\cal N})$ has information about
monodromy properties of homology cycles.
Under some constant shift of real part of $t\rightarrow t+1$,
that is, a shift of an NS-NS B-field in string,
the homology cycles mix each other
\ba
(\gamma_0 \,\,\,\gamma_1 \,\,\,\cdots \,\,\,\gamma_d )
\rightarrow 
(\gamma_0 \,\,\,\gamma_1 \,\,\,\cdots \,\,\,\gamma_d )
\cdot \exp\left({\cal N}\right)\,\,\,.
\ea
Because each cycle $\gamma_m$ is related with a D$m$-brane,
the result reflects a Chern-Simons term 
for even form ``RR-fields'' $C\wedge \exp(B)$
with $C:=\sum_{m=0}^{d} C^{(2m)}$.
Physically the meanings of these ``RR-fields'' 
are not clear
for higher dimensional ($>5$)
Calabi-Yau cases in the context of compactified string theories.
But as mathematical interest, we expect that these
``fields'' would be realized as some characteristic classes of
some vector bundle over the Calabi-Yau moduli space.
Probably mirror symmetries will be formulated 
mathematically beyond the range of type II string theories.

In contrast to the $\exp(t{\cal N})$, 
the functions $\{a_m \}$ are single-valued
with respect to ``$t$''. It reads concretely
\begin{eqnarray*}
&&\left\{
\begin{array}{ccl}
a_0 & = &1\,\,\,,\,\,\, a_1 =0\,\,\,,\\
a_n & = & S_n (0,\tilde{x}_2 , \cdots ,\tilde{x}_n )\,\,\,\,\,\,\,
\,\,\,(n=2,3,\cdots ,d)\,\,\,,
\end{array}
\right.\,\,\,\nom \\
&&\tilde{x}_m := \frac{1}{m!} \left(\frac{1}{2\pi i}\frac{\del}{\del \rho}
\right)^m \log \Biggl[\,\,
\sum_{l=0}^{\infty} \frac{\Gamma (N(l+\rho)+1)}{\Gamma (N\rho +1)}
\biggl\{ \frac{\Gamma (\rho +1)}{\Gamma (l+\rho +1)}\biggr\}^N 
(N\psi)^{-Nl}\,\,
\Biggr]\Bigg|_{\rho =0} \nom \\
&&\hspace{10cm}
(m=2,3,\cdots ,d)\,\,\,\,\,.\nom 
\end{eqnarray*}
The $S_n $ is a Schur function defined as
\ba
\exp\left(\sum_{m=1}^{\infty} w_n y^n \right)=
\sum_{n=0}^{\infty}
S_n (w_1 ,w_2, \cdots , w_n) y^n \,\,\,.\nom
\ea
These $a_n$'s  $(n\geq 2)$ are expanded as power series with respect to 
$q$. 
Furthermore all the amplitudes $c^{(l)}(\gamma_m)$'s
are put together into a 
matrix ${\Pi}$ by taking into account of fusion relations
(\ref{dfusion})
\begin{eqnarray*}
&&{\Pi}:=\left(\matrix{ {u}_{0} \cr 
{u}_{1} \cr 
\vdots \cr 
{u}_{d}}
\right)\,\,\,\,,\\
&& {u}_{l} =\frac{1}{\Kappa_{l-1}} {\partial_t}
\frac{1}{\Kappa_{l-2}} {\partial_t} \cdots {\partial_t}
\frac{1}{\Kappa_{1}} {\partial_t} \frac{1}{\Kappa_{0}} {\partial_t} 
{u}_0 \,\,\,\,\,\,\,\,\,(1\leq l \leq d)\,\,\,\,,\\
&& \left\{
\begin{array}{l}
\Kappa_{m}=  {\dis {\partial_t} \frac{1}{\Kappa_{m-1}}{\partial_t} 
\frac{1}{\Kappa_{m-2}}{\partial_t} 
\cdots
{\partial_t} \frac{1}{\Kappa_{1}}
{\partial_t} \frac{1}{\Kappa_{0}}{\partial_t} {c^{(0)}}(\gamma_{m+1})
}
\,\,\,\,\,\,\,\,(1\leq m \leq d-1)\,\,\,\,,\\
 \Kappa_0 =1 \,\,\,.
\end{array}
\right.
\end{eqnarray*}
We use a convention that a
topological metric is expressed as $\Eta_{ij} :=\ketbra{\co^{(i)}\co^{(j)}}
=N\delta_{i+j ,d}$. The symbol $\Kappa_l$ is an abbreviated form
of ${\Kappa_{1\ell}}^{\ell +1} ={\Kappa_{1\ell m}}\Eta^{m\,\ell +1}$.
The leading terms in the $q$-expansions of them are constants 
${\Kappa_{\ell}} =1+{\cal O}(q)$ or
${\Kappa_{1\ell m}}=N+{\cal O}(q)$.
The matrix ${\Pi}$ is an upper triangular one with all unit
diagonal elements.
If we introduce multiple integral matrix $I_m$ as
\ba
I_m(t)&:=&\int^t d t_0 \Kappa_t (t_0)
\int^{t_0} d t_1 \Kappa_t (t_1)
\int^{t_1} d t_2  \cdots
\int^{t_{m-2}} d t_{m-1} \Kappa_t (t_{m-1}) \,\,\,,\nom \\
\Kappa_t &:=&
\left(
\matrix{
0 & \Kappa_0 &  &  &  &  & 0 \cr
  & 0  & \Kappa_1 & & & & \cr
  & & 0  & \Kappa_2 & & & \cr
  & & & \ddots & \ddots & & \cr
  & & & & 0 & \Kappa_{d-2}& \cr
  & & & & & 0 & \Kappa_{d-1} \cr
0 & & & & & & 0
}
\right)\,\,\,\,,\nom
\ea
the ${\Pi}$ is written as a path ordered
integral form
\ba
{\Pi}&=&\sum_{m=0}^d I_m (t)\nom \\
&=& \mbox{Pexp} \left( \int_C dt \,\Kappa_t \right)\,\,\,.\label{path}
\ea
The symbol $\mbox{Pexp} (\cdots) $ means a path ordered exponential
along a contour $C$ in the K{\"a}hler parameter space with the coordinate $t$.
If we fix two end points of $C$, the result does not depend on the choice of
contour as far as the $C$ does not enclose any singular points
in the parameter spaces.
It is ensured locally by an integrable relation 
$[\del_t -\Kappa_t ,\del_t -\Kappa_t ]=0$
of a differential equation
\ba
(\del_t -\Kappa_t){\Pi}=0 \,\,\,.\nom
\ea
This is equivalent to $[\Kappa_t ,\Kappa_t ]=0$
or $\Kappa_i \Kappa_j =\Kappa_j \Kappa_i $ in components.
It is associativity relations of operators themselves.

But if the $C$ encircles some singular point, there is an arbitrariness
in defining ${\Pi}$. Then the ${\Pi}$ have
global monodromy properties. In fact there are three singular points
on the $t$-plane in this model.
The first is the infinity point 
$t=\infty$ we considered here. 
The second is the $t=t(\psi =1)$ point.
In the language in the mirror side W, it is a
conifold-like point for an arbitrary odd dimensional case
\footnote{There is a branch cut $\sim (1-\psi)^{\frac{d-1}{2}}$ 
for a period in even dimensional case.
We do not have any clear geometrical interpretation of these cuts
in contrast to vanishing cycles in odd dimensional cases. }
and curvature in its moduli space 
blows up at this point. In the neighbourhood of the point,
the topology of a vanishing cycle is a 
$d$ dimensional sphere with the radius
proportional to $(\psi -1)^{1/2}$ when we restrict ourselves
to real locus.
This sphere is homologically equivalent to
a cycle $\beta_0$, which is 
canonically dual to the fundamental one $\alpha_0$.
The $\alpha_0$ is 0-cycle in A-model 
and is wrapped by a D0-brane. The corresponding one
in the mirror side is a fundamental $d$-cycle
$\Gamma$ that is equivalent to a real $d$-dimensional torus 
$T^d$ topologically.
When we consider a patch $X_{d+2} =1$ in this case
and take a set of variables
$(X_1 ,X_2 ,\cdots , X_d)$ as independent coordinates of the W,
the cycle $\Gamma$ is represented
\ba
\Gamma:=\{(X_1 ,X_2 ,\cdots ,X_d)\in {\bf C}^d 
\,;\, |X_i|=1 \,\,(i=1,2,\cdots ,d)\}\,\,\,
.\nom
\ea
This exemplifies a torus fibered structure of the Calabi-Yau $d$-fold.
The remaining singular point is an orbifold-like point
${\dis t=-\frac{1}{2}+\frac{i}{\dis 2\tan \frac{\pi}{N}}}(\equiv t_0)$.
There is an ${\bf Z}_N$ symmetry around this point.
In the B-model side , this discrete symmetry acts on
the complex parameter $\psi \sim 0$ as a rotation
$\psi \rightarrow \exp (\frac{2\pi i}{N})\cdot \psi $.
It originates from a orbifold group of the mirror manifold W.
This symmetry separates the parameter $\psi$
into $N$ regions $\frac{2\pi}{N} m \leq \arg \psi < \frac{2\pi}{N} (m+1)$
$(m=0,1,\cdots ,N-1)$ and there are $N$ large complex structure points.
In the $t$-plane
the corresponding $N$ regions emanate from the point $t_0$.
Only one large complex structure point is mapped to a large radius limit point
with $\mbox{Im}t =+\infty$.
On the other hand, all the remaining $(N-1)$ points are 
mapped to an original point $t=0$, that is, a small radius point.
By an action of the rotation, the large radius region is
transformed into one of small radius regions.
Let us recall the form of the instanton expansion parameter
$q=\exp (2\pi i \mbox{Re} t)\cdot \exp (-2\pi \mbox{Im} t)$.
In the $N-1$ regions containing $\mbox{Im}t=0$, the expansion parameter
$q$ is large and non-perturbative effects seems to be intensive.

Around each of these three kinds of singular points, there is an associated 
monodromy transformation.
Investigation about properties of these singular points
will lead us to deeper understanding of mirror symmetries and
relations between some (gauge) 
bundles over D-brane moduli spaces of M and
mirror manifold W itself.

 \begin{figure}
     \epsfxsize=10cm
     \centerline{\epsfbox{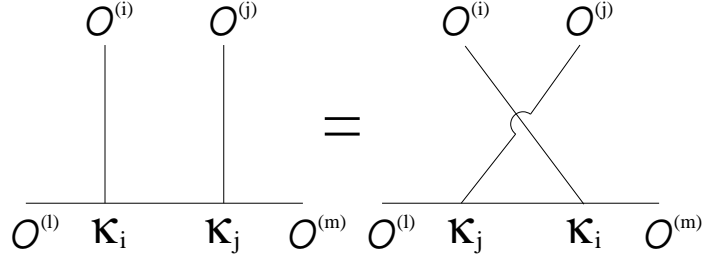}}
     \epsfxsize=10cm
     \caption{Integrable conditions are equivalent to associativity
      relations for inserted operators.}\label{associa}
 \end{figure}

We will return to the disk amplitudes.
The Eq.(\ref{path}) gives us 
relations between three point functions on sphere
and disk amplitudes in A-model for $d$-fold.
Essentially instanton corrections are contained in
the $a_l$'s. They are invariant under the discrete integer shift
of $t$.
Some explicit calculations for low dimensions are summarized in
appendix A.

Next we study another example:
\ba
&&\hspace{10mm} \mbox{M} \,;\, X_1^{d+1}+ X_2^{d+1}+\cdots + X_{d}^{d+1}
+ X_{d+1}^{2(d+2)}+ X_{d+2}^{2(d+2)}
=0 \nom \\
&& \hspace{15mm}  \mbox{in} \,\,
{\bf P}_{d+1}[\,\,\underbrace{2,2,\cdots ,2}_{\mbox{$d$ times}},1,1\,](2(d+1))
\,\,\,.\nom
\ea
K{\"a}hler moduli of this model 
is generated by two elements in $\mbox{H}^2$(M)
and we write two K{\"a}hler parameters associated with them as
$t$ and $s$.
They are mirror maps themselves
\ba 
&& t(x,y)=\frac{1}{2\pi i}\frac{g_1}{g_0}
\,\,\,,\,\, s(x,y)=\frac{1}{2\pi i}\frac{g_2}{g_0}
\,\,\,, \nom \\
%\ea
%\ba
&& g_0 (x,y) = \sum_{m,n \geq 0}
\frac{((d+1)m)! }{( m! )^d (m-2n) ! ( n! )^2} x^m y^n \,\,\,,\nom \\ 
&& g_1 (x,y) = g_0 (x,y) \log x 
+ \Biggl[ \, \sum_{m,n \geq 0}
\frac{ ((d+1)m)! }{( m! )^d (m-2n) !  ( n! )^2 } \, x^m y^n \nom \\ 
&& \hspace{1.5cm}
\times [{(d+1)\Psi ((d+1)m+1)-d \Psi (m+1)- \Psi (m-2n+1)}] 
 \, \Biggr]\,\,\,,\nom \\ 
&& g_2 (x,y)
= g_0 (x,y) \log y 
+ \Biggl[ \,
\sum_{m,n \geq 0}\frac{ ((d+1)m)! }{( m! )^d (m-2n) ! ( n! )^2}\, 
x^m y^n\nom \\
&& \hspace{1.5cm} \times 2[\Psi (m-2n+1)-\Psi (n+1)]
\, \Biggr]\,\,\,.\nom 
\ea
Here the $x$ and $y$ are complex moduli parameters for a mirror W 
in the B-model.
Analytical subgroups of vertical homology classes are characterized by 
a set of canonical basis
\ba
\{\gamma^{(0)} ,\gamma^{(1)}_1 ,\gamma^{(1)}_2 ,
\cdots \gamma^{(d-1)}_1 ,\gamma^{(d-1)}_2 ,
\gamma^{(d)}\}\,\,\,.\nom
\ea
The $\gamma^{(l)}$ belongs to some combination
of homology groups $\oplus_{m=0}^{l}\mbox{H}_{2m}$(M).

Let us consider disk amplitudes $c^{(0)}(\gamma)$ 
associated with the operator $\co^{(0)}$.
They are represented as components of a row vector $u_0$
\ba
u_0 &:=& \left(
\begin{array}{ccccccc}
c^{(0)}(\gamma^{(0)}) & c^{(0)}(\gamma^{(1)}_1) 
& c^{(0)}(\gamma^{(1)}_2) 
&\cdots 
& c^{(0)}(\gamma^{(d-1)}_1) 
& c^{(0)}(\gamma^{(d-1)}_2) 
& c^{(0)}(\gamma^{(d)})
\end{array}
\right) \nom \\
%&=& \left(
%\begin{array}{ccccccc}
%1 & \omega^{(1)}_1
%& \omega^{(1)}_2
%&\cdots 
%& \omega^{(d-1)}_1
%& \omega^{(d-1)}_2
%& 2 \omega^{(d)}_1 +\omega^{(d)}_2
%\end{array}
%\right) \nom \\
&=& \left(
  \begin{array}{ccccccccc}
\tilde{a}_0 & \tilde{a}_1 & \tilde{c}_1 & \tilde{a}_2 &  \tilde{c}_2 
& \cdots & 
 \tilde{a}_{d-1} &  \tilde{c}_{d-1} &
2 \tilde{a}_d + \tilde{c}_d
\end{array}
\right) \cdot \exp (t {\cal N}^{(1)} +s {\cal N}^{(2)} )\,\,\,.\nom 
\ea
The sets of functions $\tilde{a}$ and $\tilde{c}$ are defined
by recursion equations
\ba
&& \tilde{a}_k :=\hat{a}_k \Big|_{\rho_1 =\rho_2 =0}\,\,\,,\,\,\,
 \tilde{c}_k :=\hat{c}_k \Big|_{\rho_1 =\rho_2 =0}\,\,\,,\nom \\
&&\left\{
\begin{array}{l}
\dis \hat{a}_0 =1 \,\,,\,\,\hat{a}_1 =0 \,\,\,\,,\\
\dis \hat{a}_k =\frac{1}{k} \left({\cal D}_{\rho_1} \hat{a}_{k-1}
+T \hat{a}_{k-2} \right) \,\,\,\,\,\,\,\,(k=2,3,\cdots)\,\,\,,
\end{array}
\right. \nom \,\,\,,\nom \\
%\ea
%\ba
&&\left\{
\begin{array}{l}
\dis \hat{c}_0 =0 \,\,,\,\,\hat{c}_1 =0 \,\,\,\,,\\
\dis \hat{c}_{k+2} =\frac{1}{k+1} \left({\cal D}_{\rho_1} \hat{c}_{k+1}
+T \hat{c}_{k} +S \hat{a}_{k} \right) \,\,\,\,\,\,\,(k=0,1,\cdots)\,\,\,,
\end{array}
\right.\nom \\
%\ea
%\ba
&&T:= D_{\rho_1}^2 \log V(x,y,;\rho_1 ,\rho_2 ) \,\,\,,\nom \\
&&S:= D_{\rho_1} D_{\rho_2}  \log V(x,y,;\rho_1 ,\rho_2 ) \,\,\,,\nom \\
&&V(x,y,;\rho_1 ,\rho_2 ):=
\sum_{m,n\geq 0} \frac{\Gamma [(d+1)(m+ \rho_1 )+1]}
{\Gamma [(d+1) \rho_1 +1]} \nom \\
&&\hspace{25mm}
\times \left[ \frac{\Gamma [1+ \rho_1 ]}{\Gamma [ m+1+ \rho_1 ]}\right]^d 
\times \left[ \frac{\Gamma [1+ \rho_2 ]}{\Gamma [ n+1+ \rho_2 ]}\right]^2 \nom 
\\
&&\hspace{25mm} \times  \left[ \frac{\Gamma [1+ \rho_1 -2 \rho_2 ]}
{\Gamma [m-2n+1+ \rho_1 -2 \rho_2  ]}\right]\,x^my^n \,\,\,.\nom
\ea
We use the symbol
${\cal D}_{\rho}$ as $\dis \frac{1}{2\pi i}\frac{\del}{\del \rho}$.
These two sets of functions $\tilde{a}$'s and $\tilde{c}$'s
contain information about quantum corrections in the disk amplitudes.
The vector $u_0$ have non-trivial monodromies around $t=\infty$ and $s=\infty$
and their properties are characterized by two matrices ${\cal N}^{(1)}$
and ${\cal N}^{(2)}$
\ba
&&{\cal N}^{(1)}:=
\bordermatrix{
 & \overbrace{}^1 & \overbrace{}^2 & \overbrace{}^2 & \overbrace{}^2 &
  \cdots & \overbrace{}^2 & \overbrace{}^1 \cr
1 \left\{ \right. & 0 & {e_1^{(1)}} & & &  & { O} &  \cr
2 \left\{ \right. &   & 0 & I &   &   &   &  \cr
2 \left\{ \right. &   &   & 0 & I &   &   &  \cr
          \vdots &   &   &   & \ddots & \ddots  &   &  \cr
2 \left\{ \right. &   &   &   &  & 0 & I  &  \cr
2 \left\{ \right. &   &   &   &  &   & 0  & {e^{(2)}_1 } \cr
1 \left\{ \right. &   & { O}  &   &  &   &    &  0
}\,\,\,,\nom \\
%\ea
%
%\ba
&&{\cal N}^{(2)}:=
\bordermatrix{
 & \overbrace{}^1 & \overbrace{}^2 & \overbrace{}^2 & \overbrace{}^2 &
  \cdots & \overbrace{}^2 & \overbrace{}^1 \cr
1 \left\{ \right. & 0 & {e_2^{(1)}} & & &  & { O} &  \cr
2 \left\{ \right. &   & 0 & I' &   &   &   &  \cr
2 \left\{ \right. &   &   & 0 & I' &   &   &  \cr
          \vdots &   &   &   & \ddots & \ddots  &   &  \cr
2 \left\{ \right. &   &   &   &  & 0 & I'  &  \cr
2 \left\{ \right. &   &   &   &  &   & 0  & {e^{(2)}_2 } \cr
1 \left\{ \right. &   & { O}  &   &  &   &    &  0
}\,\,\,,\nom \\
%\ea
%
%\ba
&& \kankaku e^{(1)}_1 =\left(\matrix{1 & 0}\right)\,\,,\,\,\,\,\,\,
e^{(2)}_1 =\left(\matrix{2 \cr 1}\right)\,\,,\nom \\
&&\kankaku  e^{(1)}_2 =\left(\matrix{0 & 1}\right)\,\,,\,\,\,\,\,\,
e^{(2)}_2 =\left(\matrix{1 \cr 0}\right)\,\,,\nom \\
&& \kankaku I=\left(\matrix{1 & 0 \cr 0 & 1}\right)\,\,,\,\,\,\,\,\,
 I' =\left(\matrix{0 & 1 \cr 0 & 0}\right)\,\,.\nom
\ea
These ${\cal N}^{(1)}$ and
${\cal N}^{(2)}$ are commutable with each other 
$[ {\cal N}^{(1)} ,{\cal N}^{(2)}]=0$
and
satisfy nilpotent relations
\ba
\left({\cal N}^{(1)}\right)^{d+1} =0 \,\,\,\,,\,\,\,\,
\left({\cal N}^{(2)}\right)^{2} =0 \,\,\,.
\ea
When one shifts $t$ and $s$ as $t\rightarrow t+m^{(1)}$ and
$s\rightarrow s+m^{(2)}$ 
$( {}^\exists m^{(1)}, {}^\exists m^{(2)}\in {\bf Z})$,
the set of canonical homology basis is transformed
\ba
&&(\gamma^{(0)} \,\,\,\gamma^{(1)}_1 \,\,\,\gamma^{(1)}_2 \,\,\,
\cdots\,\,\, \gamma^{(d-1)}_1\,\,\,\gamma^{(d-1)}_2 \,\,\,
\gamma^{(d)})\nom \\
&&\rightarrow
(\gamma^{(0)} \,\,\,\gamma^{(1)}_1 \,\,\,\gamma^{(1)}_2 \,\,\,
\cdots\,\,\, \gamma^{(d-1)}_1\,\,\,\gamma^{(d-1)}_2 \,\,\,
\gamma^{(d)})
\cdot \exp\left(m^{(1)}{\cal N}^{(1)} +m^{(2)}{\cal N}^{(2)}\right)
\,\,\,,\nom
\ea
and various D-branes with different dimensions mix one another.
It implies a Chern-Simons interaction term between
``RR-fields'' and an NS-NS 2-form B similarly to the previous example.
We make one remark here: The $0$-cycle $\gamma^{(0)}$ is invariant
under the monodromy transformations and is associated with a D0-brane.
It corresponds to a unique fundamental cycle $\Gamma$
in the B-model moduli space. The $\Gamma$ is the same cycle as that in
previous example and is topologically equivalent to $T^d$.
These torus structures seem to be universal for
any $d$-dimensional Calabi-Yau manifolds (at least
for Calabi-Yau manifolds realized as some (complete intersection of) 
zero loci of ambient toric varieties).

We shall go back to the disk amplitudes.
Let $\co^{(1)}_t$, $\co^{(1)}_s$ be
A-model operators coupled with parameters $t$, $s$ 
respectively
associated with two elements in the $\mbox{H}^2$(M).
When we switch on only marginal background operators in the A-model,
fusion relations for this mode are expressed linearly as
\ba
&&\co^{(1)}_i \co^{(0)}=\co^{(1)}_i\,\,\,,\nom \\
&&\co^{(1)}_i \co^{(l)}_j ={\Kappa^{(l)\,k}_{ij}} \co^{(l+1)}_k\,\,\,,\nom \\
&&\co^{(1)}_i \co^{(d)}=0\,\,\,,\nom 
\ea
where the $i,j,k$ run over $t$ or $s$.
We can define two kinds of fusion matrices $\Kappa_t$ and $\Kappa_s$
\ba
&&{\Kappa_i}:=\bordermatrix{
 & \overbrace{}^1 & \overbrace{}^2 & \overbrace{}^2 & \overbrace{}^2 &
 \cdots & \overbrace{}^2 & \overbrace{}^1 \cr
1\left\{ \right. & 0 & \Kappa^{(0)}_i & & & & & O \cr
2\left\{ \right. & & 0 & \Kappa^{(1)}_i & & & & \cr
2\left\{ \right. & & & 0 & \Kappa^{(2)}_i & & & \cr
\vdots           & & & & \ddots & \ddots & & \cr
2\left\{ \right. & & & & & 0 & \Kappa^{(d-2)}_i & \cr
2\left\{ \right. & & & & & & 0 & \Kappa^{(d-1)}_i \cr
1\left\{ \right. & O & & & & & & 0 
}\,\,\,,\\
&&\Kappa^{(0)}_t :=\left(\matrix{1 & 0}\right)
\,\,\,,\,\,\,\Kappa^{(0)}_s :=\left(\matrix{0 & 1}\right)\,\,\,,\nom \\
&&\Kappa^{(d-1)}_t :=\left(\matrix{1 \cr 0}\right)
\,\,\,,\,\,\,\Kappa^{(d-1)}_s :=\left(\matrix{0 \cr 1}\right)\,\,\,,\nom \\
&&{\left(\Kappa^{(l)}_i\right)_j}^k :={\Kappa^{(l)\,k}_{ij}}
={\Kappa^{(l)}_{ijm}}\Eta^{mk}\,\,\,\,\,\,\,(l=1,2,\cdots ,d-2)
\,\,\,,\nom \\
&&\Eta_{ij}:=\ketbra{\co^{(1)}_i \co^{(d-1)}_j}=
(d+1)\cdot \left(\matrix{2 & 1 \cr 1 & 0}\right)\,\,\,.\nom
\ea
All the amplitudes $c^{(l)}(\gamma^{(m)})$ are constructed 
by transforming the fusion structure into 
a disk amplitude matrix ${\Pi}$.
We can express the results 
in a compact form by applying the same recipe developed in the previous example
\ba
&& {\Pi}={\rm Pexp}\left(\int_{\cal C} dt \,\Kappa_t +ds \,\Kappa_s
\right)\,\,\,,\nom \\
&&\Kappa_t := \del_t {\sf F}\,\,\,,\,\,\,
\Kappa_s :=\del_s {\sf F}\,\,\,,\nom \\
%\ea
%\ba
&&{\sf F}:=\bordermatrix{
 & \overbrace{}^1 & \overbrace{}^2 & \overbrace{}^2 & \overbrace{}^2 &
 \cdots & \overbrace{}^2 & \overbrace{}^1 \cr
1\left\{ \right. & 0 & M^{(0)} & & & & & O \cr
2\left\{ \right. & & 0 & M^{(1)} & & & & \cr
2\left\{ \right. & & & 0 & M^{(2)} & & & \cr
\vdots           & & & & \ddots & \ddots & & \cr
2\left\{ \right. & & & & & 0 & M^{(d-2)} & \cr
2\left\{ \right. & & & & & & 0 & M^{(d-1)} \cr
1\left\{ \right. & O & & & & & & 0 
}\,\,\,.\nom 
\ea
Block components in the matrix ${\sf F}$ are defined as
\ba
&&M^{(0)}= \left(\matrix{t & s}\right) \,\,\,,\,\,\,
M^{(d-1)}= \left(\matrix{t \cr s}\right) \,\,\,,\nom \\
&&M^{(l)}= \left(\matrix{t & s \cr 0 & t}\right)+ A^{(l)}_0
\,\,\,\,(l=1,2,\cdots ,d-2)\,\,\,,\nom \\
%\ea
%\ba
&&\left\{
\begin{array}{ccl}
A^{(1)}_0 & = & \dis \left(\matrix{\del_t \tilde{a}_2 & \del_t \tilde{c}_2 \cr
\del_s \tilde{a}_2 & \del_s \tilde{c}_2 }\right) \,\,\,,\nom \\
A^{(1)}_n & = & \dis \left(\matrix{ \tilde{a}_{n+1} + \del_t \tilde{a}_{n+2} & 
\tilde{c}_{n+1} + \del_t \tilde{c}_{n+2} \cr \del_s  \tilde{a}_{n+2} &
 \tilde{a}_{n+1} + \del_s \tilde{c}_{n+2}
}\right)\,\,\,\,\,\,(n=1,2,\cdots)  \,\,\,,\nom \\
 A^{(m+1)}_n &=& (I+\del_t A^{(m)}_0)^{-1} \cdot 
\left(A^{(m)}_n +\del_t A^{(m)}_{n+1}
\right)\,\,\,\,\,\,(m\geq 1) \,\,\,.\nom 
\end{array}
\right.\nom
\ea
In $d$-dimensional cases, we have treated the one modulus and 
two moduli models concretely. But the method developed here is not
restricted to these cases and the structure of mixing between homology cycles 
and monodromy properties are expected to be universal for arbitrary $d$-folds.
In Appendix B, we show 
results for general Fermat-type Calabi-Yau $d$-fold case.

\subsection{Geometrical meaning of A-model Amplitudes}\label{geom}

Now let us concentrate on the topological meaning of these
A-model disk amplitudes.
The amplitude
is defined as a path integral form
with an operator $\co^{(l)}$ and a
homology cycle $\Gamma_m$ with a definite (real) dimension $2m$
\ba
\braket{\co^{(l)}(P)}{\Gamma_m}:=
\int_{\Sigma} {\cal D}[X,\psi_0] \,e^{-S_{eff}} 
e^{t \co^{(1)}} \co^{(l)}(P)\,\,\,.\nom
\ea
Configuration of bosonic fields
 is dominated by holomorphic instantons from Riemann
surface $\Sigma$ to the target manifold M.
That is defined as a set of the complex structure $J$ of $\Sigma$
and holomorphic maps $X$.
Also the boundary of $\Sigma$ is mapped into the homology cycle
$\Gamma_m \subset \mbox{M}$.
The $S_{eff}$ is an effective action for fermionic zero-modes and
is expressed as the Mathai-Quillen formula
\ba
S_{eff}:=\left(\psi_{L\,0}^{\bar{\imath}},{\cal R}_{\bar{\imath}j}
\psi_{R\,0}^{j}\right)\,\,\,.\nom 
\ea
Here the inner product is defined by hermitian metrics and
the ${\cal R}$ is a curvature form of a vector bundle ${\cal V}$
over the holomorphic maps.
Fiber of ${\cal V}$ is spanned by cokernel of 
a covariant derivative $\nabla$.
After integration of 1-form parts of the zero-modes $\psi_0$, 
the $S_{eff}$ leads us to a Chern
class $c({\cal V})$.
The remaining zero modes govern moduli space of stable maps
$\overline{{\cal M}}_{l,m;n}$.
It is a set of one market point $P$ on $\Sigma$ and
holomorphic maps $X$ with degree $n$ 
\ba
\overline{\cal M}_{l,m;n} &:=& \{
X: \Sigma \rightarrow \mbox{M}
\,\,\,(\mbox{hol. map with}\,\deg X=n) \,\,\,,\,\,\,\nom \\
&&X(\del\Sigma)\subset \Gamma_m \,\,\,,\,\,\,
 X(P) \in P.D.(e_l) 
\}\,\,\,.\nom
\ea
We introduce an embedding $\iota \,;\,\Gamma_m \hookrightarrow M$.
Then the amplitude is represented 
\ba
\braket{ \co^{(l)}(P) }{ {\Gamma}_m }
&=&%\frac{1}{l!}
\int_{ {\Gamma}_{m} } \iota^{\ast}(e^l) 
+\sum_{n=1}^{\infty} \left(
\int_{ \overline{\cal M}_{l,m;n} } c({\cal V}_{l,m;n}) \phi^{\ast} (e_l) 
\right) q^n \,\,\,,\nom \\
q&:=& \exp (2\pi i\, t)\,\,\,.
\ea
The ${\cal V}_{l,m;n}$ is a vector bundle over
the moduli space $\overline{\cal M}_{l,m;n}$ with $\mbox{coker}\nabla$
as its fiber.
The $\phi$ is an evaluation map of $\overline{\cal M}$ and acts on the
set $\{ P,X \}$ as
\ba
\phi (\{ P, X \}):=X(P) \,\,\,.
\ea
In particular the degree zero part can be evaluated by
K{\"a}hler form $e$.
We now make a remark here: We have fixed the degree of the
homology cycle. But there are mixings between cycles with
different dimensions when we discuss
calculations in the mirror technique.
This phenomenon can be interpreted as shifts of K{\"a}hler form $e$
by cohomology elements $\zeta \in \mbox{H}^2$(M) ($e \rightarrow e+\zeta$).
Its modification is equivalent to some constant shifts of
K{\"a}hler parameters, which  lead us to monodromy transformations.

%\newpage

\section{Extension to Fano Manifolds}
\cleqn

In previous sections, we considered Calabi-Yau cases.
By using mirror symmetries, we can calculate the amplitudes
in the A-model indirectly. But the existence of B-model is essential 
there and that necessarily restricts us to manifolds with
$c_1 (\mbox{TM})=0$, that is,  Calabi-Yau spaces.
If we can study the calculation within the range of A-model only,
we remove the condition $c_1 =0$ for manifolds.
In this section, we develop a method
to analyse the A-model for manifolds with $c_1 (\mbox{TM})>0$
as an attempt to treat general K{\"a}hler manifolds.

\subsection{Fusion Relations for Operators}

Let us first recall the fusion relations considered in the 
previous sections.
Each A-model operator $\co^{(l)}$ is associated with
a cohomology element $e^l \in \mbox{H}^{2l}$(M).
Operator products of these are written down
\ba
\co^{(l)}\cdot\co^{(m)}=\sum_{n} {\Kappa_{lm}}^n \co^{(n)}\,\,\,.\nom
\ea
Here the product ``$\cdot $'' means equality of both sides
when we evaluate them in the A-model correlator
\ba
\ketbra{\co^{(l)}\cdot\co^{(m)} \cdots }
=\sum_{n} {\Kappa_{lm}}^n \ketbra{\co^{(n)} \cdots }\,\,\,.\nom 
\ea
When we take a derivative of a disk amplitude 
$c^{(l)}(\gamma)=\braket{\co^{(l)}}{\gamma}$, fusion relations
give us a set of differential equations
\ba
\frac{\del}{\del t^k }\braket{\co^{(l)}}{\gamma}&=&
\braket{\co^{(k)} \cdot \co^{(l)}}{\gamma}\nom \\
&=& \sum_{m} {\Kappa_{kl}}^m
\braket{\co^{(m)}}{\gamma}\,\,\,.\nom 
\ea
We can collect these amplitudes in one matrix ${\Pi}$
by choosing an appropriate set of homology basis $\{ \gamma_n \}$
and the fusion relations are transformed into a
set of differential equations for ${\Pi}$
\ba
&&{\Pi}:=\bordermatrix{
 & & {\scriptstyle \gamma_n} & \cr
 & & & \cr
{\scriptstyle \co^{(m)}} & & \braket{\co^{(m)}}{\gamma_n } & \cr
 & & &  }\,\,\,,\nom\\
%\ea
%\ba
&&\frac{\del}{\del t^k}{{\Pi}}={\Kappa_k}{\Pi}\,\,\,,
\label{diffeq} \\
&&{\left(\Kappa_k \right)_l}^m := {\Kappa_{kl}}^m \,\,\,.\nom 
\ea
Here we introduce matrices $\Kappa_k $ whose $(l,m)$ component
is a fusion coupling ${\Kappa_{kl}}^m$.
If one can solve this equation (\ref{diffeq}),
the disk amplitudes are obtained.
When is this equation solvable?
The integrable conditions are needed to solve
simultaneous differential equations.
They are expressed as commutativity of differential
operators $\dis \frac{\del}{\del t^l}-\Kappa_l $
\ba
\left[
\frac{\del}{\del t^l}-\Kappa_l \,,\,\frac{\del}{\del t^m}-\Kappa_m
\right]=0 \,\,\,.\label{comm}
\ea
In the cases for Fano manifolds, three point couplings are
derivatives of some prepotentials $F$
\ba
&&{\Kappa_{kl}}^m 
:= \Eta^{mn} \del_{t^k} \del_{t^l} \del_{t^n} F \,\,\,,\label{kappa} \\
&& \Eta_{kl}:=\ketbra{\co^{(k)}\co^{(l)} } \,\,\,,\nom \\
&& \Eta^{mn} \Eta_{nl} ={\delta^m}_l \,\,\,.\nom
\ea
In such cases, identities $\del_{t^l}\Kappa_m =\del_{t^m}\Kappa_l $
are satisfied automatically and
integrable conditions are turned into formulae
\ba
\left[\Kappa_l , \Kappa_m
\right]=0 \,\,\,.
\ea
That is the associativity relations for two inserted operators
$\co^{(l)}$ and $\co^{(m)}$
\ba
&&{\Kappa_{l n}}^r {\Kappa_{m r}}^s =
{\Kappa_{m n}}^r {\Kappa_{l r}}^s %\\
%\mbox{or }&&\del_{l}\del_{n}\del_{r}F
%\,\Eta^{r r'}\del_{m}\del_{r'}\del_{s}F
%=\del_{m}\del_{n}\del_{r}F
%\,\Eta^{r r'}\del_{l}\del_{r'}\del_{s}F \,\,\,.
\,\,\,.\label{f-associa}
\ea
These determine instanton parts of prepotentials completely 
\cite{rational1, rational2, rational3}.
Under these conditions (\ref{kappa})(\ref{f-associa}),
we can integrate the equations (\ref{diffeq})
and obtain the amplitudes for Fano manifold cases
\ba
&&{\Pi}=\mbox{Pexp}\left(
\int_{C} d t^l \,\Kappa_{l}
\right)\,\,\,,\nom \\
&&{(\Kappa_{l})_m}^n :={\Kappa_{lm}}^n\,\,\,.\nom
\ea
The contour $C$ could be chosen appropriately.
But we do not know the global structure of
moduli spaces for Fano cases in contrast to Calabi-Yau spaces.
There is no associated B-models and we do not have
any kind of periods whose complex moduli parameters 
are analytically continued over all the
moduli spaces.
In the next subsection, we analyse these amplitudes ${\Pi}$
for several cases concretely.

\subsection{Several Examples}

We take three concrete examples,
$CP^N$ $(N=1,2,3)$, Grassmann manifold $Gr(2,4)$, $M_{4,3}$ (zero locus of a
hypersurface with degree $3$ in $CP^4$)
and analyze the ${\Pi}$ of them.
For simplicity, we switch off all background 
sources $t^l$'s of operators $\co^{(l)}$
except for marginal ones and concentrate on analytic subgroups generated
by products of K{\"a}hler forms.

As a first example, we consider the projective space $CP^N$.
The $CP^N$ is (complex) $N$ dimensional space with 
1st Chern class $c_1 (\mbox{T}CP^N) =(N+1)e$.
The $e$ is a K{\"a}hler form.
%Euler number is $\chi (CP^N) =N+1$.%
Its non-vanishing Hodge numbers are $h^{l,l}=1$ $(l=0,1,\cdots N)$
and all cohomology elements are generated by products of a K{\"a}hler
form $e$. A corresponding operator $\co^{(l)}$ is defined
for each $e^l \in \mbox{H}^{l,l}$ $(l=0,1,\cdots ,N)$.

When one switches on only K{\"a}hler parameter $t_1=x$
with all others off $t_0 =t_2 =\cdots =t_N =0$,
a fusion coupling $\Kappa_1 $ associated with $\co^{(1)} $ 
is given as
\ba
&&\Kappa_1 =
\left(
\matrix{
0 & 1 &   &   &   & 0 \cr
  & 0 & 1 &   &   &  \cr
  &   & 0 & 1 &   &  \cr
  &   &   &  \ddots & \ddots  & \cr
  &   &   &   & 0 & 1 \cr
q &   &   &   &   & 0 }
\right)\,\,\,,\nom \\
&&{(\Kappa_1)_{\ell}}^m :={\Kappa_{1\ell}}^m=
\Kappa_{1\ell n}\Eta^{nm}\,\,\,.\nom
\ea
Here the topological metric $\Eta_{ij}$ is given as
$\Eta_{ij}:=\ketbra{\co^{(i)}\co^{(j)}}=N\cdot \delta_{i+j,N}$.
(All the other couplings $\Kappa_{\ell}$ 
associated with $\co^{(\ell )}$'s
are constructed as products of the $\Kappa_1$ in this background,
that is, $\Kappa_{\ell} =(\Kappa_1)^{\ell}$.)
Then the ${\Pi}$ is obtained 
\ba
{\Pi}=\mbox{Pexp}\left(\int_C dx\,\Kappa_1 \right)\,\,\,.
\ea
It is expanded by a parameter ``$q:=\exp(x)$'' in an infinite series
\ba
&&{\Pi}=\sum_{l=0}^{\infty}a_l q^l \,\,\,,\nom \\
&&a_0:=\exp
\left(
\matrix{
0 & x &   &   &   & 0 \cr
  & 0 & x &   &   &  \cr
  &   & 0 & x &   &  \cr
  &   &   &  \ddots & \ddots  & \cr
  &   &   &   & 0 & x \cr
0 &   &   &   &   & 0 }
\right)\,\,\,.\nom
\ea
The leading coefficient
$a_0$ is the classical part of the amplitude and is determined 
only by geometrical data of $CP^N$.
Quantum corrections are encoded in the expansion coefficients
$a_l$ $(l=1,2,\cdots)$ and we focus these parts for lower dimensional cases.\\
\mbox{}\\
$\bullet$ \underline{$CP^{1}$ model}\\

The cohomology group is generated by two elements
$1$ and $e$ and we introduce parameters $t_0 $ and $t_1 $ for
each operator $\co^{(0)}$, $\co^{(1)}$.
The prepotential $F$ of the $CP^1$ model is represented as
\ba
F= \frac{1}{2} t_0^2 t_1 + e^{t_1}\,\,\,.\nom
\ea
Fusion couplings are given as
\ba
\Kappa_0 =\left(
\matrix{1 & 0 \cr 0 & 1 }\right)\,\,\,,\,\,\,
\Kappa_1 =\left(
\matrix{0 & 1 \cr q & 0 }\right)\,\,\,.\nom
\ea
When one turns off the parameter $t_0$, 
the disk amplitude ${\Pi}$ is
written concretely
\begin{eqnarray*}
&& {\Pi} =\left( \matrix{1 & x \cr 0 & 1}\right)
+ \sum_{l=1}^{\infty} 
\left( \matrix{{u_{11}}^l & {u_{12}}^l \cr 
{u_{21}}^l & {u_{22}}^l \cr}
\right) q^l\,\,\,,
\,\,\, \\
&&  {u_{11}}^l =\frac{1}{(l!)^2}\,\,\,,\,\,\,
{u_{21}}^l =\frac{l}{(l!)^2}\,\,\,,\\
&&  {u_{12}}^l =\frac{l}{(l!)^2}x +\alpha_l \,\,\,,\,\,\,
{u_{22}}^l =\frac{1}{(l!)^2} x + \frac{\alpha_l}{l}-\frac{1}{l\cdot (l!)^2}
\,\,\,.
\end{eqnarray*}
A series of numbers $\alpha_l$ is defined 
by a recursion formula
\begin{eqnarray*}
%&&
\left\{
\begin{array}{l}
\alpha_{l+1} =\dis \frac{1}{l(l+1)} \alpha_l -\frac{2l+1}{l}\cdot 
\frac{1}{\left\{ (l+1)! \right\}^2} \,\,\,\,\,\,\,\,\,\,\,\,(l=1,2,\cdots)
\,\,\,,\\
\alpha_1 = -1 \,\,\,.
\end{array}
\right.
\end{eqnarray*}
The component $u_{21}^l$ is non-vanishing and
the ${\Pi}$ is not triangular matrix.
It is contrasted with Calabi-Yau cases.
If we want to consider effects by $t_0$,
all we have to do is to  multiply a matrix $\exp( t_0 \Kappa_0)$
to the above ${\Pi}$ from the left.\\
\mbox{}\\
$\bullet$ \underline{$CP^2$ model}\\

There are three elements $1$, $e$, and $e^2$ in cohomology groups
$\mbox{H}^{\ast}( CP^2 )$.
Background sources $t_0$, $t_1$, and $t_2$ couple with
associated operators $\co^{(0)}$, $\co^{(1)}$, $\co^{(2)}$.
Then a prepotential of $CP^2$ is realized in the following form
%\ba
%F=\frac{1}{2}t_0^2 t_2 +\frac{1}{2}t_0 t_1^2 
%+\sum_{d\geq 1} N_d \frac{t_2^{3d-1} e^{d\,t_1}}{(3d-1)!}\,\,\,.\nom
%\ea
\ba
F=t_0^2 t_2 +t_0 t_1^2 
+\sum_{d\geq 1} N_d \,\frac{t_2^{3d-1} e^{d\,t_1}}{(3d-1)!}\,\,\,.\nom
\ea
The set of numbers $N_d$ is determined by associativity relations
of fusion couplings.
With switching off $t_0$ and $t_2$ and setting $t_1 =x\neq 0$,
a fusion coupling induced by $\co^{(1)}$ is given as
\ba
&&\Kappa_1 =\left(
\matrix{0 & 1 & 0 \cr 0 & 0 & 1 \cr q & 0 & 0}\right)\,\,\,,\nom\\
&&{(\Kappa_1)_{\ell}}^m :={\Kappa_{1\ell}}^m\,\,\,.\nom
\ea
By using this, an explicit formula for ${\Pi}$ is
obtained in the $CP^2$ model
\begin{eqnarray*}
&&{\Pi} =\left(\matrix{1 & x & \dis \frac{x^2}{2} \cr
0 & 1 & x \cr 0 & 0 & 1}\right)
+{\sum_{l=1}^{\infty}}
\left(\matrix{\dis {u_{11}}^l & \dis {u_{12}}^l & \dis {u_{13}}^l \cr
\dis {u_{21}}^l & \dis {u_{22}}^l & \dis {u_{23}}^l \cr
\dis {u_{31}}^l & \dis {u_{32}}^l & \dis {u_{33}}^l }\right)\cdot q^l \,\,\,,\\
&&{u_{11}}^l = \prod^{l}_{m=1} \frac{1}{m^3} \,\,\,,\,\,\,
{u_{21}}^l = l\cdot \left(\prod^{l}_{m=1} \frac{1}{m^3} \right)\,\,\,,\,\,\,
{u_{31}}^l = l^2 \cdot \left(\prod^{l}_{m=1} \frac{1}{m^3}\right) \,\,\,,\\
&&{u_{12}}^l = \left( \prod^{l}_{m=1} \frac{1}{m^3} \right)
x +\left(\frac{\alpha_l}{l^2}-\frac{2}{l}\cdot \prod^{l}_{m=1} \frac{1}{m^3}
 \right)\,\,\,,\\
&&{u_{22}}^l = l\cdot \left(\prod^{l}_{m=1} \frac{1}{m^3} \right) x
+\left(\frac{\alpha_l}{l}- \prod^{l}_{m=1} \frac{1}{m^3}
 \right)\,\,\,,\\
&&{u_{32}}^l = l^2 \cdot \left(\prod^{l}_{m=1} \frac{1}{m^3}\right)x
+ \alpha_l \,\,\,,\\
&&{u_{13}}^l = \left( \prod^{l}_{m=1} \frac{1}{m^3} \right)
\frac{x^2}{2}
+\left(\frac{\alpha_l}{l^2}-\frac{2}{l}\cdot \prod^{l}_{m=1} \frac{1}{m^3}
 \right) x +
\left(\frac{\beta_l}{l^2}-
\frac{2}{l^3}{\alpha_l}+\frac{3}{l^2}\cdot \prod^{l}_{m=1} \frac{1}{m^3}
 \right) \,\,\,,\\
&&{u_{23}}^l = l\cdot \left(\prod^{l}_{m=1} \frac{1}{m^3} \right) \frac{x^2}{2}
+\left(\frac{\alpha_l}{l}- \prod^{l}_{m=1} \frac{1}{m^3}
 \right) x  +
\left(\frac{\beta_l}{l}-
\frac{1}{l^2}{\alpha_l}+\frac{1}{l}\cdot \prod^{l}_{m=1} \frac{1}{m^3}
 \right) \,\,\,,\\
&&{u_{33}}^l = l^2 \cdot \left(\prod^{l}_{m=1} \frac{1}{m^3}\right)
\frac{x^2}{2}
+ \alpha_l x + \beta_l \,\,\,.
\end{eqnarray*}
Two sets of numbers $\{\alpha_l\}$ and $\{\beta_l\}$
are calculated recursively by the following equations
\begin{eqnarray*}
&&
\left\{
\begin{array}{l}
\dis
\alpha_{l+1}=\frac{1}{l^2 (l+1)} \alpha_l -\frac{3l+2}{l (l+1)^2}
\cdot \left(\prod^{l}_{m=1} \frac{1}{m^3} \right) \,\,\,,\\
\dis \alpha_1 =-1 \,\,\,,
\end{array}
\right.\\
&&
\left\{
\begin{array}{l}
\dis \beta_{l+1}=\frac{1}{l^2 (l+1)} \beta_l -\frac{3l+2}{l^3 (l+1)^2} 
{\alpha_l}+\frac{6 l^2 +8 l+3}{l^2 (l+1)^3}
\cdot \left(\prod^{l}_{m=1} \frac{1}{m^3} \right) \,\,\,,\\
\dis \beta_1 =1 \,\,\,.
\end{array}
\right.
\end{eqnarray*}

%%%%%%%%%%%%%%%%%%%%%%%%%%%%%%%%%%%%%%%%%%%%%%%%%%%%%%%%%%
\mbox{}\\
$\bullet$ \underline{$CP^3$ model}\\

For the $CP^3$ case, a prepotential is represented as
%\ba
%F=\frac{1}{2}t_0^2 t_3 +t_0 t_1 t_2 +
%\frac{1}{6} t_1^3 
%+\sum_{d\geq 1}\sum_{l=0}^{2d}  
%N_{l,d} \frac{t_2^{4d-2l} t_3^l }{(4d-2l)!l!} e^{d\,t_1}\,\,\,.\nom
%\ea
\ba
F=\frac{3}{2}t_0^2 t_3 +3 t_0 t_1 t_2 +
\frac{1}{2} t_1^3 
+\sum_{d\geq 1}\sum_{l=0}^{2d}  
N_{l,d} \frac{t_2^{4d-2l} t_3^l }{(4d-2l)!l!} e^{d\,t_1}\,\,\,.\nom
\ea
The moduli parameters $t_m$ $(m=0,1,2,3)$ are associated with
cohomology elements $e^m$ and couple to A-model operators $\co^{(m)}$.
When turning off all parameters except for $t_1 =x$, we find a
fusion coupling for $\co^{(1)}$
\ba
\Kappa_1 =\left(
\matrix{0 & 1 & 0 & 0 \cr 0 & 0 & 1 & 0 \cr 0 & 0 & 0 & 1
\cr q & 0 & 0 & 0}\right)\,\,\,.\nom
\ea
Applying the same method as that in the previous examples,
we find a disk amplitude ${\Pi}$ for this model
\begin{eqnarray*}
&&{\Pi} =\left(\matrix{1 & \dis x & \dis \frac{x^2}{2} 
& \dis \frac{x^3}{6} \cr
0 & 1 & \dis x & \dis \frac{x^2}{2} \cr 0 & 0 & 1 & \dis x \cr
0 & 0 & 0 & 1 }\right)
+{\sum_{l=1}^{\infty}}
\left(\matrix{\dis {u_{11}}^l & \dis {u_{12}}^l & \dis {u_{13}}^l 
& \dis {u_{14}}^l \cr
\dis {u_{21}}^l & \dis {u_{22}}^l & \dis {u_{23}}^l 
& \dis {u_{24}}^l \cr
\dis {u_{31}}^l & \dis {u_{32}}^l & \dis {u_{33}}^l 
& \dis {u_{34}}^l
\cr
\dis {u_{41}}^l & \dis {u_{42}}^l & \dis {u_{43}}^l 
& \dis {u_{44}}^l
}\right)\cdot q^l \,\,\,,\\
&&{u_{11}}^l = \prod^{l}_{m=1} \frac{1}{m^4} \,\,\,,\,\,\,
{u_{21}}^l = l\cdot \left(\prod^{l}_{m=1} \frac{1}{m^4} \right)\,\,\,,\\
&&{u_{31}}^l = l^2 \cdot \left(\prod^{l}_{m=1} \frac{1}{m^4}\right) 
\,\,\,,\,\,\,
{u_{41}}^l = l^3 \cdot \left(\prod^{l}_{m=1} \frac{1}{m^4}\right) \,\,\,,\\
&&{u_{12}}^l =  \left(\prod^{l}_{m=1} \frac{1}{m^4} \right)
x +\left(\frac{\alpha_l}{l^3}- \frac{3}{l} \prod^{l}_{m=1} \frac{1}{m^4}
 \right) \,\,\,,\\
&&{u_{22}}^l =  l\cdot \left(\prod^{l}_{m=1} \frac{1}{m^4} \right)
x +\left(\frac{\alpha_l}{l^2}- 2  \prod^{l}_{m=1} \frac{1}{m^4}
 \right)\,\,\,, \\
&&{u_{32}}^l =  l^2 \cdot \left(\prod^{l}_{m=1} \frac{1}{m^4} \right)
x +\left(\frac{\alpha_l}{l}- l \prod^{l}_{m=1} \frac{1}{m^4}
 \right)\,\,\,, \\
&&{u_{42}}^l =  l^3 \cdot \left(\prod^{l}_{m=1} \frac{1}{m^4} \right)
x + {\alpha_l}\,\,\,, \\
&&{u_{13}}^l = \left( \prod^{l}_{m=1} \frac{1}{m^4} \right)
\frac{x^2}{2}
+\left(\frac{\alpha_l}{l^3}- \frac{3}{l} \prod^{l}_{m=1} \frac{1}{m^4}
 \right) x + \left(\frac{\beta_l}{l^3}-
\frac{3}{l^4}{\alpha_l}+\frac{6}{l^2}\cdot \prod^{l}_{m=1} \frac{1}{m^4}
 \right)\,\,\,, \\
&&{u_{23}}^l = l\cdot \left( \prod^{l}_{m=1} \frac{1}{m^4} \right)
\frac{x^2}{2}
+\left(\frac{\alpha_l}{l^2}- 2 \prod^{l}_{m=1} \frac{1}{m^4}
 \right) x + \left(\frac{\beta_l}{l^2}-
\frac{2}{l^3}{\alpha_l}+\frac{3}{l}\cdot \prod^{l}_{m=1} \frac{1}{m^4}
 \right)\,\,\,, \\
&&{u_{33}}^l = l^2 \cdot \left( \prod^{l}_{m=1} \frac{1}{m^4} \right)
\frac{x^2}{2}
+\left(\frac{\alpha_l}{l}- l \prod^{l}_{m=1} \frac{1}{m^4}
 \right) x + \left(\frac{\beta_l}{l}-
\frac{1}{l^2}{\alpha_l}+ \prod^{l}_{m=1} \frac{1}{m^4}
 \right)\,\,\,, \\
&&{u_{43}}^l = l^3 \cdot \left( \prod^{l}_{m=1} \frac{1}{m^4} \right)
\frac{x^2}{2} +{\alpha_l} x + {\beta_l}\,\,\,,\\
&&{u_{14}}^l = \left( \prod^{l}_{m=1} \frac{1}{m^4} \right)
\frac{x^3}{6}
+\left(\frac{\alpha_l}{l^3}- \frac{3}{l} \prod^{l}_{m=1} \frac{1}{m^4}
 \right) \frac{x^2}{2} + \left(\frac{\beta_l}{l^3}-
\frac{3}{l^4}{\alpha_l}+\frac{6}{l^2}\cdot \prod^{l}_{m=1} \frac{1}{m^4}
 \right) x\\
&&\,\,\,\,\,\,\,
+\left(\frac{\gamma_l}{l^3}-\frac{3}{l^4}{\beta_l}+\frac{6}{l^5}{\alpha_l}
-\frac{10}{l^3}
\prod^{l}_{m=1} \frac{1}{m^4} \right)\,\,\,, \\
&&{u_{24}}^l = l \cdot \left( \prod^{l}_{m=1} \frac{1}{m^4} \right)
\frac{x^3}{6}
+\left(\frac{\alpha_l}{l^2}- 2 \prod^{l}_{m=1} \frac{1}{m^4}
 \right) \frac{x^2}{2} + \left(\frac{\beta_l}{l^2}-
\frac{2}{l^3}{\alpha_l}+\frac{3}{l}\cdot \prod^{l}_{m=1} \frac{1}{m^4}
 \right) x\\
&&\,\,\,\,\,\,\,
+\left(\frac{\gamma_l}{l^2}-\frac{2}{l^3}{\beta_l}+\frac{3}{l^4}{\alpha_l}
-\frac{4}{l^2}
\prod^{l}_{m=1} \frac{1}{m^4} \right)\,\,\,, \\
&&{u_{34}}^l = l^2 \cdot \left( \prod^{l}_{m=1} \frac{1}{m^4} \right)
\frac{x^3}{6}
+\left(\frac{\alpha_l}{l}- l \prod^{l}_{m=1} \frac{1}{m^4}
 \right) \frac{x^2}{2} + \left(\frac{\beta_l}{l}-
\frac{1}{l^2}{\alpha_l}+ \prod^{l}_{m=1} \frac{1}{m^4}
 \right) x\\
&&\,\,\,\,\,\,\,
+\left(\frac{\gamma_l}{l}-\frac{1}{l^2}{\beta_l}+\frac{1}{l^3}{\alpha_l}
-\frac{1}{l}
\prod^{l}_{m=1} \frac{1}{m^4} \right)\,\,\,, \\
&&{u_{44}}^l = l^3 \cdot \left( \prod^{l}_{m=1} \frac{1}{m^4} \right)
\frac{x^3}{6} +{\alpha_l}\frac{x^2}{2} + {\beta_l} x +{\gamma_l}\,\,\,.
\end{eqnarray*}
The following three recursion formulae determine
three sets of numbers $\{\alpha_l\}$, $\{\beta_l\}$, $\{\gamma_l\}$
\begin{eqnarray*}
&&
\left\{
\begin{array}{l}
\dis
\alpha_{l+1}=\frac{1}{l^3 (l+1)} \alpha_l -
\left[\frac{1}{{(l+1)}^2}+\frac{3}{l (l+1)}\right]
\cdot \left(\prod^{l}_{m=1} \frac{1}{m^4} \right)\,\,\,, \cr
\dis \alpha_1 =-1 \,\,\,,
\end{array}
\right.\\
&&
\left\{
\begin{array}{l}
\dis
\beta_{l+1}=\frac{1}{l^3 (l+1)} \beta_l -
\left[\frac{1}{l^3 {(l+1)}^2}+\frac{3}{ l^4 (l+1)}\right]{\alpha_l}\\
\,\,\,\,\,
\dis +\left[\frac{1}{{(l+1)}^3}+\frac{3}{l {(l+1)}^2}+\frac{6}{l^2 {(l+1)}}
\right]
\cdot \left(\prod^{l}_{m=1} \frac{1}{m^4} \right)\,\,\,, \cr
\dis \beta_1 =1 \,\,\,,
\end{array}
\right.\\
&&
\left\{
\begin{array}{l}
\dis
\gamma_{l+1}=\frac{1}{l^3 (l+1)} \gamma_l -
\left[\frac{1}{l^3 {(l+1)}^2}+\frac{3}{ l^4 (l+1)}\right]{\beta_l}\\
\,\,\,\,\,
\dis +\left[\frac{1}{l^3 {(l+1)}^3}+\frac{3}{l^4 {(l+1)}^2}
+\frac{6}{l^5 {(l+1)}}
\right]{\alpha_l}\\
\,\,\,\,\, \dis-\left[\frac{1}{{(l+1)}^4}+\frac{3}{l {(l+1)}^3}
+\frac{6}{l^2 {(l+1)}^2}+\frac{10}{l^3 {(l+1)}}\right]
\cdot \left(\prod^{l}_{m=1} \frac{1}{m^4} \right)\,\,\,, \cr
\dis \gamma_1 =-1 \,\,\,.
\end{array}
\right.\\
\end{eqnarray*}
It is straightforward to extend the calculations here to
general $CP^N$ cases.
For this $CP^N$ case, each cohomology class with a fixed degree
is one dimensional space and all the cohomology elements
are generated by only one
K{\"a}hler form $e$.
As a second example, we will consider a Grassmann manifold
$Gr(2,4)$ that has a two dimensional cohomology group $\mbox{H}^4(Gr(2,4))$.\\
%\newpage
\mbox{}\\
$\bullet$ \underline{$Gr(2,4)$ model}\\

The $Gr(2,4)$ is 
realized as a $2$ dimensional subspace in ${\bf C}^4$.
It is a complex $4$ dimensional manifold and
has a 1st Chern class $c_1(TGr(2,4))=4 e$.
The $e$ is a K{\"a}hler form of the manifold.
Also Hodge numbers are known to be
$h^{0,0}=h^{1,1}=h^{3,3}=h^{4,4}=1$
and $h^{2,2}=2$.
The total cohomology classes are $6$ dimensional spaces and
there are six operators $\co^{(0)}$, $\co^{(1)}$, $\co^{(2)}_1$,
$\co^{(2)}_2$, $\co^{(3)}$, $\co^{(4)}$ associated with them.
We shall introduce moduli parameters $t_0$, $t_1$, 
$t_{2,1}$, $t_{2,2}$, $t_3$, and $t_4$ corresponding to
the above observables.
Topological metrics $\Eta_{(\ell ,i)(m,j)}
:=\ketbra{ \co^{(\ell)}_i \co^{(m)}_j}$ 
are non-zero for $\Eta_{(0)(4)}=$ $\Eta_{(1)(3)}=$
$\Eta_{(2,1)(2,1)}=$ $\Eta_{(2,2)(2,2)}=4$.
Then a prepotential of $Gr(2,4)$ is expressed as
%\ba
%F&=&\frac{1}{2}t_0^2 t_4 +\frac{1}{2} t_0 t_{2,1}^2 
%+\frac{1}{2} t_0 t_{2,2}^2 +
%t_0^2 t_1 t_3 +\frac{1}{2} t_1^2 t_{2,1} +
%\frac{1}{2} t_1^2 t_{2,2} \nom\\
%&&+\sum_{d\geq 1}\sum_{k,l,m\geq 0}  
%N_{k,l,m,d} 
%\frac{t_{2,1}^{4d+1-k-2l-3m} t_{2,2}^k t_3^l t_4^m }
%{(4d+1-k-2l-3m)!k!l!m!} e^{d\,t_1}\,\,\,.\nom
%\ea
\ba
F&=&{2}t_0^2 t_4 +{2} t_0 t_{2,1}^2 
+{2} t_0 t_{2,2}^2 +
4 t_0 t_1 t_3 +{2} t_1^2 t_{2,1} +
{2} t_1^2 t_{2,2} \nom\\
&&+\sum_{d\geq 1}\sum_{k,l,m\geq 0}  
N_{k,l,m,d} \,
\frac{t_{2,1}^{4d+1-k-2l-3m}\, t_{2,2}^k \, t_3^l \, t_4^m }
{(4d+1-k-2l-3m)!k!l!m!} \, e^{d\,t_1}\,\,\,.\nom
\ea
When one switches on only marginal parameter $t_1=x\neq 0$ in the
background,
fusion couplings for these operators $\co^{(m)}$'s
are calculated
\begin{eqnarray*}
\begin{array}{lcllcl}
\Kappa_{0} & = &\left(
\matrix{ 1 & 0 & 0 & 0 & 0 & 0 \cr 0 & 1 & 0 & 0 & 0 & 0 \cr 0 & 0 & 1 & 0 & 0
   & 0 \cr 0 & 0 & 0 & 1 & 0 & 0 \cr 0 & 0 & 0 & 0 & 1 & 0 \cr 0 & 0 & 0 & 0
   & 0 & 1 \cr  }
\right) \,\,\,,\,\,\,&
\Kappa_{1} &= & \left(
\matrix{ 0 & 1 & 0 & 0 & 0 & 0 \cr 0 & 0 & 1 & 1 & 0 & 0 \cr 0 & 0 & 0 & 0 & 1
   & 0 \cr 0 & 0 & 0 & 0 & 1 & 0 \cr q & 0 & 0 & 0 & 0 & 1 \cr 0 & q & 0 & 0
   & 0 & 0 \cr  }\right)  \,\,\,,\\
 & & & & & \\
\Kappa_{2,1}& =& \left(
\matrix{ 0 & 0 & 1 & 0 & 0 & 0 \cr 0 & 0 & 0 & 0 & 1 & 0 \cr 0 & 0 & 0 & 0 & 0
   & 1 \cr q & 0 & 0 & 0 & 0 & 0 \cr 0 & q & 0 & 0 & 0 & 0 \cr 0 & 0 & 0 & q
   & 0 & 0 \cr  } \right) \,\,\,,\,\,\,&
\Kappa_{2,2} & = & \left(
\matrix{ 0 & 0 & 0 & 1 & 0 & 0 \cr 0 & 0 & 0 & 0 & 1 & 0 \cr q & 0 & 0 & 0 & 0
   & 0 \cr 0 & 0 & 0 & 0 & 0 & 1 \cr 0 & q & 0 & 0 & 0 & 0 \cr 0 & 0 & q & 0
   & 0 & 0 \cr  } \right) \,\,\,,\\
 & & & & & \\
\Kappa_{3} & = & \left(
\matrix{ 0 & 0 & 0 & 0 & 1 & 0 \cr q & 0 & 0 & 0 & 0 & 1 \cr 0 & q & 0 & 0 & 0
   & 0 \cr 0 & q & 0 & 0 & 0 & 0 \cr 0 & 0 & q & q & 0 & 0 \cr 0 & 0 & 0 & 0
   & q & 0 \cr  }\right)  \,\,\,,\,\,\,&
\Kappa_{4} & = & \left(
\matrix{ 0 & 0 & 0 & 0 & 0 & 1 \cr 0 & q & 0 & 0 & 0 & 0 \cr 0 & 0 & 0 & q & 0
   & 0 \cr 0 & 0 & q & 0 & 0 & 0 \cr 0 & 0 & 0 & 0 & q & 0 \cr 0 & 0 & 0
   & 0 & 0 & 0 \cr  } \right) \,\,\,.
\end{array}
\end{eqnarray*}
Here we introduce a variable $q:=\exp (x)$.
By performing the similar calculations for the previous examples,
a disk amplitude for this $Gr(2,4)$ model is written down concretely
\begin{eqnarray*}
&&{\Pi} =
%\left(\matrix{1 & \dis x & \dis \frac{x^2}{2}  & \dis \frac{x^2}{2}
%& \dis \frac{x^3}{6} & \dis \frac{x^4}{12} \cr
%0 & 1 & \dis x & \dis x & \dis {x^2}& \dis \frac{x^3}{6}
% \cr 0 & 0 & 1 & 0 & \dis x & \dis \frac{x^2}{2} \cr
%0 & 0 & 0 & 1 & \dis x & \dis \frac{x^2}{2} \cr 
%0 & 0 & 0 & 0 & 1 & \dis x \cr
%0 & 0 & 0 & 0 & 0 & 1
%}\right)
\left(\matrix{\dis 1 & x & \dis \frac{x^2}{2} {\bf e}^t
& \dis  \frac{x^3}{6} & \dis  \frac{x^4}{12} \cr
\dis 0 & \dis 1 & \dis x {\bf e}^t
& \dis {x^2} & \dis \frac{x^3}{6} \cr
\dis 0 {\bf e} & \dis 0 {\bf e} & \dis 1 {\bf e}
{\bf e}^t & \dis x {\bf e}& \dis  \frac{x^2}{2} {\bf e} \cr
\dis 0 & \dis 0 & \dis 0 {\bf e}^t
& \dis 1 & \dis x \cr
\dis 0 & \dis 0 & \dis 0 {\bf e}^t
& \dis 0 & \dis 1
}\right)
+{\sum_{l=1}^{\infty}}
\left(\matrix{\dis {u_{11}}^l & \dis {u_{12}}^l & \dis {u_{13}}^l {\bf e}^t
& \dis {u_{14}}^l & \dis {u_{15}}^l \cr
\dis {u_{21}}^l & \dis {u_{22}}^l & \dis {u_{23}}^l {\bf e}^t
& \dis {u_{24}}^l & \dis {u_{15}}^l \cr
\dis {u_{31}}^l {\bf e} & \dis {u_{32}}^l {\bf e} & \dis {u_{33}}^l {\bf e}
{\bf e}^t & \dis {u_{34}}^l {\bf e}& \dis {u_{35}}^l {\bf e} \cr
\dis {u_{41}}^l & \dis {u_{42}}^l & \dis {u_{43}}^l {\bf e}^t
& \dis {u_{44}}^l& \dis {u_{45}}^l \cr
\dis {u_{51}}^l & \dis {u_{52}}^l & \dis {u_{53}}^l {\bf e}^t
& \dis {u_{54}}^l& \dis {u_{55}}^l 
}\right)\cdot q^l\,\,\,,\\
&& {\bf e}:=\left(\matrix{1 \cr 1}\right)\,\,\,,
\end{eqnarray*}
where
each component ${u_{ij}}^l$ in the series expansion is given as
\begin{eqnarray*}
&&{u_{11}}^l = 2^l \cdot \prod^{l}_{m=1} \frac{2m-1}{m^5} \,\,\,,\,\,\,
{u_{21}}^l = 2^l \cdot l\cdot \left(\prod^{l}_{m=1} 
\frac{2m-1}{m^5} \right)\,\,\,,\\
&&{u_{31}}^l = 2^{l-1}\cdot l^2 \cdot \left(\prod^{l}_{m=1} 
\frac{2m-1}{m^5}\right) \,\,\,,\,\,\,
{u_{41}}^l =  2^{l-1}\cdot l^3 \cdot 
\left(\prod^{l}_{m=1} \frac{2m-1}{m^5}\right) \,\,\,,\\
&&{u_{51}}^{l=1} =0\,\,\,,\\
&&{u_{51}}^{l+1} =  2^{l}\cdot \frac{l}{l+1} \cdot 
\left(\prod^{l}_{m=1} \frac{2m-1}{m^5}\right)\,\,\,\,(l=1,2,\cdots ) \,\,\,,\\
&&{u_{12}}^{l} =2^l \cdot \left(\prod^{l}_{m=1} 
\frac{2m-1}{m^5} \right) x+\frac{2}{l^3} \alpha_l -\frac{3}{l}\cdot 2^l \cdot
\left(\prod^{l}_{m=1} \frac{2m-1}{m^5} \right)\,\,\,,\\
&&{u_{22}}^{l} =2^l \cdot l\cdot \left(\prod^{l}_{m=1} 
\frac{2m-1}{m^5} \right) x+\frac{2}{l^2} \alpha_l -{2}\cdot  2^l \cdot
\left(\prod^{l}_{m=1} \frac{2m-1}{m^5} \right)\,\,\,,\\
&&{u_{32}}^{l} =2^{l-1} \cdot l^2 \cdot \left(\prod^{l}_{m=1} 
\frac{2m-1}{m^5} \right) x+\frac{1}{l} \alpha_l -  2^{l-1} \cdot l \cdot
\left(\prod^{l}_{m=1} \frac{2m-1}{m^5} \right)\,\,\,,\\
&&{u_{42}}^{l} =2^{l-1} \cdot l^3 \cdot \left(\prod^{l}_{m=1} 
\frac{2m-1}{m^5} \right) x+ \alpha_l \,\,\,,\\
&&{u_{52}}^{l=1} =1\,\,\,,\\
&&{u_{52}}^{l+1} =2^{l} \cdot \frac{l}{l+1} \cdot \left(\prod^{l}_{m=1} 
\frac{2m-1}{m^5} \right) x+\frac{2}{l^2 (l+1)} 
\alpha_l \\
&& \,\,\,\,\,
-  2^{l} \cdot \left[\frac{2}{l+1}+\frac{l}{{(l+1)}^2}\right] \cdot
\left(\prod^{l}_{m=1} \frac{2m-1}{m^5} \right)\,\,\,\,(l=1,2,\cdots )\,\,\,,\\
%\end{eqnarray*}
%\begin{eqnarray*}
&&{u_{13}}^{l} =2^l \cdot \left(\prod^{l}_{m=1} 
\frac{2m-1}{m^5} \right) \frac{x^2}{2}
+\Biggl\{\frac{2}{l^3} \alpha_l -\frac{3}{l}\cdot 2^l \cdot
\left(\prod^{l}_{m=1} \frac{2m-1}{m^5} \right)\Biggr\} x\\
&&\,\,\,\,\, -\frac{6}{l^4} \alpha_l +\frac{2}{l^3} \beta_l  
+\frac{6}{l^2} \cdot 2^{l} \cdot 
\left(\prod^{l}_{m=1} \frac{2m-1}{m^5}\right) \,\,\,,
\\
&&{u_{23}}^{l} =2^l \cdot l\cdot \left(\prod^{l}_{m=1} 
\frac{2m-1}{m^5} \right) \frac{x^2}{2}
+\Biggl\{ \frac{2}{l^2} \alpha_l -{2}\cdot  2^l \cdot
\left(\prod^{l}_{m=1} \frac{2m-1}{m^5} \right) \Biggr\} x\\
&&\,\,\,\,\, -\frac{4}{l^3} \alpha_l +\frac{2}{l^2} \beta_l
+\frac{3}{l}\cdot  2^{l} \cdot 
\left(\prod^{l}_{m=1} \frac{2m-1}{m^5}\right) \,\,\,,\\
&&{u_{33}}^{l} =2^{l-1} \cdot l^2 \cdot \left(\prod^{l}_{m=1} 
\frac{2m-1}{m^5} \right) \frac{x^2}{2}
+\Biggl\{ \frac{1}{l} \alpha_l -  2^{l-1} \cdot l \cdot
\left(\prod^{l}_{m=1} \frac{2m-1}{m^5} \right)\Biggr\} x \\
&&\,\,\,\,\, -\frac{1}{l^2} \alpha_l +\frac{1}{l} \beta_l +
2^{l-1}\cdot 
\left(\prod^{l}_{m=1} \frac{2m-1}{m^5}\right) \,\,\,,\\
&&{u_{43}}^{l} =2^{l-1} \cdot l^3 \cdot \left(\prod^{l}_{m=1} 
\frac{2m-1}{m^5} \right) \frac{x^2}{2}
+ \alpha_l x +\beta_l \,\,\,,\\
&&{u_{53}}^{l=1} =-1+x\,\,\,,\\
&&{u_{53}}^{l+1} =2^{l} \cdot \frac{l}{l+1} \cdot \left(\prod^{l}_{m=1} 
\frac{2m-1}{m^5} \right) \frac{x^2}{2}
+\Biggl\{ \frac{2}{l^2 (l+1)} 
\alpha_l \\
&& \,\,\,\,\,
-  2^{l} \cdot \left[\frac{2}{l+1}+\frac{l}{{(l+1)}^2}\right] \cdot
\left(\prod^{l}_{m=1} \frac{2m-1}{m^5} \right)\Biggr\} x \\
&&\,\,\,\,\, -\Biggl[ \frac{2}{l^2 (l+1)^2}+\frac{4}{l^3 (l+1)^3}
\Biggr] \alpha_l +\frac{2}{l^2 (l+1)} \beta_l \\
&&\,\,\,\,\, 
+\Biggl[ \frac{l}{ (l+1)^3 }+\frac{2}{ (l+1)^2 }+\frac{3}{l (l+1) }
\Biggr]\cdot  2^{l}\cdot 
\left(\prod^{l}_{m=1} \frac{2m-1}{m^5}\right) \,\,\,,\\
%\end{eqnarray*}
%\begin{eqnarray*}
&&{u_{14}}^{l} =2^l \cdot \left(\prod^{l}_{m=1} 
\frac{2m-1}{m^5} \right) \frac{x^3}{6}
+\Biggl\{\frac{2}{l^3} \alpha_l -\frac{3}{l}\cdot 2^l \cdot
\left(\prod^{l}_{m=1} \frac{2m-1}{m^5} \right)\Biggr\} \frac{x^2}{2}\\
&&\,\,\,\,\, +\Biggl\{-\frac{6}{l^4} \alpha_l +\frac{2}{l^3} \beta_l  
+\frac{6}{l^2} \cdot 2^{l} \cdot 
\left(\prod^{l}_{m=1} \frac{2m-1}{m^5}\right) \Biggr\}x \\
&&\,\,\,\,\, + \frac{12}{l^5} \alpha_l 
-\frac{6}{l^4} \beta_l +\frac{2}{l^3}\gamma_l
-\frac{10}{l^3}\cdot 2^{l}\cdot
\left(\prod^{l}_{m=1} \frac{2m-1}{m^5}\right) \,\,\,,
\\
&&{u_{24}}^{l} =2^l \cdot l\cdot \left(\prod^{l}_{m=1} 
\frac{2m-1}{m^5} \right) \frac{x^3}{6}
+\Biggl\{ \frac{2}{l^2} \alpha_l -{2}\cdot  2^l \cdot
\left(\prod^{l}_{m=1} \frac{2m-1}{m^5} \right) \Biggr\} \frac{x^2}{2}\\
&&\,\,\,\,\, +\Biggl\{ -\frac{4}{l^3} \alpha_l +\frac{2}{l^2} \beta_l
+\frac{3}{l}\cdot  2^{l} \cdot 
\left(\prod^{l}_{m=1} \frac{2m-1}{m^5}\right) \Biggr\} x \\
&&\,\,\,\,\, + 
\frac{6}{l^4}\alpha_l -\frac{4}{l^3}\beta_l
+\frac{2}{l^2}\gamma_l -\frac{4}{l^2}\cdot
2^{l}\cdot
\left(\prod^{l}_{m=1} \frac{2m-1}{m^5}\right) \,\,\,,
\\
&&{u_{34}}^{l} =2^{l-1} \cdot l^2 \cdot \left(\prod^{l}_{m=1} 
\frac{2m-1}{m^5} \right) \frac{x^3}{6}
+\Biggl\{ \frac{1}{l} \alpha_l -  2^{l-1} \cdot l \cdot
\left(\prod^{l}_{m=1} \frac{2m-1}{m^5} \right)\Biggr\} \frac{x^2}{2} \\
&&\,\,\,\,\, +\Biggl\{ -\frac{1}{l^2} \alpha_l +\frac{1}{l} \beta_l +
2^{l-1}\cdot 
\left(\prod^{l}_{m=1} \frac{2m-1}{m^5}\right) \Biggr\} x \\
&&\,\,\,\,\, +
\frac{1}{l^3}\alpha_l -\frac{1}{l^2}\beta_l +\frac{1}{l}\gamma_l
-\frac{1}{l}\cdot
2^{l-1}\cdot 
\left(\prod^{l}_{m=1} \frac{2m-1}{m^5}\right) \,\,\,,
\\
&&{u_{44}}^{l} =2^{l-1} \cdot l^3 \cdot \left(\prod^{l}_{m=1} 
\frac{2m-1}{m^5} \right) \frac{x^3}{6}
+ \alpha_l \frac{x}{2} +\beta_l {x} +\gamma_l \,\,\,,\\
&&{u_{54}}^{l=1} =2-2x+x^2 \,\,\,,\\
&&{u_{54}}^{l+1} =2^{l} \cdot \frac{l}{l+1} \cdot \left(\prod^{l}_{m=1} 
\frac{2m-1}{m^5} \right) \frac{x^3}{6}
+\Biggl\{ \frac{2}{l^2 (l+1)} 
\alpha_l \\
&& \,\,\,\,\,
-  2^{l} \cdot \left[\frac{2}{l+1}+\frac{l}{{(l+1)}^2}\right] \cdot
\left(\prod^{l}_{m=1} \frac{2m-1}{m^5} \right)\Biggr\} \frac{x^2}{2} \\
&&\,\,\,\,\, +\Biggl\{
-\Biggl[ \frac{2}{l^2 (l+1)^2}+\frac{4}{l^3 (l+1)^3}
\Biggr] \alpha_l +\frac{2}{l^2 (l+1)} \beta_l \\
&&\,\,\,\,\, 
+\Biggl[ \frac{l}{ (l+1)^3 }+\frac{2}{ (l+1)^2 }+\frac{3}{l (l+1) }
\Biggr]\cdot  2^{l}\cdot 
\left(\prod^{l}_{m=1} \frac{2m-1}{m^5}\right) \Biggr\}x\\
&&\,\,\,\,\,
+\left[
\frac{2}{l^2 (l+1)^3 }+
\frac{4}{l^3 (l+1)^2 }+
\frac{6}{l^4 (l+1) }
\right]\alpha_l
-\left[
\frac{2}{l^2 (l+1)^2 }+
\frac{4}{l^3 (l+1) }
\right]\beta_l \\
&&\,\,\,\,\,
+\frac{2}{l^2 (l+1)}\gamma_l
-\left[
\frac{l}{ (l+1)^4}+
\frac{2}{ (l+1)^3}+
\frac{3}{l (l+1)^2}+
\frac{4}{l^2 (l+1) }
\right]\cdot 2^{l}\cdot 
\left(\prod^{l}_{m=1} \frac{2m-1}{m^5}\right) \,\,\,,\\
%\end{eqnarray*}
%\begin{eqnarray*}
&&{u_{15}}^{l} =2^l \cdot \left(\prod^{l}_{m=1} 
\frac{2m-1}{m^5} \right) \frac{x^4}{24}
+\Biggl\{\frac{2}{l^3} \alpha_l -\frac{3}{l}\cdot 2^l \cdot
\left(\prod^{l}_{m=1} \frac{2m-1}{m^5} \right)\Biggr\} \frac{x^3}{6}\\
&&\,\,\,\,\, +\Biggl\{-\frac{6}{l^4} \alpha_l +\frac{2}{l^3} \beta_l  
+\frac{6}{l^2} \cdot 2^{l} \cdot 
\left(\prod^{l}_{m=1} \frac{2m-1}{m^5}\right) \Biggr\}\frac{x^2}{2} \\
&&\,\,\,\,\, + \Biggl\{ \frac{12}{l^5} \alpha_l 
-\frac{6}{l^4} \beta_l +\frac{2}{l^3}\gamma_l
-\frac{10}{l^3}\cdot 2^{l}\cdot
\left(\prod^{l}_{m=1} \frac{2m-1}{m^5}\right) \Biggr\}x
\\
&&\,\,\,\,\, -\frac{20}{l^6}\alpha_l
+\frac{12}{l^5}\beta_l -\frac{6}{l^4}\gamma_l +\frac{2}{l^3}\delta_l
+\frac{15}{l^4}\cdot
2^{l}\cdot 
\left(\prod^{l}_{m=1} \frac{2m-1}{m^5}\right) \,\,\,,
\\
&&{u_{25}}^{l} =2^l \cdot l\cdot \left(\prod^{l}_{m=1} 
\frac{2m-1}{m^5} \right) \frac{x^4}{24}
+\Biggl\{ \frac{2}{l^2} \alpha_l -{2}\cdot  2^l \cdot
\left(\prod^{l}_{m=1} \frac{2m-1}{m^5} \right) \Biggr\} \frac{x^3}{6}\\
&&\,\,\,\,\, +\Biggl\{ -\frac{4}{l^3} \alpha_l +\frac{2}{l^2} \beta_l
+\frac{3}{l}\cdot  2^{l} \cdot 
\left(\prod^{l}_{m=1} \frac{2m-1}{m^5}\right) \Biggr\} \frac{x^2}{2} \\
&&\,\,\,\,\, + \Biggl\{
\frac{6}{l^4}\alpha_l -\frac{4}{l^3}\beta_l
+\frac{2}{l^2}\gamma_l -\frac{4}{l^2}\cdot
2^{l}\cdot
\left(\prod^{l}_{m=1} \frac{2m-1}{m^5}\right) \Biggr\}x\\
&&\,\,\,\,\, -\frac{8}{l^5}\alpha_l
+\frac{6}{l^4}\beta_l -\frac{4}{l^3}\gamma_l +\frac{2}{l^2}\delta_l
+\frac{5}{l^3}\cdot
2^{l}\cdot 
\left(\prod^{l}_{m=1} \frac{2m-1}{m^5}\right) \,\,\,,
\\
&&{u_{35}}^{l} =2^{l-1} \cdot l^2 \cdot \left(\prod^{l}_{m=1} 
\frac{2m-1}{m^5} \right) \frac{x^4}{24}
+\Biggl\{ \frac{1}{l} \alpha_l -  2^{l-1} \cdot l \cdot
\left(\prod^{l}_{m=1} \frac{2m-1}{m^5} \right)\Biggr\} \frac{x^3}{6} \\
&&\,\,\,\,\, +\Biggl\{ -\frac{1}{l^2} \alpha_l +\frac{1}{l} \beta_l +
2^{l-1}\cdot  
\left(\prod^{l}_{m=1} \frac{2m-1}{m^5}\right) \Biggr\} \frac{x^2}{2} \\
&&\,\,\,\,\, +\Biggl\{
\frac{1}{l^3}\alpha_l -\frac{1}{l^2}\beta_l +\frac{1}{l}\gamma_l
-\frac{1}{l}\cdot
2^{l-1}\cdot 
\left(\prod^{l}_{m=1} \frac{2m-1}{m^5}\right) \Biggr\}x
\\
&&\,\,\,\,\, 
 -\frac{1}{l^4}\alpha_l
+\frac{1}{l^3}\beta_l -\frac{1}{l^2}\gamma_l +\frac{1}{l}\delta_l
+\frac{1}{l^2}\cdot
2^{l-1}\cdot 
\left(\prod^{l}_{m=1} \frac{2m-1}{m^5}\right) \,\,\,,
\\
&&{u_{45}}^{l} =2^{l-1} \cdot l^3 \cdot \left(\prod^{l}_{m=1} 
\frac{2m-1}{m^5} \right) \frac{x^4}{24}
+ \alpha_l \frac{x^3}{6} +\beta_l \frac{x^2}{2} +\gamma_l x+\delta_l \,\,\,,\\
&&{u_{55}}^{l=1} =-2+2x-x^2+\frac{x^3}{3}\,\,\,,\\
&&{u_{55}}^{l+1} =2^{l} \cdot \frac{l}{l+1} \cdot \left(\prod^{l}_{m=1} 
\frac{2m-1}{m^5} \right) \frac{x^4}{24}
+\Biggl\{ \frac{2}{l^2 (l+1)} 
\alpha_l \\
&& \,\,\,\,\,
-  2^{l} \cdot \left[\frac{2}{l+1}+\frac{l}{{(l+1)}^2}\right] \cdot
\left(\prod^{l}_{m=1} \frac{2m-1}{m^5} \right)\Biggr\} \frac{x^3}{6} \\
&&\,\,\,\,\, +\Biggl\{
-\Biggl[ \frac{2}{l^2 (l+1)^2}+\frac{4}{l^3 (l+1)^3}
\Biggr] \alpha_l +\frac{2}{l^2 (l+1)} \beta_l \\
&&\,\,\,\,\, 
+\Biggl[ \frac{l}{ (l+1)^3 }+\frac{2}{ (l+1)^2 }+\frac{3}{l (l+1) }
\Biggr]\cdot  2^{l}\cdot 
\left(\prod^{l}_{m=1} \frac{2m-1}{m^5}\right) \Biggr\}\frac{x^2}{2}\\
&&\,\,\,\,\,
+\Biggl\{\left[
\frac{2}{l^2 (l+1)^3 }+
\frac{4}{l^3 (l+1)^2 }+
\frac{6}{l^4 (l+1) }
\right]\alpha_l
-\left[
\frac{2}{l^2 (l+1)^2 }+
\frac{4}{l^3 (l+1) }
\right]\beta_l \\
&&\,\,\,\,\,
+\frac{2}{l^2 (l+1)}\gamma_l
-\left[
\frac{l}{ (l+1)^4}+
\frac{2}{ (l+1)^3}+
\frac{3}{l (l+1)^2}+
\frac{4}{l^2 (l+1) }
\right]\cdot 2^{l}\cdot 
\left(\prod^{l}_{m=1} \frac{2m-1}{m^5}\right) \Biggr\}x\\
&& \dis \,\,\,\,\,
-\left[
\frac{2}{l^2 (l+1)^4 }+
\frac{4}{l^3 (l+1)^3 }+
\frac{6}{l^4 (l+1)^2 }+
\frac{8}{l^5 (l+1) }
\right]\alpha_l \\
&& \dis \,\,\,\,\,
+\left[
\frac{2}{l^2 (l+1)^3 }+
\frac{4}{l^3 (l+1)^2 }+
\frac{6}{l^4 (l+1) }
\right]\beta_l \\
&& \dis \,\,\,\,\,
-\left[
\frac{2}{l^2 (l+1)^2 }+
\frac{4}{l^3 (l+1) }
\right]\gamma_l %\\
%&& \dis \,\,\,\,\,
+
%\left[
\frac{2}{l^2 (l+1) }
%\right]
\delta_l \\
&& \dis \,\,\,\,\,
+\left[
\frac{l}{ (l+1)^5 }+
\frac{2}{ (l+1)^4 }+
\frac{3}{l (l+1)^3 }+
\frac{4}{l^2 (l+1)^2 }+
\frac{5}{l^3 (l+1) }
\right]\cdot
2^{l}\cdot 
\left(\prod^{l}_{m=1} \frac{2m-1}{m^5}\right) \,\,\,.
\end{eqnarray*}
The four sets of numbers $\alpha_l$, $\beta_l$, $\gamma_l$,
$\delta_l$ in the above expressions are defined recursively by the
following formulae
\begin{eqnarray*}
&&\left\{
\begin{array}{l}
\dis\alpha_{l+1}=\left[\frac{2}{l^2 {(l+1)}^2}+\frac{2}{l^3 (l+1)}
\right] \alpha_l  \\
\dis \,\,\,\,\,\,- 2^l \cdot 
\left[\frac{2l}{(l+1)^3}+\frac{3}{(l+1)^2}+\frac{3}{l (l+1)}\right]\cdot 
\left(\prod^{l}_{m=1} \frac{2m-1}{m^5} \right)\,\,\,,\\
\dis \alpha_{l=1}=0\,\,\,,\\
\end{array}
\right.
\\
%\end{eqnarray*}
%\begin{eqnarray*}
&&\left\{
\begin{array}{l}
\dis\beta_{l+1}=\left[
\frac{2}{l^2 (l+1)^2 }+\frac{2}{l^3 (l+1) }
\right] \beta_l  \\
 \dis \,\,\,\,\,\, -\left[\frac{2}{l^2 (l+1)^3 }+\frac{3}{l^3 (l+1)^2}
+\frac{3}{l^4 (l+1)}\right] \alpha_l \\
\dis  \,\,\,\,\,\,
+\left[
\frac{3l}{ (l+1)^4 }+\frac{5}{ (l+1)^3 }+\frac{6}{l (l+1)^2 }+
\frac{6}{l^2 (l+1) }
\right] \cdot 2^{l}\cdot
\left(\prod^{l}_{m=1} \frac{2m-1}{m^5}\right) \,\,\,,\\
\dis \beta_{l=1}=0\,\,\,,\\
\end{array}
\right.
\\
%\end{eqnarray*}
%\begin{eqnarray*}
&&\left\{
\begin{array}{l}
\dis\gamma_{l+1}=\left[
\frac{2}{l^2 (l+1)^2 }+\frac{2}{l^3 (l+1) }
\right] \gamma_l  \\
 \dis \,\,\,\,\,\, -\left[\frac{2}{l^2 (l+1)^3 }+\frac{3}{l^3 (l+1)^2}
+\frac{3}{l^4 (l+1)}\right] \beta_l \\
\dis  \,\,\,\,\,\,
+\left[
\frac{6}{l^2 (l+1)^4 }+\frac{10}{l^3 (l+1)^3 }+\frac{12}{l^4 (l+1)^2 }+
\frac{12}{l^5 (l+1) }
\right] \alpha_l \\
\dis  \,\,\,\,\,\,
-\left[
\frac{4l}{ (l+1)^5 }+
\frac{7}{ (l+1)^4 }+
\frac{9}{l (l+1)^3 }+
\frac{10}{l^2 (l+1)^2 }+
\frac{10}{l^3 (l+1) }
\right]
\cdot 2^{l}\cdot
\left(\prod^{l}_{m=1} \frac{2m-1}{m^5}\right) \,\,\,,\\
\dis \gamma_{l=1}=0\,\,\,,\\
\end{array}
\right.
\\
%\end{eqnarray*}
%\begin{eqnarray*}
&&\left\{
\begin{array}{l}
\dis\delta_{l+1}=\left[
\frac{2}{l^2 (l+1)^2 }+\frac{2}{l^3 (l+1) }
\right] \delta_l  \\
 \dis \,\,\,\,\,\, -\left[\frac{2}{l^2 (l+1)^3 }+\frac{3}{l^3 (l+1)^2}
+\frac{3}{l^4 (l+1)}\right] \gamma_l \\
\dis  \,\,\,\,\,\,
+\left[
\frac{6}{l^2 (l+1)^4 }+\frac{10}{l^3 (l+1)^3 }+\frac{12}{l^4 (l+1)^2 }+
\frac{12}{l^5 (l+1) }
\right] \beta_l \\
\dis \,\,\,\,\,\,
 -\left[
\frac{8}{l^2 (l+1)^5 }+
\frac{14}{l^3 (l+1)^4 }+
\frac{18}{l^4 (l+1)^3 }+
\frac{20}{l^5 (l+1)^2 }+
\frac{20}{l^6 (l+1) }
\right]\alpha_l \\
\dis  \,\,\,\,\,\,
+\left[
\frac{5l}{ (l+1)^6 }+
\frac{9}{ (l+1)^5 }+
\frac{12}{l (l+1)^4 }+
\frac{14}{l^2 (l+1)^3 }+
\frac{15}{l^3 (l+1)^2 }+
\frac{15}{l^4 (l+1) }
\right]\\
\dis  \,\,\,\,\,\,
\times 2^{l}\cdot
\left(\prod^{l}_{m=1} \frac{2m-1}{m^5}\right) \,\,\,,\\
\dis \delta_{l=1}=0\,\,\,.\\
\end{array}
\right.
\end{eqnarray*}
It is straightforward to turn on all the other background parameters
$\{ t_l \}$ in the above analysis
but the calculation is tedious. 

\newpage

%%%%%%%%%%%%%%%%%%%%%%%%%%%%%%%%%%%%%%%%%%%%%%%%%%%%%%%%%%
$\bullet$ \underline{$M_{4,3}$ model}\\

As a final example we consider a degree $3$ hypersurface $M_{4,3}$ in 
$CP^4$. It is a complex $3$ dimensional manifold with
a positive 1st Chern class $c_1 (T M_{4,3})=2 e$.
The $e$ is a K{\"a}hler form of this $M_{4,3}$.
The non-vanishing Hodge numbers are
$h^{0,0}=h^{1,1}=h^{2,2}=h^{3,3}=1$ and
$h^{2,1}=h^{1,2}=5$.
To avoid complexities in calculations, 
we focus on the vertical parts $\oplus_{\ell =0}^{3}
\mbox{H}^{\ell ,\ell } $ of cohomology classes
here.
Then there are associated four operators $\co^{(m)}$ $(m=0,1,2,3)$
and non-vanishing components of
topological metric lie on off-diagonal parts 
$\Eta_{lm}:=\ketbra{\co^{(l)}\co^{(m)}}=3\cdot \delta_{l+m,3}$.
The prepotential for $M_{4,3}$ is similar to that in the $CP^3$ case
%\ba
%F=\frac{1}{2}t_0^2 t_3 +t_0 t_1 t_2 +
%\frac{1}{6} t_1^3 
%+\sum_{d\geq 1}\sum_{l=0}^{d}  
%N_{l,d} \frac{t_2^{2d-2l} t_3^l }{(2d-2l)!l!} e^{d\,t_1}\,\,\,.\nom
%\ea
\ba
F=\frac{3}{2}t_0^2 t_3 +3 t_0 t_1 t_2 +
\frac{1}{2} t_1^3 
+\sum_{d\geq 1}\sum_{l=0}^{d}  
N_{l,d} \,\frac{t_2^{2d-2l} \, t_3^l }{(2d-2l)!l!}\, e^{d\,t_1}\,\,\,.\nom
\ea
A fusion coupling of the $\co^{(1)}$ is represented when
we turn off all parameters except for $t_1 =x$
\begin{eqnarray*}
&&\Kappa_{1}=\left(
\matrix{ 0 & 1 & 0 & 0 \cr 6\,q & 0 & 1 & 0 \cr 0 & 15\,q & 0 & 1 \cr 
  36\,{q^2} & 0 & 6\,q & 0 \cr  }  \right)\,\,\,.\nom
\end{eqnarray*}
By performing multiple integrals of products of $\Kappa_1$,
we find the disk amplitude ${\Pi}$ for this $M_{4,3}$
\begin{eqnarray*}
&&{\Pi} =
\left(\matrix{ 1 & \dis x &\dis  {{{x^2}}\over 2} & 
\dis {{{x^3}}\over 6} \cr 0 & 1 & x & \dis 
{{{x^2}}\over 2} \cr 0 & 0 & 1 & x \cr 0 & 0 & 0 & 1 \cr  }  \right) 
+{\sum_{l=1}^{\infty}}
\left(\matrix{\dis {u_{11}}^l & \dis {u_{12}}^l & \dis {u_{13}}^l 
& \dis {u_{14}}^l  \cr
\dis {u_{21}}^l & \dis {u_{22}}^l & \dis {u_{23}}^l 
& \dis {u_{24}}^l  \cr
\dis {u_{31}}^{l}  & \dis {u_{32}}^l  & \dis {u_{33}}^l 
& \dis {u_{34}}^l  \cr
\dis {u_{41}}^l & \dis {u_{42}}^l & \dis {u_{43}}^l
& \dis {u_{44}}^l 
}\right)\cdot q^l\,\,\,.
\end{eqnarray*}
The $q$ is defined as $q:=\exp (x)$.
Each component ${u_{ij}}^l$ is a polynomial of $x$ whose degree is at most $4$
\begin{eqnarray*}
&&{u_{11}}^l = \frac{v^{(1)}_l }{l}\,\,\,,\,\,\,
{u_{21}}^l = v^{(1)}_l \,\,\,,\,\,\,
{u_{31}}^{l+1} = \frac{15 v^{(1)}_{l} + v^{(2)}_{l+1}}{l+1}\,\,\,,\,\,\,
{u_{41}}^l = v^{(2)}_l \,\,\,,\\
&&{u_{12}}^l = \frac{v^{(1)}_l }{l} \cdot {x}+
\left( \frac{\alpha^{(1)}_l }{l}-
\frac{v^{(1)}_l }{l^2}\right)\,\,\,,\\
&&{u_{22}}^l = v^{(1)}_l {x}
+\alpha^{(1)}_l \,\,\,,\\
&&{u_{32}}^{l+1} = \left(
\frac{15 v^{(1)}_{l} + v^{(2)}_{l+1}}{l+1}
\right) {x}\\
&& \,\,\,\,\,\, +\left(
\frac{15 \alpha^{(1)}_l
+\alpha^{(2)}_{l+1}}{l+1}
-\frac{15 v^{(1)}_l
+v^{(2)}_{l+1}}{(l+1)^2}
\right)\,\,\,,\\
&&{u_{42}}^l = v^{(2)}_l \cdot {x}
+\alpha^{(2)}_l \,\,\,,\\
%\end{eqnarray*}
%\begin{eqnarray*}
&&{u_{13}}^l = \frac{v^{(1)}_l}{l} \cdot \frac{x^2}{2}+
\left( \frac{\alpha^{(1)}_l }{l}-
\frac{v^{(1)}_l }{l^2}\right){x}
+\left(\frac{\beta^{(1)}_l }{l}-\frac{\alpha^{(1)}_l }{l^2}
+\frac{v^{(1)}_l }{l^3} \right)\,\,\,,\\
&&{u_{23}}^l = v^{(1)}_l \frac{x^2}{2}
+\alpha^{(1)}_l {x} +\beta^{(1)}_l \,\,\,,\\
&&{u_{33}}^{l+1} = \left(
\frac{15 v^{(1)}_{l} + v^{(2)}_{l+1}}{l+1}
\right)
\frac{x^2}{2}
%\\&& \,\,\,\,\,\, 
+\left(
\frac{15 \alpha^{(1)}_l
+\alpha^{(2)}_{l+1}}{l+1}
-\frac{15 v^{(1)}_l
+v^{(2)}_{l+1}}{(l+1)^2}
\right)
{x}\\
&&\,\,\,\,\,\, +\left(
\frac{15 {\beta^{(1)}_l}
+{\beta^{(2)}_{l+1}}}{l+1}
-\frac{15 {\alpha^{(1)}_l}
+{\alpha^{(2)}_{l+1}}}{(l+1)^2 }
+\frac{15 {v^{(1)}_l}
+{v^{(2)}_{l+1}}}{(l+1)^3 }
\right)\,\,\,,\\
&&{u_{43}}^l = v^{(2)}_l \cdot \frac{x^2}{2}
+\alpha^{(2)}_l {x} +\beta^{(2)}_l \,\,\,,\\
%\end{eqnarray*}
%\begin{eqnarray*}
&&{u_{14}}^l = \frac{v^{(1)}_l }{l} \cdot \frac{x^3}{6}+
\left( \frac{\alpha^{(1)}_l }{l}-
\frac{v^{(1)}_l }{l^2}\right)\frac{x^2}{2}
+\left(\frac{\beta^{(1)}_l }{l}-\frac{\alpha^{(1)}_l }{l^2}
+\frac{v^{(1)}_l }{l^3} \right)x\\
&& \,\,\,\,\,\, +\left(
\frac{\gamma^{(1)}_l }{l}-\frac{\beta^{(1)}_l }{l^2}
+\frac{\alpha^{(1)}_l }{l^3}-\frac{v^{(1)}_l }{l^4}
\right) \,\,\,,\\
&&{u_{24}}^l = v^{(1)}_l \frac{x^3}{6}
+\alpha^{(1)}_l \frac{x^2}{2} +\beta^{(1)}_l x
+\gamma^{(1)}_l \,\,\,,\\
&&{u_{34}}^{l+1} = \left(
\frac{15 v^{(1)}_{l} + v^{(2)}_{l+1}}{l+1}
\right)
\frac{x^3}{6}\\
&& \,\,\,\,\,\, +\left(
\frac{15 \alpha^{(1)}_l
+\alpha^{(2)}_{l+1}}{l+1}
-\frac{15 v^{(1)}_l
+v^{(2)}_{l+1}}{(l+1)^2}
\right)
\frac{x^2}{2}\\
&&\,\,\,\,\,\, +\left(
\frac{15 {\beta^{(1)}_l}
+{\beta^{(2)}_{l+1}}}{l+1}
-\frac{15 {\alpha^{(1)}_l}
+{\alpha^{(2)}_{l+1}}}{(l+1)^2 }
+\frac{15 {v^{(1)}_l}
+{v^{(2)}_{l+1}}}{(l+1)^3 }
\right){x}\\
&&\,\,\,\,\,\,
+\left(\frac{15 {\gamma^{(1)}_l}
+{\gamma^{(2)}_{l+1}}}{l+1}
-\frac{15 {\beta^{(1)}_l}
+{\beta^{(2)}_{l+1}}}{(l+1)^2}
+\frac{15 {\alpha^{(1)}_l}
+{\alpha^{(2)}_{l+1}}}{(l+1)^3}
-\frac{15 {v^{(1)}_l}
+{v^{(2)}_{l+1}}}{(l+1)^4}\right)\,\,\,,\\
&&{u_{44}}^l = v^{(2)}_l \cdot \frac{x^3}{6}
+\alpha^{(2)}_l \frac{x^2}{2} +\beta^{(2)}_l x
+\gamma^{(2)}_l \,\,\,.
\end{eqnarray*}
The numbers $v_l$, $\alpha_l$, $\beta_l$ and $\gamma_l$ are
obtained by recursive relations
\begin{eqnarray*}
&&\left\{
\begin{array}{l}
%&&
\dis v^{(1)}_{l=1}=6 \,\,\,,\\
%\dis v^{(1)}_{l+1}=6\cdot \left(
%\prod_{m=1}^{l} \frac{3(3m+1)(3m+2)}{m (m+1)^3 }
%\right)  \,\,\,\,(l=1,2,\cdots )
\dis v^{(1)}_{l+1}=2 \cdot 3^{l+1} \cdot \left(
\prod_{m=1}^{l} \frac{(3m+1)(3m+2)}{m (m+1)^3 }
\right)  \,\,\,\,(l=1,2,\cdots )\,\,\,,
\end{array}
\right.\\
&&\left\{
\begin{array}{l}
%&&
\dis v^{(2)}_{l=1}=0\,\,\,,\,\, v^{(2)}_{l=2}=18\,\,\,,\\
%\dis v^{(2)}_{l+1}=18 \cdot \left(
%\prod_{m=2}^{l} \frac{3(3m-1)(3m-2)}{(m-1)^2 m (m+1) }
%\right)\,\,\,\,(l=2,3,\cdots )
\dis v^{(2)}_{l+1}=2\cdot 3^{l+1} \cdot  \left(
\prod_{m=2}^{l} \frac{(3m-1)(3m-2)}{(m-1)^2 m (m+1) }
\right)\,\,\,\,(l=2,3,\cdots )\,\,\,,
\end{array}
\right.
\\
%\end{eqnarray*}
%\begin{eqnarray*}
&&\left\{
\begin{array}{l}
%&&
\dis \alpha^{(1)}_{l+1}
=\left[\frac{15}{(l+1)^2} 
+\frac{6}{l(l+1)} 
\right] \alpha^{(1)}_l 
+\frac{1}{(l+1)^2} \alpha^{(2)}_{l+1} \\
%&&
\dis \,\,\,\,\,\, -\left[\frac{30}{(l+1)^3 }+
\frac{6}{l (l+1)^2 }+
\frac{6}{l^2 (l+1) }
\right]v^{(1)}_l -\frac{2}{(l+1)^3 }v^{(2)}_{l+1} \,\,\,,\\
%&&
\dis \alpha^{(2)}_{l+2}
=\left[\frac{90}{(l+1)(l+2)} 
+\frac{36}{l(l+2)} 
\right] \alpha^{(1)}_l 
+\frac{6}{(l+1)(l+2)} \alpha^{(2)}_{l+1} \\
%&&
\dis \,\,\,\,\,\, -\left[\frac{90}{(l+1) (l+2)^2 }+
\frac{90}{(l+1)^2 (l+2) }+
\frac{36}{l (l+2)^2 }+
\frac{36}{l^2 (l+2) }
\right]v^{(1)}_l \\
%&&
\dis \,\,\,\,\,\, -\left[
\frac{6}{(l+1)^2 (l+2) }+
\frac{6}{(l+1) (l+2)^2 }
\right]v^{(2)}_{l+1} \,\,\,,\\
%&&
\dis \alpha^{(1)}_{l=1}=9 \,\,\,,\\
%&&
\dis \alpha^{(2)}_{l=1}=0\,\,\,,\,\, \alpha^{(2)}_{l=2}=36\,\,\,,
\end{array}
\right.\\
%\end{eqnarray*}
%\begin{eqnarray*}
&&\left\{
\begin{array}{l}
%&&
\dis \beta^{(1)}_{l+1}
=\left[\frac{15}{(l+1)^2} 
+\frac{6}{l(l+1)} 
\right] \beta^{(1)}_l 
+\frac{1}{(l+1)^2} \beta^{(2)}_{l+1} \\
%&&
\dis \,\,\,\,\,\, -\left[\frac{30}{(l+1)^3 }+
\frac{6}{l (l+1)^2 }+
\frac{6}{l^2 (l+1) }
\right]\alpha^{(1)}_l 
-\frac{2}{(l+1)^3 }\alpha^{(2)}_{l+1} \\
%&&
 \dis \,\,\,\,\,\, +\left[\frac{45}{(l+1)^4 }+
\frac{6}{l (l+1)^3 }+
\frac{6}{l^2 (l+1)^2 }+
\frac{6}{l^3 (l+1) }
\right]v^{(1)}_l 
+\frac{3}{(l+1)^4 }v^{(2)}_{l+1} \,\,\,,\\
%&&
\dis \beta^{(2)}_{l+2}
=\left[\frac{90}{(l+1)(l+2)} 
+\frac{36}{l(l+2)} 
\right] \beta^{(1)}_l 
+\frac{6}{(l+1)^2} \beta^{(2)}_{l+1} \\
%&&
\dis \,\,\,\,\,\, -\left[\frac{90}{(l+1) (l+2)^2 }+
\frac{90}{(l+1)^2 (l+2) }+
\frac{36}{l (l+2)^2 }+
\frac{36}{l^2 (l+2) }
\right]\alpha^{(1)}_l \\
%&&
\dis \,\,\,\,\,\, -\left[
\frac{6}{(l+1) (l+2)^2 }+
\frac{6}{(l+1)^2 (l+2) }
\right]\alpha^{(2)}_{l+1} \\
%&&
\dis \,\,\,\,\,\, +\Biggl[\frac{90}{(l+1) (l+2)^3 }+
\frac{90}{(l+1)^2 (l+2)^2 }+
\frac{90}{(l+1)^3 (l+2) }\\
%&&
\dis \,\,\,\,\,\, +\frac{36}{l (l+2)^3 }+
\frac{36}{l^2 (l+2)^2 }+
\frac{36}{l^3 (l+2) }
\Biggr]v^{(1)}_l \\
%&&
\dis \,\,\,\,\,\, +\left[
\frac{6}{(l+1) (l+2)^3 }+
\frac{6}{(l+1)^2 (l+2)^2 }+
\frac{6}{(l+1)^3 (l+2) }
\right]v^{(2)}_{l+1} \,\,\,,\\
%&&
\dis \beta^{(1)}_{l=1}=-18 \,\,\,,\\
%&&
\dis \beta^{(2)}_{l=1}=6\,\,\,,\,\, \beta^{(2)}_{l=2}=-45 \,\,\,,
\end{array}
\right.
\\
%\end{eqnarray*}
%\begin{eqnarray*}
&&\left\{
\begin{array}{l}
%&&
\dis \gamma^{(1)}_{l+1}
=\left[\frac{15}{(l+1)^2} 
+\frac{6}{l(l+1)} 
\right] \gamma^{(1)}_l 
+\frac{1}{(l+1)^2} \gamma^{(2)}_{l+1} \\
%&&
\dis
\,\,\,\,\,\, -\left[\frac{30}{(l+1)^3 }+
\frac{6}{l (l+1)^2 }+
\frac{6}{l^2 (l+1) }
\right]\beta^{(1)}_l 
-\frac{2}{(l+1)^3 }\beta^{(2)}_{l+1} \\
%&&
\dis
\,\,\,\,\,\, +\left[\frac{45}{(l+1)^4 }+
\frac{6}{l (l+1)^3 }+
\frac{6}{l^2 (l+1)^2 }+
\frac{6}{l^3 (l+1) }
\right]\alpha^{(1)}_l 
+\frac{3}{(l+1)^4 }\alpha^{(2)}_{l+1} \\
%&&
\dis
\,\,\,\,\,\, -\left[\frac{60}{(l+1)^5 }+
\frac{6}{l (l+1)^4 }+
\frac{6}{l^2 (l+1)^3 }+
\frac{6}{l^3 (l+1)^2 }+
\frac{6}{l^4 (l+1) }
\right]v^{(1)}_l 
-\frac{4}{(l+1)^5 }v^{(2)}_{l+1} \,\,\,,\\
%&&
\dis
\gamma^{(2)}_{l+2}
=\left[\frac{90}{(l+1)(l+2)} 
+\frac{36}{l(l+2)} 
\right] \gamma^{(1)}_l 
+\frac{6}{(l+1)(l+2)} \gamma^{(2)}_{l+1} \\
%&&
\dis
 \,\,\,\,\,\,-\Biggl[\frac{90}{(l+1) (l+2)^3 }+
\frac{90}{(l+1)^2 (l+2)^2 }+
\frac{90}{(l+1)^3 (l+2) }\\
%&&
\dis
\,\,\,\,\,\, +
\frac{36}{l (l+2)^2 }+
\frac{36}{l^2 (l+2) }
\Biggr]\beta^{(1)}_l \\
%&&
\dis
 \,\,\,\,\,\,-\left[\frac{6}{(l+1) (l+2)^2 }+
\frac{6}{(l+1)^2 (l+2) }+
\right]
\beta^{(2)}_{l+1} \\
%&&
\dis
\,\,\,\,\,\, +\Biggl[\frac{90}{(l+1) (l+2)^3 }+
\frac{90}{(l+1)^2 (l+2)^2 }+
\frac{90}{(l+1)^3 (l+2) }\\
%&&
\dis
\,\,\,\,\,\, +
\frac{36}{l (l+2)^3 }+
\frac{36}{l^2 (l+2)^2 }+
\frac{36}{l^3 (l+2) }
\Biggr]\alpha^{(1)}_l \\
%&&
\dis
\,\,\,\,\,\, +\left[
\frac{6}{(l+1) (l+2)^3 }+
\frac{6}{(l+1)^2 (l+2)^2 }+
\frac{6}{(l+1)^3 (l+2) }
\right]
\alpha^{(2)}_{l+1} \\
%&&
\dis
\,\,\,\,\,\, -\Biggl[
\frac{90}{(l+1) (l+1)^4 }+
\frac{90}{(l+1)^2 (l+1)^3 }+
\frac{90}{(l+1)^3 (l+1)^2 }+
\frac{90}{(l+1)^4 (l+1) }\\
%&&
\dis
\,\,\,\,\,\,+
\frac{36}{l (l+2)^4 }+
\frac{36}{l^2 (l+2)^3 }+
\frac{36}{l^3 (l+2)^2 }+
\frac{36}{l^4 (l+2) }
\Biggr]v^{(1)}_l \\
%&&
\dis
\,\,\,\,\,\,-\left[
\frac{6}{(l+1) (l+2)^4 }+
\frac{6}{(l+1)^2 (l+2)^3 }+
\frac{6}{(l+1)^3 (l+2)^3 }+
\frac{6}{(l+1)^4 (l+2) }
\right]v^{(2)}_{l+1} \,\,\,,\\
%&&
\dis \gamma^{(1)}_{l=1}=21 \,\,\,,\\
%&&
\dis \gamma^{(2)}_{l=1}=-6\,\,\,,\,\, \gamma^{(2)}_{l=2}=\frac{63}{2}\,\,\,.
\end{array}
\right.
\end{eqnarray*}
For Fano cases, the expansion coefficients are polynomials of K{\"a}hler 
parameters whose degrees are finite independent of instanton degree.
Its structure resembles closely that in the Calabi-Yau cases.
A constant shift of K{\"a}hler parameter mixes components of
the ${\Pi}$ with each other.
But it is not clear that the matrix ${\Pi}$ factorizes into
a monodromy part and a genuine instanton correction part for 
Fano manifolds cases.
Also we lack for information about
total monodromy properties of moduli spaces in the Fano cases.

Next when we turn on a parameter $t_0$, the amplitude ${\Pi}$
is modified. A corresponding observable is a puncture operator $P$
and plays important role in topological gravity.
But the $\Kappa_0$ is always identity matrix in all the
cases.
In order to couple the $P$ to our cases,
all we have to do is to change the amplitude ${\Pi}$
into $\exp (t_0 \mbox{ $1\hspace{-7pt}1 $})\cdot {\Pi}$.
This alternation might be an operation by which
world sheet gravity $\sigma_n$
comes to couple with the system.
More precise studies about these things will appear
elsewhere.

%\newpage

%%%%%%%%%%%%%%%%%%%%%%%%%%%%%%%%%%%%%%%%%%%%%%%%%%%%%%%%%%%%%%%%%%%%%%%

\section{Conclusions and Discussions}
\cleqn

In this article, we investigated the disk amplitudes in topological 
A-models and developed a method to calculate the contributions from 
world sheet instantons with boundaries.
We study fusion structures of A-model operators and find that
these correlators satisfy a set of differential equations characterized
by three point functions  $\Kappa_l$.
The integrable conditions in this open string amplitudes 
are equivalent to associativities of
operators and are realized as commutativity of an arbitrary pair
of fusion matrices. For Fano manifold cases, these commutativities
make us to obtain all instanton corrections in tree level of closed string
theory. The disk amplitudes are collected into one matrix ${\Pi}$
and are written as path ordered exponentials of 
integrals of these fusion couplings. Probably it is possible to
interpret the ${\Pi}$ as a kind of Baker-Akhiezer functions 
or $\tau$-functions of the integrable systems.
Local deformations of the integral contours do not affect the results.
But the amplitudes have global monodromy properties.

In this paper, for simplicity, 
we switch off all perturbation operators except for marginal ones
$\co^{(1)}$ associated to K{\"a}hler forms.
However we would like to emphasize that our results are applicable to
the completely general backgrounds.

We treat several examples.
As a first case, we investigate Calabi-Yau $3$-fold. The
A-model disk amplitudes are expressed by using K{\"a}hler parameters and
a prepotential $F$. In particular, the correlators associated with
$\co^{(0)}$ are period integrals themselves if they are transformed into the
B-model side by mirror maps.
We observe that the prepotential formula in the toric analysis can be 
interpreted geometrically as a sum of products of disk amplitudes 
associated with canonical homology basis. 
A pair of cycles is glued on two disk boundaries and the product is
weighted with a homology intersection number of these cycles.
The $F$ is essentially a sum of contributions from disk amplitudes with 
various homology cycles as their boundaries.

Next we consider Calabi-Yau $d$-fold cases concretely.
For each case, the amplitude is given as an upper triangular matrix and
it is factorized into two parts, that is, monodromy 
part and instanton part.
The former is constructed from 
some commutable nilpotent matrices ${\cal N}$'s.
All eigenvalues of the monodromy matrices $\exp({\cal N})$ are units.
The number of the ${\cal  N}$ is equal to the
dimension of K{\"a}hler moduli spaces, or 2nd Betti number $b_2 $.
The latter part is composed of a set of single-valued functions 
under monodromy transformations around large radius limit points.
In other words, that is invariant by some constant shifts of K{\"a}hler  
parameters. 
It contains essentially information on
instanton corrections with disklike topology.
Its expansion coefficients are interpreted as integrals of
Chern classes of vector bundles ${\cal V}$ over stable maps 
$\overline{\cal M}$ with
one inserted operator on $\Sigma$.
The region of that integration is over the fixed homology cycles.
But the shifts of K{\"a}hler forms mix these instanton parts
with each other. 
In the language of D-branes,
the set of homology cycles with even dimensions
are wrapped by D-branes. They are susy cycles 
\cite{BBS, HL} with minimum volumes
and
``RR-fields'' associated to the branes could be realized as Chern classes
of some bundle over Calabi-Yau moduli spaces.
The structure of the monodromy transformations implies 
existence of some Chern-Simons terms for these ``RR-fields''.
It is consistent with one of the T-duality transformation
$B\rightarrow B+\zeta$  ($ \zeta \in \mbox{H}^2$(M)).
Especially it is remarkable that the form of the monodromy matrix 
is completely fitted to the form of mixings of homology cycles
wrapped by D-branes.

In the context of string compactifications, the internal space
with more than ten dimension seems to be meaningless.
But as mathematical interest, 
there seems to be no obstruction to extend the 
``D-branes'' to higher dimensional cases through geometrical characterization.
In various dimensional cases, mirror symmetries are analysed so far
\cite{GMP, KS3, KS4, KS5, KS2, JN}.
Their results illustrate existence of those symmetries in higher dimensional 
cases.
Also analyses here imply that mirror symmetries might 
be generalized to the cases including RR-fields and
(flat) gauge fields on the D-branes.

Furthermore 
our results show that the torus fibered structure of Calabi-Yau $d$-fold
is universal for any dimensions.
The $0$-cycle in the A-model side 
is wrapped by a D0-brane.
This cycle $\alpha_0$ 
corresponds to a unique $d$-cycle $\hat{\alpha}_0$ dual to
a cohomology element $\mbox{H}^{d,0}$(W) of the mirror W.
The $\hat{\alpha}_0$ is the fundamental cycle in the toric language
of mirror (B-model) calculations and
is homologically equivalent to a $d$ dimensional torus.
It supports the conjecture that
any Calabi-Yau $d$-fold is realized as some torus fibered space,
which is
proposed by \cite{SYZ}.
But base manifold in this fibered space is not clear in our analyses.
More precise studies are needed to clarify the global
structure of the fibered Calabi-Yau spaces.
It will yield reconstruction of
mirror partner W from the moduli spaces of the original
manifold M.
We wish to present a more detailed study on this subject in future.

There are still some open problems.
In order to complete our analysis of boundary states in string context,
we have to consider non-topological states. 
They might have more information on properties of susy cycles and be
related with moduli spaces of D-branes.

The second problem is a relation with U-duality.
The mirror symmetry is said to be a kind of generalized 
T-duality. In fact the mixings of homology cycles are controlled by
integral shifts of B-fields, which is a part of T-dual transformation.
If we consider an S-duality in some context, the mirror symmetries here might
be lifted to a U-duality.
That symmetry exchanges stringy fundamental states with
solitonic ones (D-branes).
In our previous papers \cite{KS6, KS7},
open string higher loop corrections play essential roles in
analysing mass formulae of BPS states. 
Also those corrections are essentially estimated by tadpole graphs
which are described by disk amplitudes, that is, boundary states.
Possibly this problem might be treated in M-theory.

For our analyses in Fano cases, there are some non-trivial monodromy
properties in the disk amplitudes. They reflect mixings of homology cycles
with different dimensions. But the mixings do not have the same structure as 
that in Calabi-Yau case.
The $0$-dimensional cycle wrapped by D0-brane corresponds to a 
$d$ dimensional fundamental homology cycle in the B-model side 
for Calabi-Yau cases. How should we interpret the $0$-cycle for 
Fano cases because of lacking for the associated B-model.

The last is the meanings of other singular points on moduli spaces.
The large radius limit points contain information on
mixings of cycles wrapped by D-branes.
What information is hidden at other singular points?
For Calabi-Yau $3$-folds, conifold singularities 
are related to Black Hole's physics 
unexpectedly \cite{coni}. Is there similar physics at the
other singular points? 
We do not have any answer about this now.

Our analysis is still limited,
but we hope it will give some insights to the studies of
moduli spaces of Calabi-Yau manifolds and their mirror symmetries.

~

~

%\section*{Acknowledgement}

\newpage

%\appendix
\section*{Appendix A}
\section*{ Expansion Coefficients for Calabi-Yau $d$-folds}\label{d-dim}
We consider $d$ dimensional Calabi-Yau manifold M
\ba
M\,;\, X_{1}^{d+2}+X_{2}^{d+2}+\cdots + X_{d+1}^{d+2} +X_{d+2}^{d+2} =0
\,\,\,.\nom
\ea
Essentially instanton corrections are encoded in
a set of functions $\{ a_n \}$ 
\begin{eqnarray*}
&&\left\{
\begin{array}{ccl}
a_0 & = &1\,\,\,,\,\,\, a_1 =0\,\,\,,\\
a_n & = & S_n (0,\tilde{x}_2 , \cdots ,\tilde{x}_n )\,\,\,\,\,\,\,
\,\,\,(n=2,3,\cdots ,d)\,\,\,,
\end{array}
\right.\,\,\,\nom \\
&&\tilde{x}_m := \frac{1}{m!} \left(\frac{1}{2\pi i}\frac{\del}{\del \rho}
\right)^m \log \Biggl[\,\,
\sum_{l=0}^{\infty} \frac{\Gamma (N(l+\rho)+1)}{\Gamma (N\rho +1)}
\biggl\{ \frac{\Gamma (\rho +1)}{\Gamma (l+\rho +1)}\biggr\}^N 
(N\psi)^{-Nl}\,\,
\Biggr]\Bigg|_{\rho =0} \nom \\
&&\hspace{10cm}
(m=2,3,\cdots ,d)\,\,\,\,\,.\nom 
\end{eqnarray*}
%\ba
%t(\psi)=\frac{N}{2\pi i}\Biggl[\,\,\log (N\psi)^{-1} 
%+\frac{\dis\sum_{n=1}^{\infty}
%\frac{(Nn)!}{(n!)^N}\left(\sum_{l=n+1}^{Nn}\frac{1}{l}\right) 
%\cdot (N\psi)^{-n} }
%{\dis \sum_{m=0}^{\infty} \frac{(Nm)!}{(m!)^N}\cdot (N\psi)^{-m}}
%\,\,\Biggr]\,\,\,.
%\ea
The $S_n $ is a Schur function defined as
\ba
\exp\left(\sum_{m=1}^{\infty} w_n y^n \right)=
\sum_{n=0}^{\infty}
S_n (w_1 ,w_2, \cdots , w_n) y^n \,\,\,.\nom
\ea
These $a_n$'s  $(n\geq 2)$ are expanded as power series with respect to 
$q=\exp (2\pi i\,t)$. For all dimensions, $a_0 =1$, $a_1=0$ are satisfied.
We summarize these $q$-expansions of the $\{ a_n \}$ $(n=2,3,\cdots ,d)$
for lower dimensional cases.

The manifolds for $d=1,2$ are special cases.
In complex one dimensional torus case, the disk amplitude is calculated as
\ba
\Pi =\left(\matrix{1 & t \cr 0 & 1}\right) \,\,\,.\nom
\ea
For a K3 case, the amplitude contains no quantum corrections
\ba
\Pi =\left(\matrix{1 & t & \dis \frac{t^2}{2}
\cr 0 & 1 & t \cr 0 & 0 & 1}\right) \,\,\,.\nom
\ea

\newpage
\begin{eqnarray*}
&&\bullet\, \underline{\mbox{3 dimensional case}}\\
&&a_2=575\,q + {{975375\,{q^2}}\over 4} + {{1712915000\,{q^3}}\over 9}
\\&& \kankaku + 
  {{3103585359375\,{q^4}}\over {16}}  + 229305888887648\,{q^5} \\&&\kankaku + 
  297899690589234450\,{q^6} \\&&\kankaku 
+ {{20243246069160012225125\,{q^7}}\over {49}} \\&&\kankaku + 
  {{38464733280000707788879375\,{q^8}}\over {64}} \\&&\kankaku + 
  {{73459946418796448525169406250\,{q^9}}\over {81}} \\&&\kankaku + 
  1408576329956909429553448731160\,{q^{10}} %\\&& 
%+  {{270968494790252125528892214150615100\,{q^{11}}}\over {121}} \\&& + 
%  {{21777464187851606420870973035874654875\,{q^{12}}}\over 6} \\&& + 
%  {{1010395276839075565274075350036933479430000\,{q^{13}}}\over {169}} \\&& + 
%  {{1956998424413111258767909636912318686822953125\,{q^{14}}}\over
%  {196}} \\&&
% +   {{151865734266049048402846066336032882657245075975\,{q^{15}}}\over 9}
 + {{{\rm O}(q)}^{11}}\,\,,\\
&&a_3=
-1150\,q - {{975375\,{q^2}}\over 4}  - {{3425830000\,{q^3}}\over {27}} 
\\&&\kankaku - 
  {{3103585359375\,{q^4}}\over {32}}  - {{458611777775296\,{q^5}}\over 5} 
\\&&\kankaku - 
  99299896863078150\,{q^6} \\&&\kankaku - 
{{40486492138320024450250\,{q^7}}\over {343}} \\&&\kankaku - 
  {{38464733280000707788879375\,{q^8}}\over {256}} \\&&\kankaku - 
  {{146919892837592897050338812500\,{q^9}}\over {729}} \\&&\kankaku - 
  281715265991381885910689746232\,{q^{10}} %\\&& 
%-  {{541936989580504251057784428301230200\,{q^{11}}}\over {1331}} \\&& - 
%  {{21777464187851606420870973035874654875\,{q^{12}}}\over {36}} \\&& - 
%  {{2020790553678151130548150700073866958860000\,{q^{13}}}\over
%  {2197}} \\&& 
% - {{1956998424413111258767909636912318686822953125\,{q^{14}}}\over
% {1372}} 
%\\&& - 
%  {{60746293706419619361138426534413153062898030390\,{q^{15}}}\over {27}}
 + {{{\rm O}(q)}^{11}}\,\,,\\
\end{eqnarray*}
\newpage
\begin{eqnarray*}
&&\bullet\, \underline{\mbox{4 dimensional case}}\\
&&a_2=10080\,q + 73483200\,{q^2} + 1042526047360\,{q^3}  \\&&\kankaku + 
  19619298683429760\,{q^4} + {{2159313463640684990016\,{q^5}}\over 5} 
\\&&\kankaku + 
  10503727932089702308803840\,{q^6} \\&&\kankaku + 
  {{1916318246261017803528501768960\,{q^7}}\over 7} \\&&\kankaku + 
  7506486378661143816276636705580800\,{q^8} \\&&\kankaku + 
  {{1925818389017402099739567134263974922720\,{q^9}}\over 9} \\&&\kankaku + 
  6289794812252109825519839548497264776243328\,{q^{10}} 
%\\&& + 
%  {{22934089107505163463124928413202144656532359228800\,{q^{11}}}\over 
%    {121}} 
%\\&& + 5830144097730770059759382463748028133895164298099200\,{q^{12}} \\&& + 
%  {{30835335017614537595881319704210867613000525937374719487040\,
%      {q^{13}}}\over {169}} 
%\\&& + {{4056336631328900139203621472901544393416877581
%       9529654491368960\,{q^{14}}}\over 7} \\&& + 
%  {{931948361275786759599970027486620761619233414702909975887338185472\,
%      {q^{15}}}\over 5} 
+ {{{\rm O}(q)}^{11}}\,\,,\\
&&a_3=
 -10080\,q  - 11340000\,{q^2} + {{438895264640\,{q^3}}\over 3} \\&&\kankaku + 
  5676562588304160\,{q^4} + {{4094181652239685185984\,{q^5}}\over {25}} 
\\&&\kankaku + 
  4614422515155056774029440\,{q^6} \\&&\kankaku + 
  {{6470070606934642619706952592640\,{q^7}}\over {49}} \\&&\kankaku + 
  3863116083202577087773569076265760\,{q^8} \\&&\kankaku + 
  {{9355876893021790537527739478921992631840\,{q^9}}\over {81}} \\&&\kankaku + 
  3521863577835487363968251940437175654809280\,{q^{10}} 
%+ {{145416988702847145220074707675790268237974095053440\,{q^{11}}}\over 
%    {1331}} \\&& + {{37848615603572017001657577710988403163887013353911680\,
%    {q^{12}}}\over {11}}
% \\&& + {{26536523089495965803351577484533515241016214272
%       98092348938560\,{q^{13}}}\over {24167}} \\&& + 
%  {{24841062715526784202429562877480145093483341323083013976864280320\,
%      {q^{14}}}\over {7007}} 
%\\&& + {{413426908719038383206351990554744204454074425
%       506367381823433131405568\,{q^{15}}}\over {3575}} \\&& 
+ {{{\rm O}(q)}^{11}}\,\,,\\
&&a_4=
-50803200\,{q^2}  - 740710656000\,{q^3}  - 13208552898508800\,{q^4} 
\\&&\kankaku - 
  274370680772336332800\,{q^5} 
\\&&\kankaku - 6338295071425859161389056\,{q^6} \\&&\kankaku - 
  158065860086155715545404887040\,{q^7} \\&&\kankaku - 
  4174032361462124886410787618250752\,{q^8} \\&&\kankaku - 
  115205378241303863208140231152926277632\,{q^9} \\&&\kankaku - 
  {{82331184334279500333868824265916634002480128\,{q^{10}}}\over {25}}
%\\&& - 
%  96857903784765441050882846121235768546511075328\,{q^{11}} \\&& - 
%  {{3176045475797901621366737694175104956179377660618809344\,{q^{12}}}\over 
%    {1089}} 
%\\&& - {{10837774483199058645482843201032074792387085017696215101440\,
%      {q^{13}}}\over {121}} 
%\\&& - {{2802915504034346987123442665727432583118300386
%       571797590149326811136\,{q^{14}}}\over {1002001}} \\&& - 
%  {{9062422359291916419325313356377891679925103833194255105398740097380352\,
%      {q^{15}}}\over {102245}} 
+ {{{\rm O}(q)}^{11}}\,\,,\\
\end{eqnarray*}
\newpage
\begin{eqnarray*}
&&\bullet\,\underline{\mbox{5 dimensional case}}\\
&&a_2=144256\,q + 17462862648\,{q^2} \\&&\kankaku 
+ {{39251149387190104\,{q^3}}\over 9} \\&&\kankaku + 
  1482599929918063737550\,{q^4} \\&&\kankaku + 
  {{14968967753047894532544812256\,{q^5}}\over {25}} \\&&\kankaku + 
  270025284718395377923251679826570\,{q^6} %\\&& + 
%  131503750437751933439242023738208726056\,{q^7} \\&& + 
%  {{135554074424604150657150885485204342654470503\,{q^8}}\over 2} \\&& + 
%  {{2955426111853526056153774265356676885790192256875232\,{q^9}}\over {81}} 
%\\&& + 
%  {{508295502996605034479093666961490107917468361463314477648\,{q^{10}}}\over 
%    {25}} 
%\\&& + {{140987438450734387065957529306825676994336825082168964058187382
%       4\,{q^{11}}}\over {121}} \\&& + 
%  {{123020789869142081374687719044482597169324326445239100198574687013997\,
%      {q^{12}}}\over {18}} 
%\\&& + {{69088463336543068382778291617163787576617133658
%       2378109067704760163244898816\,{q^{13}}}\over {169}} \\&& + 
%  2486524889734583090762572657510446323278166007707495133015410263864426865268
%    774\,{q^{14}} 
%\\&& + {{38360228190531723403937162633447914364755687235486859189
%       175073314397581857645657166456\,{q^{15}}}\over {25}} 
%\\&& 
+ {{{\rm O}(q)}^{7}}\,\,,\\
&&a_3=
-44541\,q + {{76625259837\,{q^2}}\over 8} \\&& \kankaku + 
  {{11064424981812685\,{q^3}}\over 3} \\&&\kankaku + 
  {{283587971949297078031031\,{q^4}}\over {192}} \\&&\kankaku + 
  {{81375921128111302728858579834\,{q^5}}\over {125}} \\&&\kankaku + 
  {{111620758226133549844474689849549037\,{q^6}}\over {360}} %\\&& + 
%  {{5486268147560615237858135528173983147076\,{q^7}}\over {35}} \\&& + 
%  {{212528117910854368703978480558554635577688561457\,{q^8}}\over {2560}} 
%\\&& + 
%{{3695844405890421418868655665709612359086831327607467\,{q^9}}\over {81}} 
%\\&& + 
%  {{116295913239557792463731813808280003736164130789433063621079\,
%     {q^{10}}}\over {4500}} 
%\\&& + {{998707727499498249577116495520470800721617710
%       47764039660413977787\,{q^{11}}}\over {6655}} \\&& + 
%  {{253714644972980822878361008924150092089450255076259593494482057403308970
%       3\,{q^{12}}}\over {285120}} \\&& + 
%  {{53104310994227740609020242940201362249618571401003789642820862545471776281
%       7263\,{q^{13}}}\over {98865}} \\&& + 
%  {{36217685625567400057355636745879263010518021932785405747917590078910156995
%       8529399\,{q^{14}}}\over {110}} \\&& + 
%  {{46021301898967963020265697977805744858762098991971830860763141282382200108
%       1935986803436\,{q^{15}}}\over {225}} 
%\\&& 
+ {{{\rm O}(q)}^{7}}\,\,,\\
&&a_4=
-154889\,q  - {{286339506551\,{q^2}}\over {16}}  - 
  {{270354099090739391\,{q^3}}\over {81}} \\&&\kankaku - 
  {{1994922586698227118215983\,{q^4}}\over {2304}} \\&&\kankaku - 
  {{171342020016473054406883272264\,{q^5}}\over {625}} \\&&\kankaku - 
  {{640866103338137787756334143777874909\,{q^6}}\over {6480}} %\\&&\kankaku - 
%  {{34453578488555273894972417779592426307331\,{q^7}}\over {882}} \\&& - 
%  {{337332720218267249535976581251968815358294037843\,{q^8}}\over {20480}} 
%\\&& - 
%  {{239052498846413521608012407995823203696342222939762597\,{q^9}}\over 
%    {32805}} 
%\\&& - {{135250754152664559868197278011207070047280894304889691324900
%       3\,{q^{10}}}\over {405000}} 
%\\&& - 
%  {{8271114178521912362682726401228059355165726369518724873335268148959\,
%      {q^{11}}}\over {5270760}} \\&& - 
%  {{769178264969731189603165472924167027958713324892228064127763628181998706
%       3\,{q^{12}}}\over {10264320}} \\&& - 
%  {{18348232445011706678267205691917767573130130538439185214876249171443249162
%       526907\,{q^{13}}}\over {50895702}} \\&& - 
%  {{32421849529431227709140155739729242112534092181602166067724100333188534812
%       24857032109\,{q^{14}}}\over {18738720}} \\&& - 
%  {{20527009916386902000163235219461709571895170507271299485076843097435863416
%       827804539790566109\,{q^{15}}}\over {250965000}} 
%\\&& 
+ {{{\rm O}(q)}^{7}}\,\,,\\
&&a_5=
309778\,q + {{119861158263\,{q^2}}\over {16}}  - 
  {{42310805608216586\,{q^3}}\over {243}} \\&&\kankaku - 
  {{493093891453918167040411\,{q^4}}\over {1536}} \\&&\kankaku - 
  {{1671979601542630148578508055791\,{q^5}}\over {9375}} \\&&\kankaku - 
  {{14884051464652702471234056058626752111\,{q^6}}\over {162000}} %\\&& - 
%  {{443564654150528475801375169753225408332339569\,{q^7}}\over 
%{9261000}} \\&& - 
%  {{157624936367161460194559310083271686610792579336367\,{q^8}}\over 
% {6144000}} 
%\\&& - {{83460006452630993996056686395355815298917960650676495753\,
%      {q^9}}\over {5904900}} 
%\\&& - {{129435770913985894194951919492529203775339856
%       988207428729942163\,{q^{10}}}\over {16200000}} 
%\\&& - 
%  {{160733332249911928607408704048213514987484792552877114464597147591086341\,
%      {q^{11}}}\over {34787016000}} \\&& - 
%  {{16926302083565808638731240268453533224810904477606245446917429008266535727
%       0021\,{q^{12}}}\over {62099136000}} \\&& - 
%  {{35723620514797041625175067464224452288835187119720956524982026785240510423
%       895980349959\,{q^{13}}}\over {21834256158000}} \\&& - 
%  {{12345298996788725033493048848117237292079717721705989065143248181716236977
%       6882417995252865087\,{q^{14}}}\over {123799227552000}} \\&& - 
%  {{23876176588741788173252590888548115378482859283215689722799621654320297849
%       584679248507496609827919\,{q^{15}}}\over {38759034600000}} 
%\\&& 
+ {{{\rm O}(q)}^{7}}\,\,,
\end{eqnarray*}
\newpage
\begin{eqnarray*}
&&\bullet\,\underline{\mbox{6 dimensional case}}\\
&&a_2
=1998080\,q + 4174645462848\,{q^2} + {{173351535803486093312\,{q^3}}\over 9} 
\\&&\kankaku + 
  125084361408235060148824528\,{q^4} \\&&\kankaku + 
  {{4925700694401049779531460267401216\,{q^5}}\over 5} \\&&\kankaku + 
  8787523560602575025843757375030064656384\,{q^6} %\\&& + 
%  {{598710010380887137306323561214170625376292175872\,{q^7}}\over 7} \\&& + 
%  888033363633420938178456772465669848834611341018942068\,{q^8} \\&& + 
%  {{784987011830728492475679071603395672258050962915763306539823104\,
%      {q^9}}\over {81}} 
%\\&& + {{27508713109686880591704616585214762649689521893525
%       87280851292088014848\,{q^{10}}}\over {25}} 
%\\&& + 
%  {{15612845723973537424696009118924762328336433232803363145682473833132143594
%       7008\,{q^{11}}}\over {121}} \\&& + 
%  {{13987636896837097730656936703814092688003059054184257099904274297807726510
%       6385480704\,{q^{12}}}\over 9} \\&& + 
%  {{32360584744261121501847564226619556040686849972592536188501713111445924877
%       149052716018958336\,{q^{13}}}\over {169}} \\&& + 
%  {{16836737512703252345062109122781716005138568430891089070172657233786218443
%       775832114403847724728320\,{q^{14}}}\over 7} \\&& + 
%  {{76806966994451124374119626060800648419401333601300694616627478802587580306
%       8461493713137215082069772730368\,{q^{15}}}\over {25}} 
%\\&& 
+ {{{\rm O}(q)}^{7}}\,\,,\\
&&a_3=
691712\,q + 5177687949504\,{q^2} \\&&\kankaku + 
  {{806514791977225060352\,{q^3}}\over {27}} \\&&\kankaku + 
  213763669278793913475191576\,{q^4} \\&&\kankaku + 
  {{222265154400541781032704441312755712\,{q^5}}\over {125}} \\&& \kankaku + 
  16428039454518721043990933106854902034432\,{q^6} %\\&& + 
%  {{40146679772385769585146353966859233767727173730304\,{q^7}}\over {245}} 
%\\&& + 
%  {{8662059808790590586667126319184325929325622003625954031\,{q^8}}\over 5} 
%\\&& + 
%  {{69877367856114954369982272411746181073308776765911444353161273344\,
%      {q^9}}\over {3645}} 
%\\&& + {{825300587243333137718620490014885052866526867088
%       12983694646290849216512\,{q^{10}}}\over {375}} 
%\\&& + 
%  {{17329526450922217465835050301336891638011637581229093721694201949425203308
%       658688\,{q^{11}}}\over {6655}} \\&& + 
%  {{46924009829577697345442800861657808887342205947957753654896114120825283661
%       766566938112\,{q^{12}}}\over {1485}} \\&& + 
%  {{28402525169258853608955863967180743914861946094690732310574572508399367584
%       722270814013617864704\,{q^{13}}}\over {72501}} \\&& + 
%  {{17332653288054923648894355674296796592392795358213110314333201913853876931
%       0171879920639972481789722624\,{q^{14}}}\over {35035}} \\&& + 
%  {{11347855785810634116394649849144455273278650676857362170542279178526656240
%       92041632993219683815214221851885568\,{q^{15}}}\over {17875}} 
%\\&& 
+ {{{\rm O}(q)}^{7}}\,\,,\\
&&a_4=
-3381504\,q - 3474331329168\,{q^2} \\&&\kankaku - 
  {{135215010362246119424\,{q^3}}\over {27}} \\&&\kankaku + 
  11035025851838372817917911\,{q^4} \\&&\kankaku + 
  {{196308541191778504551758653861994496\,{q^5}}\over {625}} \\&&\kankaku + 
  {{190686799407710048293521223981506934567936\,{q^6}}\over {45}} %\\&& + 
%  {{88582296219159938819914296059218421577308377186304\,{q^7}}\over {1715}} 
%\\&& + 
%  {{49588284306865699199052480127360518434110148161196848947\,{q^8}}\over 
%    {80}} 
%\\&& + {{8191231692586445665241878227050820195804345383455663379300892262
%       4\,{q^9}}\over {10935}} %\\&& + 
%  {{172076988897655035591461480350135102288798306828631006937351296472618496\,
%      {q^{10}}}\over {1875}} 
%\\&& + {{835771629549217458271684957985491929856490044
%       08992628434353547572494157589250048\,{q^{11}}}\over {73205}} \\&& + 
%  {{32114958758597978129846133934679714601111822568896688326236041743078595777
%       9458638664896\,{q^{12}}}\over {22275}} \\&& + 
%  {{43508443416212639513684922511709283867673106050554895757526100579968358495
%       31021270932595715473408\,{q^{13}}}\over {23562825}} \\&& + 
%  {{29376383481145074237120620432633131202023424398328708648308309997439129837
%       93834710513872545660480159744\,{q^{14}}}\over {1226225}} \\&& + 
%  {{84330017077758088956741975271111179675757667232079736567084711239278655120
%       86040400485458089833675540400177152\,{q^{15}}}\over {268125}} 
%\\&& 
+ {{{\rm O}(q)}^{7}}\,\,,\\
&&a_5=
3381504\,q - 2758789385912\,{q^2} - 
  {{1621921777812870897664\,{q^3}}\over {81}} \\&&\kankaku - 
  {{14849774715749017836223605677\,{q^4}}\over {108}} \\&&\kankaku - 
  {{3338292965542097491928389912718314496\,{q^5}}\over {3125}} \\&&\kankaku - 
  {{56095153127006342643814305613940936094080768\,{q^6}}\over {6075}} %\\&& - 
%  {{702968388071439386468130382018147005059492732486746112\,{q^7}}\over 
%    {8103375}} 
%\\&& - {{9743292565146133875472135939797158224848043283639434273686
%       1\,{q^8}}\over {112000}} \\&& - 
%  {{158258813257162425734838258725993053305553478435533125901960919225216\,
%      {q^9}}\over {17222625}} 
%\\&& - 
%  {{16144934998526753390353177286104281731140003602051987030911127055747198519
%       36\,{q^{10}}}\over {15946875}} \\&& - 
%  {{43944666942061325519763959944533472850487292716115570981763784637562622042
%       30925770752\,{q^{11}}}\over {3804829875}} \\&& - 
%  {{20731897014709207310244275507467243666178315463369917566314173689224034898
%       359001546952873928\,{q^{12}}}\over {1528220925}} \\&& - 
%  {{51490534081363349548807196958610730053799805977631862407286933215008763887
%       244120899146919262325150683136\,{q^{13}}}\over
%       {315232073281125}}
% \\&& - 
%  {{47605933600280672153093991460477351539885461288701327029388488275998095230
%       096238157169983344308342108475828224\,{q^{14}}}\over {23695945898625}} 
%   \\&& 
%- {{39125801690632050743873860408147509851941874311636565472750609191451528
%       3951297971167934098375928729663319950388625408\,{q^{15}}}\over 
%    {15543987834375}} 
%\\&& 
+ {{{\rm O}(q)}^{7}}\,\,,\\
&&a_6=
6995748257792\,{q^2} + 24640041160313671680\,{q^3} \\&&\kankaku + 
  {{3340137952685596994282796032\,{q^4}}\over {27}} \\&&\kankaku + 
  791276138613199166063943691707648\,{q^5} \\&&\kankaku + 
  {{544155970530490018382447436413165338674240832\,{q^6}}\over {91125}} 
%\\&& + 
%  {{285350073113785615439955127122770266788765638770688\,{q^7}}\over 
%    {5625}} 
%\\&& + {{14416042869681001099314808487982333377295334309644564787568\,
%      {q^8}}\over {30625}} 
%\\&& + {{11598703982762537510535783620960029828871566566
%       981788487653917817168\,{q^9}}\over {2480625}} 
%\\&& + 
%  {{82073575276296175590802522461306191224945218180871039171612897962904867961
%       596\,{q^{10}}}\over {1674421875}} \\&& + 
%  {{13314966774282455555338876013317360706558205643686039670754399016386964235
%       56803584\,{q^{11}}}\over {2480625}} \\&& + 
%  {{10754965848694733704592652691173664227966114571959538166420938452676087018
%       477065354193118966647\,{q^{12}}}\over {1765095168375}} \\&& + 
%  {{69899755234928543019067103890872714064301681514528857260725968989522465036
%       229679308081346925652079104\,{q^{13}}}\over {980608426875}} \\&& + 
%  {{43482366973627577467830226049125526987796493527037266424634137410926865437
%       568817877692450714561790538436143223552\,{q^{14}}}\over 
%    {50827803952550625}} 
%\\&& + {{1632892328317969924601643341490309425224253537577
%       51224707491917578406074305618658468198841098328184827688347366191562752
%       \,{q^{15}}}\over {15559531822209375}} 
%\\&& 
+ {{{\rm O}(q)}^{7}}\,\,,
\end{eqnarray*}
\newpage
\begin{eqnarray*}
&&\bullet\,\underline{\mbox{7 dimensional case}}\\
&&a_2
=28165644\,q + 1084202044892451\,{q^2} \\&&\kankaku + 
99728961580840263059520\,{q^3} \\&&\kankaku + 
  {{53757309252523450995715769878563\,{q^4}}\over 4} %\\&& + 
%  {{56284299353658113927262984269308065138144\,{q^5}}\over {25}} %\\&& + 
%  434419807125679239667173319517868586633913649936\,{q^6} %\\&& + 
%  {{4536964855203039928912892482460467950430913011761083268480\,{q^7}}\over 
%    {49}} 
%\\&& + {{3399698134612242456861064866609428655600127676447804557125992366
%       43\,{q^8}}\over {16}} 
%\\&& + 51626395023829761907727999146965854098607858729
%    43521769337910770566379800\,{q^9} 
%\\&& + 
%  {{32818121409562270792371932826304642124037196243498226468001850980264415922
%       389395576\,{q^{10}}}\over {25}} \\&& + 
%  {{41914725890887314215780087836547950304401765041328186887478423061022966019
%       540646801575171968\,{q^{11}}}\over {121}} \\&& + 
%  9427058978295988243591276060834889120601714065439204111456876958622052586457
%    3326696857435638210244\,{q^{12}} \\&& + 
%  {{34231033712991426267082661378692688990511783392947309890317231231446139720
%       2861199717430645891284349448509312\,{q^{13}}}\over {13}} \\&& + 
%  {{36850922071512896001352143155795735994276664122515260585535183729532838304
%       2984075865204809256948001987598173035554656\,{q^{14}}}\over
%       {49}} 
%\\&& + 
%  {{10948947046374218400878474650185681536064756443156109842883212571112031077
%       919763645460523472333847029309483773786698896347904\,{q^{15}}}\over 5} 
%   \\&& 
+ {{{\rm O}(q)}^{5}}\,\,,\\
&&a_3=
28552500\,q + {{4236848937896589\,{q^2}}\over 2} \\&&\kankaku + 
  227690801399286677815632\,{q^3} \\&&\kankaku + 
  {{527217434543398525112024561243277\,{q^4}}\over {16}} %\\&& + 
%  5751694811419775945909944108570564437060\,{q^5} %\\&& + 
%  {{5699949648952785541195182873222599720472647018322\,{q^6}}\over 5}% \\&& + 
%  {{424644952332325095873653254977952348054827197716607689645344\,{q^7}}\over 
%    {1715}} 
%\\&& + {{51631381394447009707172785048873180236482359722060492858315154
%       751451\,{q^8}}\over {896}} 
%\\&& + 
%  {{14861177800720309695284811668685252973882882327270447958871859031666210218
%       66\,{q^9}}\over {105}} \\&& + 
%  {{63525804548684773757982129214867016190519202261334651925547549124143011484
%       25196216123\,{q^{10}}}\over {1750}} \\&& + 
%  {{44938529278160129888436035666768705529914074911498628670726403467596876183
%       602639124023653649216\,{q^{11}}}\over {46585}} \\&& + 
%  {{73937032435413146665606083831760599731226153675732960568872113661578674738
%       20109738223271272267534163\,{q^{12}}}\over {28}} \\&& + 
%  {{81422622980581352567213637833020240729983604396897067208662377852339694305
%       8011660419151723762084184812819865408\,{q^{13}}}\over {10985}} \\&& + 
%  {{10417119828293895932556236887178833781904741614618149575283669053376255054
%       81695208714977023195628138632428292403721356\,{q^{14}}}\over
%       {49}} 
%\\&& + 
%  {{70670319118560576060845837294035324283935300110905722409721424426009601237
%       528371675458228562174192144143277663654627346460232512\,{q^{15}}}\over 
%    {11375}} 
%\\&& 
+ {{{\rm O}(q)}^{5}}\,\,,\\
&&a_4=
 -51371415\,q - {{411335645719671\,{q^2}}\over {16}} \\&&\kankaku + 
  91853523325764027808458\,{q^3} \\&&\kankaku + 
  {{5064421223684923731829011918579849\,{q^4}}\over {256}} %\\&& + 
%  {{2061079333859309931846996400734118605342243\,{q^5}}\over {500}} %\\&& + 
%  {{180988090280057769217946012986418717817100967293533\,{q^6}}\over {200}} 
%\\&& + 
%  {{12615683926006946997141718008749366659513784446048323262538574\,
%      {q^7}}\over {60025}} 
%\\&& + {{14705945765919964803592762846503586991872236674
%      70913675191993100792767\,{q^8}}\over {28672}} 
%\\&& + 
%  {{14096446996400450640803051167131773594449310188223943498955025287835873625
%       4781\,{q^9}}\over {10800}} \\&& + 
%  {{96346733389339512509084115565017122898909925125736442681026841821983576738
%       7750727447337\,{q^{10}}}\over {280000}} \\&& + 
%  {{47898634773212814714033686610403953027501275259880856385434554685929147105
%       2453535448436387044343\,{q^{11}}}\over {512435}} \\&& + 
%  {{70608046055027608220682497209045244606517533878952392348016596121296442012
%       4326926779744733004355208268979\,{q^{12}}}\over {2710400}} \\&& + 
%  {{17956514101104583757670573494748532907088265136084349736335499680315587722
%       5249367672494101327235900270374494887824751\,{q^{13}}}\over 
%    {2419116700}} 
%\\&& + {{14805844994599127277060472989066322301559990060393035752
%       81039944222572118151664559309608439798332604743946310342912049002857\,
%      {q^{14}}}\over {68668600}} \\&& + 
%  {{43836299761748378910719948621127123779409903398033766285824275063205762482
%       130888967542330899939953675416655155223712613079406504788\,{q^{15}}}
%     \over {6881875}} 
%\\&& 
+ {{{\rm O}(q)}^{5}}\,\,,\\
&&a_5=
6391872\,q - 2205705362512023\,{q^2} \\&&\kankaku -
 196569129709703650356786\,{q^3} \\&&\kankaku - 
  {{733324518503477784502808861028159\,{q^4}}\over {32}} %\\&& - 
%  {{81916031215218439525066575939355571326588149\,{q^5}}\over {25000}} 
%\\&& - 
%  {{1083212711339382277697421835255211058418585339326357\,{q^6}}\over 
%    {2000}} 
%\\&& - {{20917952631849815818560574729917692583880160651504207042561231
%       1\,{q^7}}\over {2100875}} \\&& - 
%  {{711160256135285576982346142193535060816072888110542313172754915840349\,
%      {q^8}}\over {35840}} 
%\\&& - {{27290613989421542843330703106812778971745732488
%       42656755299656719972717275667919\,{q^9}}\over {648000}} \\&& - 
%  {{75198087685383541265653651039053654843960446841166952410734599726841236691
%       1429759541013\,{q^{10}}}\over {800000}} \\&& - 
%  {{61593465295259674881860630388465441879338790619982353901655077307686488086
%       387830506349185230913869\,{q^{11}}}\over {281839250}} \\&& - 
%  {{31332821723930224993855311463346902156746695277189213696834752799400709044
%       202481923241836744885755699583433\,{q^{12}}}\over {596288000}} \\&& - 
%  {{89886086233471663092076434559138069847155398295877866720467502659327165262
%       407778297680159026259637448991403683923403039\,{q^{13}}}\over 
%    {6918673762000}} 
%\\&& - {{34768430608166918771314421893327482409347650204639651
%       02974051231173228331200179819024836350891722895152888688940043180431535
%       23\,{q^{14}}}\over {105749644000}} \\&& - 
%  {{20454817255300325373880356578289440955770047771216771227951807180406451992
%       3255344800008349199345751821154024700915980152621019677000897\,{q^{15}}
%      }\over {241106490625}} 
%\\&& 
+ {{{\rm O}(q)}^{5}}\,\,,\\
&&a_6=
83567214\,q + {{70266259194158727\,{q^2}}\over {32}} \\&&\kankaku + 
  {{977590470089357596666733\,{q^3}}\over {12}} \\&&\kankaku + 
  {{6148546275722300412022432329036039\,{q^4}}\over {2048}} %\\&& - 
+ {{{\rm O}(q)}^{5}}\,\,,\\
&&a_7=
-167134428\,q - {{10921247520838407\,{q^2}}\over {32}} \\&&\kankaku + 
  {{2976368115211624055130173\,{q^3}}\over {36}} \\&&\kankaku + 
  {{58982722644231416288220304066419321\,{q^4}}\over {4096}} %\\&& + 
+ {{{\rm O}(q)}^{5}}\,\,,\\
\end{eqnarray*}
\newpage
\begin{eqnarray*}
&&\bullet\,\underline{\mbox{8 dimensional case}}\\
&&a_2
=412077600\,q + 315199135995975000\,{q^2} \\&&\kankaku + 
  629888601165740265184000000\,{q^3} \\&&\kankaku + 
  1929728022282541043288456451883593750\,{q^4} %\\&& + 
%  7555365761413714666504687499453707721041903104\,{q^5} %\\&& + 
%  34712030698542837689181530555631282093818475577918272000\,{q^6} %\\&& + 
%  {{8747407819652488511374140845171732585689410179686985130763776000000\,
%      {q^7}}\over {49}} 
%\\&& + {{19968345418938122708247803886350783203962676223729
%       49905255241737688321796875\,{q^8}}\over 2} \\&& + 
%  {{53622555188317944995780447909867783531807346828739960537736284435017136615
%       556500000000\,{q^9}}\over 9} \\&& + 
%  3743958299573381220556828449436346607431069338546046199780640015830722083826
%    7283772616568464000\,{q^{10}} \\&& + 
%  2453859292230454376912822516472146233835010180591890786298602867194924915571
%    38658960867952330703595520000\,{q^{11}} \\&& + 
%  1665602438046948613330458454162645448942539086873572753094760809186761455482
%    645769243680707135138677105940130000000\,{q^{12}} \\&& + 
%  {{19679573568091615921937186139653407732439205974640294694972646972034252390
%       21091433917677062496371001621169489612053028800000000\,{q^{13}}}\over 
%    {169}} 
%\\&& + {{584476121428528931008449955595724844833419118947582285185464108
%       53874481260829604907324164552772006867231060319846418743004568000000000
%       0\,{q^{14}}}\over 7} 
%\\&& + 611943822260031205662234503270443804942872682543
%    35993454741505920511574689173417020560880122873447208524885717645693673252
%    9433067585688207360000\,{q^{15}} 
%\\&& 
+ {{{\rm O}(q)}^{5}}\,\,,\\
&&a_3=
703340000\,q + 850719962854822500\,{q^2} \\&&\kankaku + 
  {{51874151170414241611448000000\,{q^3}}\over {27}} \\&&\kankaku + 
  {{12471282323077747047340092501522265625\,{q^4}}\over 2} %\\&& + 
%  25252139855280593326477625889562566709168446720\,{q^5} %\\&& + 
%  {{355860915675552447256772751949347336006867055552867912000\,{q^6}}\over 
%    3} 
%\\&& + {{2125592634648404919156758067778391779968013982402894893389334272000
%       00\,{q^7}}\over {343}} 
%\\&& + 
%  {{56107686318960467053075822773164421297788319872449035192206830764398722265
%       625\,{q^8}}\over {16}} \\&& + 
%  {{15394412443104529274570741417990443116125141714112783184504322424663023932
%       585120250000000\,{q^9}}\over {729}} \\&& + 
%  1336606983679396560903778220545808362671240902355737968228107921210572912714
%    60427520827232826080\,{q^{10}} \\&& + 
%  {{10662820892965602178229053858510199942218629322147889960188354005583508977
%       0428266829906613402844721338880000\,{q^{11}}}\over {121}} \\&& + 
%  {{16229488536461961822732702438364909147509723722469036058212893974529444591
%       8594213602097757198280972969338527296625000\,{q^{12}}}\over
%       {27}}
% \\&& + 
%  {{92710787714642543342784178411724928050225593550927964451849268489942190774
%       352945045165935665227757535288765661950046369600000000\,{q^{13}}}\over 
%    {2197}} 
%\\&& + {{13540023587191683390098319701023566443745714863479081881113261
%       73345059191916312346625125531124724782878253066771957496436230285490000
%       000000\,{q^{14}}}\over {4459}} \\&& + 
%  {{60942770597327108132273783146592193537611686892802185665729971933412979406
%       29263364510605938803335456471146466929200119973791635928082721920860160
%       00\,{q^{15}}}\over {273}} 
%\\&& 
+ {{{\rm O}(q)}^{5}}\,\,,\\
&&a_4 =
-569448000\,q + 478736404224952500\,{q^2} \\&&\kankaku + 
  {{16494841023888735724876000000\,{q^3}}\over 9} \\&&\kankaku + 
  {{28605286088764930942867547855692578125\,{q^4}}\over 4} %\\&& + 
%  {{159436536943890148186066491415264599484125556416\,{q^5}}\over 5} %\\&& + 
%  {{476975397868140051857147035314007145252508728147370065200\,{q^6}}\over 
%    3} 
%\\&& + {{2077115433576379613550983224494416085099867273199299514564996761600
%       000\,{q^7}}\over {2401}} 
%\\&& + 
%  {{22589349715163064909719013485738557293249391696248480375840025392485510760
%       46875\,{q^8}}\over {448}} \\&& + 
%  {{22638539665128575243151418294925949824576420104862239912623706765717565219
%       889669793750000\,{q^9}}\over {729}} \\&& + 
%  {{12603894947838284558039862858687346525248905363085679220800138344349923903
%       362586278823731198495212\,{q^{10}}}\over {63}} \\&& + 
%  {{12465735464191483073784103447457424070533395742021291010986192614189377584
%       004417049530596712381933361547136000\,{q^{11}}}\over {9317}} \\&& + 
%  {{52358201101819649975342025250493628310381859214166535019685143618051081479
%       96041601884262978519786363511941651226734375\,{q^{12}}}\over
%       {567}} 
%\\&& + 
%  {{13089202046635509220670958490880952803685010080150350382801241966782903981
%       136703752638843520622478145819524301586275037510820000000\,{q^{13}}}
%     \over {199927}} 
%\\&& + {{19277190858839023892188515207636489099368377399832380
%       03311578254269200042108143816510288557587652004186194861173410395984348
%       72897996000000000\,{q^{14}}}\over {405769}} \\&& + 
%  {{26218219625785188663513735375559539567717491569581580668536016130888681533
%       61377201325409931505255479058336790099590255582398994919281709769741930
%       49600\,{q^{15}}}\over {74529}} 
%\\&& 
+ {{{\rm O}(q)}^{5}}\,\,,\\
&&a_5 =
-813753600\,q - 977980727099722500\,{q^2} \\&&\kankaku - 
  {{106469171938587328005278000000\,{q^3}}\over {81}} \\&&\kankaku - 
  {{55187753006324839405011876860910546875\,{q^4}}\over {24}} %\\&& - 
+ {{{\rm O}(q)}^{5}}\,\,,\\
&&a_6=
2196955200\,q + 186121212133484375\,{q^2} \\&&\kankaku - 
  {{201227517927600187399838000000\,{q^3}}\over {243}} \\&&\kankaku - 
  {{6638610338058374343171115466159769921875\,{q^4}}\over {1728}} %\\&& - 
\\&&\kankaku + {{{\rm O}(q)}^{5}}\,\,,\\
&&a_7 =
-2196955200\,q + {{1935676126330835625\,{q^2}}\over 2} \\&&\kankaku + 
  {{1489516428640533589720655000000\,{q^3}}\over {729}} \\&&\kankaku + 
  {{36829673757854084975640013264359569921875\,{q^4}}\over {6912}} %\\&& + 
+ {{{\rm O}(q)}^{5}}\,\,,\\
&& a_8 =
-1639796995499520000\,{q^2} \\&&\kankaku - 1876688672317127734873500000\,{q^3} \\&&\kankaku - 
  {{282861147153465890226759614755375000000\,{q^4}}\over {81}} %\\&& - 
+ {{{\rm O}(q)}^{5}}\,\,.
\end{eqnarray*}

\newpage
\section*{Appendix B}
\section*{Disk Amplitudes for Calabi-Yau $d$-Fold}\label{example}
We consider a $d$-dimensional Fermat type Calabi-Yau manifold M
with one K{\"a}hler modulus
\ba
&&\hspace{10mm} 
\mbox{M} \,;\, X_1^{l_1}+ X_2^{l_2}+\cdots + X_{d+2}^{l_{d+2}}=0 \nom \\
&& \hspace{15mm}  \mbox{in} \,\, 
{\bf P}_{d+1}[{w_1},{w_2},\cdots ,{w_{d+2}}](D)\nom \\
&& \hspace{15mm} D:=\sum_{i=1}^{d+2} w_i \,\,,\,\,\,\,\,
l_i :=\frac{D}{w_i}\,\,,\,\,\,\,\, w_{d+2} :=1\,\,.\nom 
\ea
Products of a K{\"a}hler form $e$ generate
analytical subspaces of vertical
cohomologies $\oplus_l \mbox{H}^{l,l}_J$(M). 
Each $\mbox{H}^{l,l}_J$ with a fixed $l$ is spanned by
one element $e^l$ and there exists one corresponding
operator $\co^{(l)}$ $(l=0,1,\cdots ,d)$ in the A-model.
Fusion relation of these is the same as Eq.(\ref{dfusion})
\ba
&&\co^{(1)}\co^{(j-1)} = \Kappa_{j-1} \co^{(j)} \,\,\,\,\,\,\,\,
(1\leq j \leq d)\,\,\,,\nom\\
&& \co^{(1)}\co^{(d)} = 0\,\,\,,\nom \\
&& \Kappa_{\ell} :={\Kappa_{1\ell}}^{\ell +1}=
{\Kappa_{1\ell m}}\Eta^{m \,\ell+1}\,\,\,,\nom \\
&& \Eta_{ij} :=\ketbra{\co^{(i)}\co^{(j)}}=D\cdot \delta_{i+j ,d}\,\,\,.\nom
\ea
A K{\"a}hler parameter $t$ couples with 
the operator $\co^{(1)}$.
It is a mirror map itself and is written by using di-gamma function 
$\Psi(x):=\dis \frac{{\Gamma}'(x)}{\Gamma (x)}$
\ba
t(\psi)=\frac{D}{2\pi i}\Biggl[\,\,\log (D\psi)^{-1} 
+\frac{\dis\sum_{n=1}^{\infty}
\frac{(Dn)!}{%\dis 
\prod_{i=1}^{d+2} (w_i n)!}
\biggl(
\Psi (Dm+1)-\sum_{j=1}^{d+2}\frac{1}{l_j}\Psi (w_i n+1) 
\biggr)
\cdot (D\psi)^{-n} }
{\dis \sum_{m=1}^{\infty} \frac{(Dm)!}
{%\dis 
\prod_{i=1}^{d+2} (w_i m)!}
\cdot (D\psi)^{-m}}
\,\,\Biggr]\,\,\,.\nom
\ea
Here the $\psi$ is a complex moduli parameter of a mirror W
in the B-model.
We can choose $d+1$ homology cycles 
$\gamma_l \in \oplus_{m=0}^{l} \mbox{H}_{2m}$(M) and
express disk amplitudes $c^{(0)}(\gamma)$'s associated with $\co^{(0)}$ 
by a row vector $u_0$
\ba
u_0 &:=& \left(
\begin{array}{cccc}
c^{(0)}(\gamma_0) &c^{(0)}(\gamma_1) &\cdots & c^{(0)}(\gamma_d)
\end{array}
\right) \nom \\
&=& \left(
\begin{array}{ccccc}
\hat{a}_0 & \hat{a}_1 & \hat{a}_2 & \cdots & \hat{a}_d
\end{array}
\right) \cdot \exp (t {\cal N})\nom \\
&&\!\!\!\left\{
\begin{array}{ccl}
\hat{a}_0 & = &1\,\,\,,\,\,\, \hat{a}_1 =0\,\,\,,\\
\hat{a}_n & := & S_n (0,\tilde{y}_2 , \cdots \tilde{y}_n )\,\,\,\,\,\,\,
\,\,\,(n=2,3,\cdots ,d)\,\,\,,
\end{array}
\right.\,\,\,\nom \\
&&\tilde{y}_m := \frac{1}{m!} \left(\frac{1}{2\pi i}\frac{\del}{\del \rho}
\right)^m \log \Biggl[\,\,
\sum_{l=0}^{\infty} \frac{\Gamma (D(l+\rho)+1)}{\Gamma (D\rho +1)}
\cdot \prod_{i=1}^{d+2}
\frac{\Gamma ( w_i \rho +1)}{\Gamma ( w_i (l+\rho) +1)}
\cdot
(D\psi)^{-Dl}\,\,
\Biggr]\Bigg|_{\rho =0}\nom \\
&&\hspace{10cm}
(m=2,3,\cdots ,d)\,\,\,\,\,.\nom \\
{\cal N} &:=& \left(
\matrix{
0 & 1 &   &   &   & 0 \cr
  & 0 & 1 &   &   &  \cr
  &   & 0 & 1 &   &  \cr
  &   &   &  \ddots & \ddots  & \cr
  &   &   &   & 0 & 1 \cr
0 &   &   &   &   & 0 }
\right)\,\,\,.\nom
\ea
The $S_n$'s are Schur polynomials.
The set of homology cycles is transformed under a translation
$t\rightarrow t+1$
\ba
&&(\gamma^{(0)} \,\,\gamma^{(1)} \,\, \cdots \,\,
\gamma^{(d-1)} \,\,\gamma^{(d)} )\nom \\
&&\rightarrow 
(\gamma^{(0)} \,\,\gamma^{(1)} \,\, \cdots \,\,
\gamma^{(d-1)} \,\,\gamma^{(d)} )\cdot
\exp ({\cal N}) \,\,\,.\nom 
\ea
and these cycles with different dimensions mix each other.
The total disk amplitude is collected into
a matrix ${\Pi}$
\begin{eqnarray*}
&&{\Pi}:=\left(\matrix{ {u}_{0} \cr 
{u}_{1} \cr 
\vdots \cr 
{u}_{d}}
\right)\,\,\,\,,\\
&& {u}_{l} =\frac{1}{\Kappa_{l-1}} {\partial_t}
\frac{1}{\Kappa_{l-2}} {\partial_t} \cdots {\partial_t}
\frac{1}{\Kappa_{1}} {\partial_t} \frac{1}{\Kappa_{0}} {\partial_t} 
{u}_0 \,\,\,\,\,\,\,\,(1\leq l \leq d)\,\,\,\,,\\
&& \left\{
\begin{array}{l}
\Kappa_{m}=  {\dis {\partial_t} \frac{1}{\Kappa_{m-1}}{\partial_t} 
\frac{1}{\Kappa_{m-2}}{\partial_t} 
\cdots
{\partial_t} \frac{1}{\Kappa_{1}}
{\partial_t} \frac{1}{\Kappa_{0}}{\partial_t} {c^{(0)}}(\gamma_{m+1})
}
\,\,\,\,(1\leq m \leq d-1)\,\,\,\,,\\
 \Kappa_0 =1 \,\,\,.
\end{array}
\right.
\end{eqnarray*}

%%%%%%%%%%%%%%%%%%%%%%%%%%%%%%%%%%%%%%%%%%%%%%%%%%%%%%%%%%%%%
\end{document}